\documentclass[floatfix,amssymb,twocolumn, superscriptaddress,nofootinbib, aps]{revtex4-1}

\usepackage{amssymb,amsmath,verbatim,mathtools,needspace,enumitem,etoolbox,graphicx,physics,microtype,afterpage,bm,soul}

\usepackage[dvipsnames]{xcolor}
\definecolor{linkcolor}{rgb}{0.0,0.3,0.5}
\usepackage[unicode, colorlinks=true, linkcolor=linkcolor, citecolor=linkcolor, filecolor=linkcolor,urlcolor=linkcolor, pdfusetitle]{hyperref}
\usepackage{orcidlink}
\usepackage[all]{hypcap}
\usepackage[T1]{fontenc}
\usepackage[utf8]{inputenc}
\usepackage{tabularx}
\usepackage{float}
\renewcommand{\dd}{\mathrm{d}}
\newcommand{\ii}{\mathrm{i}}
\newcommand{\ee}{\mathrm{e}}
\newcommand{\bmr}{\bm r}
\newcommand{\bmt}{\boldsymbol{\theta}}
\newcommand{\bb}{\bm b}
\newcommand{\bo}{\bm b_{\rm o}}
\newcommand{\zsl}{z_{\rm sl}}
\newcommand{\zlo}{z_{\rm lo}}
\newcommand{\zso}{z_{\rm so}}

\newcommand{\llp}{\left [}
\newcommand{\rrp}{\right ]}
\newcommand{\lp}{\left (}
\newcommand{\rp}{\right )}
\newcommand{\nn}{\nonumber}

\newcommand{\be}{\begin{equation}}
\newcommand{\eeq}{\end{equation}}

\newcommand{\refrep}[1]{\textcolor{black}{#1}}

\begin{document}

\title{\bf Wave-optics gravitational wave lensing in modified gravity}

\author{Alice Garoffolo}
\email{aligaro@sas.upenn.edu}
\affiliation{Center for Particle Cosmology, Department of Physics and Astronomy, University of Pennsylvania 209 South 33rd Street, Philadelphia, Pennsylvania 19104, USA}

\author{Gianmassimo Tasinato}
\email{g.tasinato2208.at.gmail.com}
\affiliation{Department of Physics, Swansea University, Swansea SA2 8PP, United Kingdom, \\Dipartimento di Fisica e Astronomia, Universit\`a di Bologna, Via Irnerio 46, 40126 Bologna, Italy}

\begin{abstract}
\noindent
We initiate the study of gravitational-wave lensing in the wave-optics regime within modified gravity. We consider a phenomenological setup in which the gravitational-wave amplitude obeys a curvature-coupled propagation equation. This framework reproduces the standard general relativity (GR) behavior in the geometric-optics regime, while leading to qualitatively different infrared dynamics. In particular, the usual argument implying that the amplification factor approaches unity in the zero-frequency limit no longer applies. This is due to the persistence of curvature-induced interactions in the infrared, which modify the natural propagation basis itself. As a result, the standard Fresnel treatment ceases to be valid at sufficiently low frequency. The correct infrared regime is instead controlled by an interacting static Green's  function, with a finite-frequency completion provided by a partial-wave formulation. We show that this structure admits an equivalent distorted-wave interpretation, in which the curvature interaction is absorbed into a dressed reference propagation basis, while the residual lensing effect is encoded in finite-frequency phase shifts. We further demonstrate that these phenomena admit a natural interpretation in the language of scattering amplitudes. Wave-optics lensing can therefore probe propagation-level departures from GR that remain entirely invisible in geometric optics.
\end{abstract}

\maketitle



\section{Introduction}

General relativity (GR) has passed all experimental and observational tests to date with remarkable success \cite{Will:2014kxa}. Nevertheless, gravity remains conceptually distinct from the other fundamental interactions, and its embedding within a consistent quantum and cosmological framework is still incomplete. Problems such as the cosmological constant puzzle, the ultraviolet completion of gravity, and the nature of dark sectors continue to motivate the exploration of controlled deviations from GR and of new observational probes capable of testing them \cite{Joyce:2014kja}.
Any new methods aiming at testing properties of gravity are then welcome for improving  our understanding
of gravitational theories  and their experimental consequences. 

\smallskip

Gravitational waves (GWs) provide an especially powerful arena for this program. Their direct detection has opened a new observational window on strong-field gravity, compact objects, and cosmology \cite{Barack:2018yly}. Future ground-based and space-based detectors will extend this reach over a much broader range of frequencies and propagation distances \cite{ET:2025xjr,LISA:2024hlh}, making it increasingly important to identify observables that are directly sensitive to modifications of GW propagation.

\smallskip
In this work
we focus on gravitational-wave lensing in the wave-optics regime (WO)~\cite{Schneider:1992bmb} and extend its description in a generic beyond GR theory. Although no unambiguous detection of GW lensing has yet been established~\cite{ LIGOScientific:2021izm, LIGOScientific:2023bwz}---possibly with the exception of the candidate event GW231123~\cite{LIGOScientific:2025rsn, Goyal:2025eqo, Chan:2025kyu, Hu:2025lhv, Chakraborty:2025pxt}---lensing is expected to become increasingly relevant for future GW observations~\cite{Li:2018prc, Wierda:2021upe, Ng:2017yiu, Sereno:2010dr, Gutierrez:2025ymd, Hannuksela:2019kle}. 

In the geometric-optics (GO) regime, gravitational-wave lensing has been extensively studied, both in GR~\cite{Isaacson:1968hbi, Isaacson:1968zza, Bertacca:2017vod, Laguna:2009re} and in modified gravity scenarios, where it has been shown to provide a sensitive probe of deviations from Einstein’s theory~\cite{Ezquiaga:2020dao, Streibert:2024cuf, Menadeo:2024uoq, Garoffolo:2019mna, Flanagan:2008kz, Balaudo:2022znx, Balaudo:2023klo, Garoffolo:2020vtd, Dalang:2019rke, Dalang:2020eaj, Tasinato:2021wol, Mpetha:2022xqo}.

By contrast, the WO regime is technically more involved and has therefore been explored primarily within GR and for specific lens configurations. In this regime, when the GW wavelength becomes comparable to the characteristic scale of the lens, diffraction and interference effects become important.  Lensing is described by a frequency-dependent complex amplification factor, rather than by a set of independent geometric-optics images. The standard framework is provided by the diffraction integral, a highly oscillatory integral over the lens plane, which coherently sums the contributions of different propagation paths~\cite{Nakamura:1999uwi,Nakamura:1997sw,Takahashi:2003ix,Takahashi:2004mc, Jow:2022pux,Leung:2023lmq}. In the high-frequency limit this integral reduces to the usual geometric-optics sum over stationary images, while at finite frequency it captures diffraction, interference, and phase modulation in a single amplification factor.

The growing observational relevance of wave-optics lensing has motivated several numerical approaches for evaluating the diffraction integral efficiently~\cite{Feldbrugge:2019fjs,Diego:2019lcd,yeung2024wolensing,Villarrubia-Rojo:2024xcj,Tanaka:2023mvy,Urrutia:2024pos,Urrutia:2023mtk,Urrutia:2021qak,Jung:2017flg, Yeung:2021chy, Tambalo:2022plm}. These methods have been applied to a variety of lens models and astrophysical scenarios, including compact-object and extended lenses~\cite{Tambalo:2022wlm,Caliskan:2022hbu,Suvorov:2021uvd,Braga:2024pik,Savastano:2023spl}, rotating lenses~\cite{Bonga:2024orc,Kubota:2023dlz,Li:2022izh}, binary or multilens systems~\cite{Mehrabi:2012dy,Feldbrugge:2020ycp,Feldbrugge:2020tti,Diego:2019rzc,Ramesh:2021nnl,Yamamoto:2003cd,Amoruso:2026txw,Zumalacarregui:2026uqs}, and dark-matter-motivated lenses~\cite{Singh:2025uvp}. Wave-optics effects have also been used to constrain primordial-black-hole abundances~\cite{Sugiyama:2019dgt,Diego:2019rzc}, to assess detection prospects~\cite{Yeung:2021chy,Chan:2024qmb,Mishra:2023ddt} and perform parameter inference for future observatories such as LISA~\cite{Caliskan:2023zqm,Caliskan:2022hbu,Grespan:2023cpa,Lai:2018rto,Suyama:2025gbh}. \refrep{The interplay between unlensed waveforms in modified gravity and lensed waveforms has also been considered in~\cite{Liu:2024xxn}}.
Controlled weak-field and Born-type approaches provide complementary analytic treatments~\cite{Takahashi:2005sxa,Takahashi:2005ug,Oguri:2020ldf,Yarimoto:2024uew,Garoffolo:2022usx,Macquart:2004sh}.

\smallskip

In GR, the structure of the wave equation implies that the lensing interaction is proportional to $\omega^2$, where $\omega$ is the GW frequency. As a consequence, the low-frequency limit is effectively free, and the amplification factor, defined as the ratio between the full wave and a free spherical wave, approaches unity as $\omega\to0$. This property underlies the consistency of the standard diffraction-integral description across frequencies. In this work we show that this behavior is not generic. For a broad class of curvature-dependent modifications of the propagation equation, the effective interaction survives in the infrared, so that the low-frequency problem remains interacting. Such a phenomenon can lead to sizable  consequences for GW lensing.

\smallskip

We develop this idea in a concrete phenomenological setup described by
\be
(g^{\mu\nu} \nabla_\mu \nabla_\nu - \xi R)\phi=0 \, ,
\eeq
with $\xi$ a dimensionless coupling constant, parametrizing curvature-induced corrections to the wave equation. These corrections are suppressed in the GO limit, where the wavelength is much smaller than the curvature scale and the kinetic term dominates, but become relevant in the WO regime. Crucially, the curvature-induced term survives in the static limit, and is therefore responsible for the modified infrared behavior.

This has immediate consequences for the standard wave-optics framework. The diffraction-integral formulation relies on a Fresnel (eikonal/paraxial) approximation, which assumes that radial variations of the amplitude are subleading. In the present setup this assumption breaks down at sufficiently low frequency, precisely because the interaction remains active in the infrared. As a result, the Fresnel description is not uniformly valid, and its naive extrapolation leads to an unphysical behavior of the amplification factor, which tends to zero in the static limit, effectively removing the wave, instead of approaching a constant as in GR.

\smallskip

To determine the correct infrared behavior, one must therefore return to the full wave equation. For concreteness, we consider a singular isothermal sphere (SIS) lens. This choice is motivated both by its astrophysical relevance~\cite{Robertson:2020mfh} and by the fact that, in this case, the curvature coupling generates an inverse-square interaction. The resulting static problem can be solved analytically in terms of a partial-wave Green's  function, which provides the appropriate interacting infrared limit, and shows explicitly that the propagation is no longer described by a free spherical wave.

We then construct a finite-frequency description that interpolates between this infrared regime and the standard high-frequency limit. On the one hand, we develop a controlled expansion at small frequencies around the interacting partial-wave solution. On the other hand, we analyze the eikonal regime by defining the amplification factor with respect to a curvature-dressed reference wave, instead of a free spherical wave. In this formulation, the amplification factor reduces to the usual SIS result of GR in the eikonal limit, while the difference with respect to the standard treatment is entirely encoded in the choice of reference wave.

Finally, we show that the same structure admits an interpretation in terms of  scattering amplitudes~\cite{Vishveshwara:1970zz, Bai:2016ivl, Chi:2019owc, Bjerrum-Bohr:2014zsa, Bastianelli:2021nbs, Comberiati:2024uuc, Cangemi:2023bpe, Ivanov:2024sds, Correia:2024jgr, Caron-Huot:2025tlq, Chen:2022clh, Bautista:2021wfy, Bautista:2022wjf, Bjerrum-Bohr:2025bqg, Akpinar:2025byi, DiVecchia:2023frv, KoemansCollado:2019ggb, Aoude:2022thd, Chiodaroli:2021eug, Aoude:2022trd, Newton:1982qc, 1988sfbh.book.....F, Casals:2016soq, Leaver:1986gd, Bellazzini:2022wzv, Bini:2025ltr, Bini:2025bll, Ivanov:2025ozg}. 
The relation between lensing and scattering processes has been explored in several contexts~\cite{Turyshev:2018gjj, Li:2025lvl, Motohashi:2021zyv, Kubota:2024zkv, Kubota:2023dlz, Nambu:2015aea, Nambu:2019sqn, Pijnenburg:2024btj, Chan:2025wgz,Saketh:2025cwf}, and a one-to-one correspondence between the diffraction integral and the eikonal scattering amplitude~\cite{Glauber1959, Wallace1973, Swift1974, Levy1969, Chen1984, Byron1973} has been formulated in~\cite{CarrilloGonzalez:2025gqm}. Here we show that such relation persists in modified gravity, but must be formulated relative to the curvature-dressed propagation basis. In this language, the curvature-dependent interaction is absorbed into the reference problem, while the remaining finite-frequency effects are encoded in residual phase shifts. Related analyses of phase shifts in modified gravity and effective theories can be found in~\cite{Carrillo-Gonzalez:2021mqj, Brandhuber:2024qdn, Brandhuber:2024bnz, Nie:2024pby, AccettulliHuber:2020oou}.

\smallskip

Our results indicate that wave-optics gravitational lensing can probe modifications of GW propagation that remain essentially invisible in geometric-optics. In particular, even small curvature-dependent corrections can produce order-one distortions of the interference pattern at sufficiently low frequency.

\smallskip

Our work is organized as follows. In Sec.~\ref{sec_GR} we review the standard GR treatment of wave-optics GW lensing and emphasize why the usual infrared limit is free. In Sec.~\ref{sec_beyGRtheory} we motivate curvature-coupled propagation as a phenomenological description of beyond-GR effects and specialize to the SIS lens. In Sec.~\ref{sec:GO} we analyze the modified Fresnel integral and its geometric-optics limit. In Sec.~\ref{sec:Fresnel_breakdown} we show why the low-frequency suppression of the Fresnel result is not physical but signals the breakdown of the paraxial approximation. In Sec.~\ref{sec:static} we develop the interacting static Green's  function and its finite-frequency partial-wave completion. In Sec.~\ref{sec:distorted_reference} we reformulate the problem in terms of a distorted reference wave and connect the result to eikonal scattering.  In Sec.~\ref{sec_numest} we present numerical estimates of the resulting effects for LISA-relevant configurations.  We conclude by discussing the theoretical and phenomenological implications of our findings.

\smallskip

We use units with $c=1$. The GW frequency is $\omega$. The source position is $\bmr_s$, with $r_s=|\bmr_s|$. The lens-plane coordinate is $\bb$, with $b=|\bb|$. Distances $\zsl$, $\zlo$, $\zso$ denote source--lens, lens--observer, and source--observer separations.

\section{Standard GW lensing in GR}
\label{sec_GR}

Wave-optics gravitational lensing provides a controlled framework to describe diffraction, interference, and the transition to geometric optics in GW propagation through a lensing potential.  To fix our conventions, we start with a brief review of WO lensing of GW in GR, mainly following  the treatment of~\cite{Nakamura:1999uwi}.

\begin{figure}
    \centering
    \includegraphics[width=0.9\linewidth]{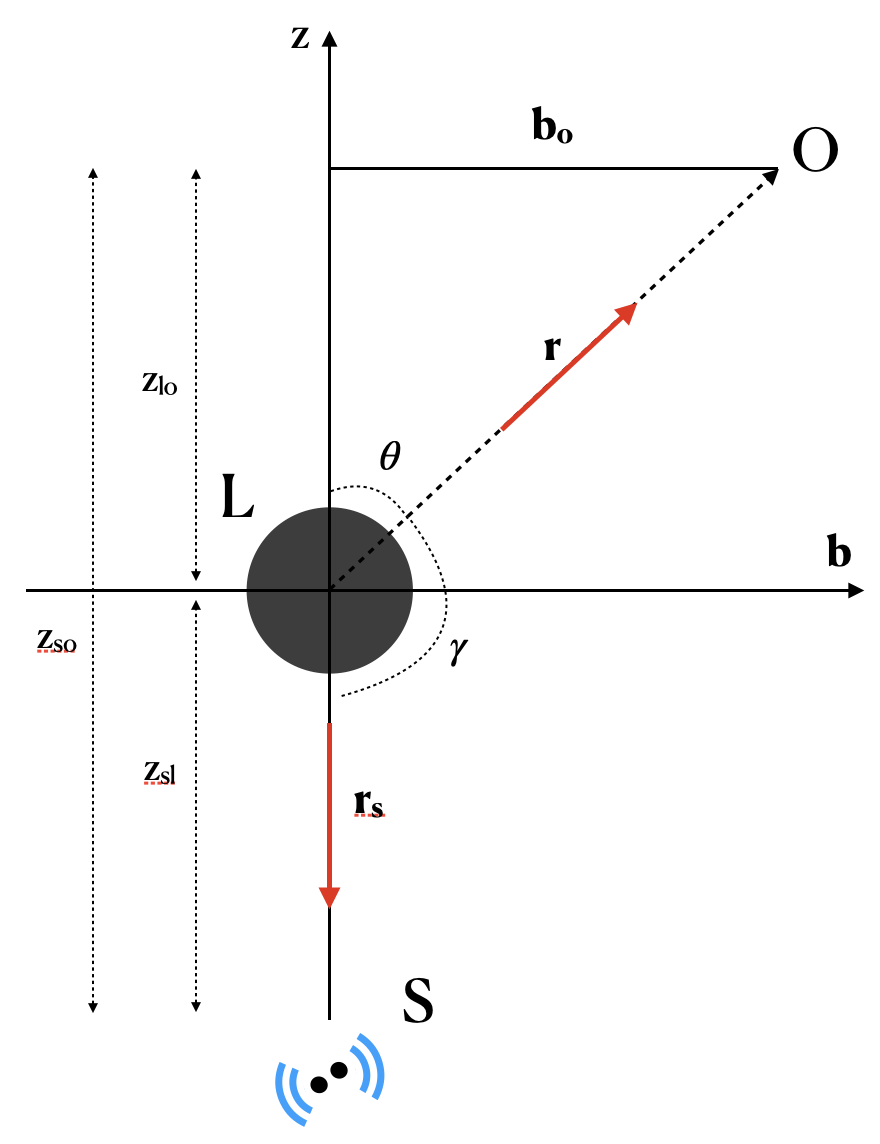}
    \caption{Lensing geometry. The lens is placed at the origin, the source lies on the negative $z$ axis, and the observer is located on the positive $z$ side at $O$. Figure inspired by~\cite{CarrilloGonzalez:2025gqm}.}
    \label{fig:LensGeom}
\end{figure}

In the standard lensing framework, the lens is modeled as a static Newtonian potential. One also usually assumes that the polarization of the GW is not significantly affected by the lensing event, so that the tensor structure can be factorized from the scalar propagation amplitude. Polarization-dependent corrections can arise beyond this approximation, for instance in Born-type treatments~\cite{Garoffolo:2022usx}, but they are suppressed in the geometric-optics limit~\cite{Isaacson:1968hbi,Isaacson:1968zza,Dalang:2021qhu}. A full treatment of polarization effects in wave optics is technically more involved, although related analyses have been carried out for black-hole lenses~\cite{Braga:2024pik,Kubota:2024zkv,Chan:2025wgz}. We therefore adopt a scalarized description of the tensor amplitude,
\be\label{dec_scal}
    h_{\mu\nu}=\phi\,e_{\mu\nu} \,,
\eeq
where the polarization tensor $e_{\mu\nu}$ is parallel transported, and all nontrivial propagation effects are encoded in $\phi$. The resulting scalar-wave equation on a weak-field background is 
\be
    \frac{1}{\sqrt{-g}}\partial_\mu \lp  \sqrt{-g}g^{\mu\nu}\partial_\nu \phi \rp=0 \,,
\eeq
with metric
\be\label{eq:Metric}
    \dd s^2 =  -(1+2U_{\rm L})\dd t^2 + (1-2U_{\rm L})\dd \bmr^2 \,.
\eeq
We use coordinates $x^\mu=(t,\bmr)$. When convenient, we write the spatial position in spherical coordinates centered on the lens as $\bmr=(r,\bmt)$, with $\bmt=(\theta,\varphi)$. We  also use cylindrical coordinates centered on the lens, $\bmr=(\bb,z)$, where $\bb$ denotes the two-dimensional transverse position in the plane orthogonal to the optical axis, as illustrated in Fig.~\ref{fig:LensGeom}.

In Eq.~\eqref{eq:Metric}, $U_{\rm L}$ is the Newtonian potential of the lens such that $|U_{\rm L}|\ll1$, which we also consider time-independent $U_{\rm L} = U_{\rm L}( \bmr )$. To first order in $U_{\rm L}$, after Fourier transforming
\be
    \phi(t,\bmr)=\ee^{-\ii\omega t}\phi_\omega(\bmr),
\eeq 
we obtain the following equation of motion
\be\label{eq:helmholtz_GR}
    \llp  \nabla^2_{\bmr}+\omega^2(1-4U_{\rm L}) \rrp \phi_\omega(\bmr)=0 \,.
\eeq
For a distant point source, we define the GW amplification factor $F$ through
\be\label{eq:F_def}
    \phi_\omega(\bmr) \equiv F( \bmr,\omega) \, \times \, \frac{ \ee^{\ii\omega| \bmr -\bmr_s|} }{ |\bmr -\bmr_s| }\,,
\eeq
thus assuming a spherical-wave for the emitted wave, compatibly with a point source at position $\bmr_s$. As  we will learn in what follows, this is one of the definitions that needs care when generalizing to setups beyond GR. The equation of the amplification factor can be found by plugging Eq.~\eqref{eq:F_def} into Eq.~\eqref{eq:helmholtz_GR}. This is most easily done, though, by centering the coordinate system on the source, rather than the lens. We denote these coordinates with tilded quantities, i.e. $\tilde{\bmr} = (\tilde r, \tilde{\bmt})$, such that $\tilde{\bmr}_S = 0$. In these \refrep{coordinates}, one finds
\be\label{eq:F_full_GR}
    \partial_{\tilde r}^2F  + 2\ii\omega\partial_{\tilde r}F  +\frac{1}{\tilde r^2}  \nabla^2_{\tilde{\bm\theta}}F = 4\omega^2 U_{\rm L}F \,.
\eeq
The Fresnel/paraxial approximation consists of assuming
\be\label{eq:fresnel_condition}
    |\partial_{\tilde r}^2F| \ll |\omega\partial_{\tilde r}F| \, ,
\eeq
together with the small-angle approximation
\be\label{eq:paraxial_condition}
    \nabla^2_{\tilde{\bm\theta}} \simeq \partial_\theta^2 + \frac{1}{\theta}\partial_\theta + \frac{1}{\theta^2}\partial_\varphi^2 \,,
\eeq
so that $\tilde{\bmt} \approx \bmt$ and $\tilde{r} \approx z + \zsl$.
Then Eq.~\eqref{eq:F_full_GR} becomes a Schrödinger-like equation, with $\tilde r$ playing the role of time,
\be\label{eq:schrodinger_like_GR}
    \ii\partial_{\tilde r}F = -\frac{1}{2\omega\tilde r^2} \nabla^2_{\tilde{\bm\theta}}F + 2\omega U_{\rm L} F \,,
\eeq
whose formal solution can be represented as a path integral. In the thin-lens approximation, all transverse integrations are Gaussian except the one over the lens plane~\cite{Nakamura:1999uwi}, yielding the standard diffraction integral form of the amplification factor
\be\label{eq:diff_GR}
    F_{\rm diff}(\bo) = \frac{\omega}{2\pi\ii} \frac{\zso} {\zsl\zlo} \int \dd^2\bb\, \exp\left[ \ii\omega t_d(\bb,\bo) - \ii\omega\psi(\bb) \right],
\eeq
where $\bb = \zsl \bmt$ is the 2D position on the lens plane, while $\bo = \zso \bmt_{\rm o}$ is the observer 2D position. 
In Eq.~\eqref{eq:diff_GR}, $t_d(\bb,\bo)$ and $\psi(\bb)$ are the geometrical and gravitational time delays, defined respectively as 
\be
    t_d(\bb,\bo) \equiv \frac{\zso}{2\zsl\zlo} \lp \bb-\frac{\zsl}{\zso}\bo \rp^2 \,,
\eeq
and
\be
    \psi(\bb) \equiv 2\int\dd z\,U_{\rm L}(\bb,z) \,,
\eeq
is the projected lensing potential.

\smallskip

The high-frequency limit of Eq.~\eqref{eq:diff_GR}, leading to the GO regime, is obtained by a stationary phase. The stationary points satisfy
\be
    \nabla_{\bb}\left[t_d(\bb,\bo)-\psi(\bb)\right]=0,
\eeq
which is the lens equation in these variables~\cite{Schneider:1992bmb}. Thus, geometric optics arises as the saddle-point approximation to the Fresnel integral.

\section{Wave-optics lensing beyond GR: general considerations}
\label{sec_beyGRtheory}
In GR, the right-hand side of Eq.~\eqref{eq:F_full_GR} vanishes for $\omega\to0$. The zero-frequency problem is therefore free, which is the origin of the standard result for the amplification factor: $F\to1$ for $\omega\to0$~\cite{Nakamura:1997sw}. We now consider the scenarios beyond GR described by the equations of motion
\be \label{eq_phengw}
    \lp \Box - \xi R  \rp \phi = 0\,,
\eeq 
and show that this property does not hold anymore.
The essential point is that modified propagation can change the infrared reference problem itself. 
 
\subsection{Propagation-level EFT motivation}
\label{sec:EFT}

Controlled departures from Einstein--Hilbert gravity are systematically
captured by a Wilsonian effective field theory (EFT) obtained by
integrating out massive degrees of freedom at some UV scale $\Lambda$.
At lowest order in a derivative expansion, the resulting action takes
the form~\cite{Donoghue:1994dn,Burgess:2003jk}
\begin{align}\label{eq:EFT_covariant}
   S_{\rm EFT} = \frac{M_{\rm Pl}^2}{2} \int \dd^4x\sqrt{-g} \Big[
   R
   &+ \frac{\alpha_1}{\Lambda^2}R^2
   + \frac{\alpha_2}{\Lambda^2}R_{\mu\nu}R^{\mu\nu} \nn\\
   &+ \frac{\alpha_3}{\Lambda^2} R_{\mu\nu\rho\sigma}R^{\mu\nu\rho\sigma}
   + \dots \Big] \,,
\end{align}
possibly supplemented by nonminimal couplings of additional matter
fields to curvature.  The coefficients $\alpha_i$ are
$\mathcal{O}(1)$ Wilson coefficients, and $\Lambda$ sets the scale of
new physics. The new contributions besides the Einstein-Hilbert term
%
%
 lead to interesting theoretical and phenomenological ramifications for low-energy
observables, including GW:
see e.g. \cite{Camanho:2014apa,Endlich:2017tqa}.


Expanding around a background $g_{\mu\nu}=\bar g_{\mu\nu}+h_{\mu\nu}$,
the quadratic action for the tensor perturbation is governed by a
kinetic operator that depends on the background curvature,
\be
S^{(2)} \sim \int\dd^4x\sqrt{-\bar g}\, h \left(
  \bar\Box+\bar{\mathcal R}
  +\frac{\bar\nabla\bar\nabla}{\Lambda^2}
  +\dots \right) h \,,
\eeq
where $\bar{\mathcal R}$ collects all independent contractions of
$\bar R$, $\bar R_{\mu\nu}$, and $\bar R_{\mu\nu\rho\sigma}$.
Already the Einstein--Hilbert term yields the curved-space
Lichnerowicz operator; the higher-curvature terms further deform this
structure. A few remarks on the individual contributions in
Eq.~\eqref{eq:EFT_covariant} are in order.
The Gauss--Bonnet combination
$R^2-4R_{\mu\nu}R^{\mu\nu}+R_{\mu\nu\rho\sigma}R^{\mu\nu\rho\sigma}$
is topological in four dimensions and does not affect the tensor
propagation equation.  The remaining independent combinations,
for example $R^2$ and $R_{\mu\nu}R^{\mu\nu}$, do contribute and
generate curvature-dependent corrections to the GW wave equation on
nontrivial backgrounds.

Using the scalarized decomposition~\eqref{dec_scal} and assuming that
the polarization tensor $e_{\mu\nu}$ is slowly varying and parallel
transported, the projected tensor equation takes the schematic scalar
form
\be
    \left( \bar\Box-\mathcal{M}_{\rm eff}^2(x,e)+\dots \right)\phi=0 \,,
\eeq
where $\mathcal{M}_{\rm eff}^2$ is built from contractions of the
background curvature with the polarization tensor.
The simplest, lowest-dimensional truncation is
\be\label{eq_mincho}
    \mathcal{M}_{\rm eff}^2 = \xi\bar R \,,
\eeq
and there are several ways in which it can arise microscopically.
One route is purely gravitational: for a generic combination of
$\alpha_i$ in Eq.~\eqref{eq:EFT_covariant}, expanding the
higher-curvature terms to quadratic order in $h_{\mu\nu}$ on a
curved background generates, after field redefinitions, an effective
curvature-dependent mass for the tensor amplitude. The coefficient
$\xi$ is then a linear combination of the $\alpha_i$, suppressed by
$\Lambda^{-2}$.  A second, perhaps more transparent, route involves a
nonminimally coupled scalar field with action
\be\label{eq:scalar_action}
   S_\sigma = -\frac{1}{2}\int\dd^4x\sqrt{-g}
   \left[(\nabla\sigma)^2 + m^2\sigma^2 + \zeta R\,\sigma^2\right]\,.
\eeq
Integrating out $\sigma$ at tree level generates, in the low-energy
effective action for the metric, a correction to the Lichnerowicz
operator of the schematic form
\be\label{eq:integrated_out}
   \delta S^{(2)} \sim \frac{\zeta}{m^2}
   \int\dd^4x\sqrt{-\bar g}\;h\,\bar R\,h + \dots\,,
\eeq
where the ellipsis denotes terms with more derivatives or higher powers
of the curvature. Equation~\eqref{eq:integrated_out} is the
direct origin of the $\xi\bar R$ correction in the tensor wave
equation, with $\xi\propto\zeta/m^2$. Note that even a conformally
coupled scalar ($\zeta=1/6$) generates such a term once $m\neq0$.
More generally, any massive field nonminimally coupled to curvature
will contribute to $\xi$ after being integrated out, so
the truncation~\eqref{eq_mincho} is a generic feature of
low-energy EFTs of gravity rather than a fine-tuned choice.
The resulting phenomenological equation is then
\be \label{eq:WaveEq}
    \left( \bar\Box - \xi \bar R \right) \phi = 0\,,
\eeq
which is the starting point of this work (hereafter we drop the bars
on background quantities).
\refrep{Alternatively, Eq.~\eqref{eq:WaveEq} may find an astrophysical realization in dark-matter environments surrounding compact objects, which have been extensively investigated in the literature~\cite{Barack:2018yly,Hui:2016ltb}.}

It is instructive to understand on which scales this correction is relevant.  For a gravitational wave of frequency $\omega$ propagating in a background with characteristic curvature radius $L_{\rm curv}$,
the curvature scalar scales as $R\sim L_{\rm curv}^{-2}$, so the $\xi
R$ term enters the wave equation with a relative suppression
\be\label{eq:power_counting}
   \frac{\xi R}{\omega^2} \sim \frac{\xi}{(\omega L_{\rm curv})^2}
\eeq
with respect to the kinetic term.  In the geometric-optics regime
$\omega L_{\rm curv}\gg 1$, this correction is negligible, consistent
with the standard result that GO lensing is insensitive to this kind of 
modification (we also find this result later, see Sec.~\ref{sec:GO}).  In the wave-optics regime,
however, $\omega$ becomes comparable to the inverse lensing scale, and
the ratio~\eqref{eq:power_counting} can be of order unity even for
parametrically small $\xi$.

We note that the parameter $\xi$ can in principle be constrained by
other observables through dedicated EFT
parametrizations of modified gravity,  to be tested with GW observations~\cite{Endlich:2017tqa,Nair:2019iur,Sennett:2019bpc}.  The aim
of this work is not to derive such bounds, but rather to identify and
characterize the qualitatively new lensing phenomenology that the
correction~\eqref{eq_phengw} generates, which can be sizeable even for
small $\xi$ due to the infrared enhancement discussed above.

\smallskip

Finally, we stress that $\phi$ in our description is not a new
fundamental scalar degree of freedom; it represents the scalar
amplitude associated with a given tensor polarization of the
gravitational wave.  In general modified-gravity scenarios, however,
Lovelock's theorem~\cite{poisson_will_2014} implies that additional propagating polarizations
are generically present.  Their effective propagation on curved
backgrounds may likewise be governed by curvature-dependent operators
of the form~\eqref{eq_phengw}.  Although we do not study extra
polarizations explicitly here, their consequences for GW lensing
constitute an interesting direction for future work, for example
extending the analysis of~\cite{LISACosmologyWorkingGroup:2026zah} 
to the context of GW lensing.

\subsection{Curvature-coupled propagation}
\label{sec:curv_SIS}

In the limit of weak-field metric as discussed in Sec.~\ref{sec_GR}, we find
\be
     R = 2\Delta U_{\rm L} \,,
\eeq
up to sign conventions. Thus, Eq.~\eqref{eq_phengw} becomes
\be\label{eq:helmholtz_MG}
    \llp  \nabla^2_{\bmr}+\omega^2(1-4U_{\rm L}) - 2\xi \Delta U_{\rm L} \rrp \phi_\omega(\bmr)=0 \,,
\eeq
at linear order in the gravitational potential. 
Following similar steps as in the GR case, one can easily find the modified equation of the amplification factor defined in Eq.~\eqref{eq:F_def}, 
\be\label{eq:F_full_MG}
    \partial_{\tilde r}^2F  + 2\ii\omega\partial_{\tilde r}F  +\frac{1}{\tilde r^2}  \nabla^2_{\tilde{\bm\theta}}F = 4\omega^2 \lp U_{\rm L} + \frac{\xi}{2\omega^2}\Delta U_{\rm L}\rp F \,.
\eeq
Within the same Fresnel/thin-lens approximation used in GR (i.e. assuming Eqs.~\eqref{eq:fresnel_condition} and~\eqref{eq:paraxial_condition} and thin-lens),  
one may arrive to the modified expression of the diffraction integral 
\be\label{eq:diff_MG}
    F^{\rm MG}_{\rm diff}(\bo) = \frac{\omega}{2\pi\ii} \frac{\zso} {\zsl\zlo} \int \dd^2\bb\, \exp\left[ \ii\omega t_d(\bb,\bo) - \ii\omega\psi_{\rm eff}(\bb) \right]\,,
\eeq
where we have defined the effective projected potential
\be\label{eq:psi_eff}
    \psi_{\rm eff}(\bb) = 2\int\dd z \llp  U_{\rm L}(\bb,z) + \frac{\xi}{2\omega^2}\Delta U_{\rm L}(\bb,z) \rrp \,.
\eeq
However, we will show that this expression of the amplification factor leads to nonphysical results. In particular,  the new term on the right-hand side of Eq.~\eqref{eq:F_full_MG} is {\it not} suppressed by $\omega^2$. Therefore, the eikonal approximation in Eq.~\eqref{eq:fresnel_condition} breaks down, and this solution $F^{\rm MG}_{\rm diff}(\bo)$ leads to inconsistent results in the small $\omega$ limit. In fact, this contribution  survives in the static limit $\omega\to0$, and it can become large at small frequencies, even for a small modified-gravity parameter $\xi$. This phenomenon --  characterizing the modified gravity setup we consider -- will play a key role in our discussion, which aims to clarify its physics.

\section{GO and WO regimes for SIS lenses}
\label{sec:GO}

In this Section we study the behavior of $F^{\rm MG}_{\rm diff}(\bo)$ of Eq.~\eqref{eq:diff_MG}, both in GO and in WO.  To produce concrete results, we focus on SIS lenses~\cite{Robertson:2020mfh}. 
We show that, in the geometrical-optics limit, the effects of modified gravity are negligibly small.  Then, we numerically compute Eq.~\eqref{eq:diff_MG} for arbitrary frequencies, and expose the infrared issue explicitly. Understanding its origin (as breakdown of the eikonal approximation), and discussing how to improve the description will be the subject of the next Section.

\subsection{The SIS lens model}
In cylindrical coordinates centered on the lens, the SIS lens potential is given by~\cite{Robertson:2020mfh,Schneider:1992bmb} 
\be \label{eq:USIS}
    U_{\rm L} = 2\sigma_v^2\ln \lp \frac{\sqrt{b^2+z^2}}{r_0} \rp = 2\sigma_v^2\ln \lp \frac{r}{r_0} \rp \,,
\eeq 
where $\sigma_v$ is the velocity dispersion (dimensionless), $b=|\bb|$ and $r_0$ is a reference distance, and its Laplacian is 
\be \label{eq:SIS_lap}
    \Delta U_{\rm L} = \frac{2\sigma_v^2}{b^2+z^2} = \frac{2\sigma_v^2}{r^2}
\eeq 
Thus, the effective gravitational time delay of Eq.~\eqref{eq:psi_eff} is 
\be\label{eq:psi_eff_SIS}
    \psi_{\rm eff}^{\rm SIS}(b) = 4\pi\sigma_v^2 \lp  b +\frac{\xi}{2\omega^2b} \rp +\text{const} \,,
\eeq
which is only a function of $b$ reflecting the rotational symmetry on the lens plane of the potential. The additive constant in the previous equation is of no physical consequence, as it can be reabsorbed in the normalization of the amplification factor. Setting $\xi = 0$ one finds the usual SIS projected potential.
Importantly, the SIS lens model leads to frequency-dependent corrections  proportional to $1/(\omega^2b)$ in the projected phase above. These corrections, which scale as inverse powers of frequency, are precisely the terms responsible for the infrared sensitivity of wave-optics lensing in the present setup.

\smallskip
The SIS  lens profile is particularly useful because the induced interaction scales as $1/r^2$, as it can be understood from Eq.~\eqref{eq:SIS_lap},  leading to an inverse-square problem with exact analytic control. More generally, however,
\be
    \Delta U_{\rm L}\propto \rho(\bm r),
\eeq
so the same mechanism persists for arbitrary matter distributions, as this interaction term does not vanish for $\omega \to 0$. Different lens models modify the detailed form of the Green's  function and of the finite-frequency corrections, but not the existence of a subtle small-frequency limit.
The discussion above should therefore be viewed as a minimal phenomenological framework designed to isolate a specific infrared effect: the possibility that curvature-induced propagation terms remain non-negligible in the $\omega\to0$ limit. Hence, Eq.~\eqref{eq_phengw} 
provides a simple analytically tractable realization of amplified small-frequency effects in
WO treatments of modified gravity models.

\smallskip
Having chosen a lens model allows us to define dimensionless quantities. The SIS model is characterized by one parameter: the lens's velocity dispersion $\sigma_v$. The corresponding Einstein angular scale is
\be\label{eq:thetaE}
    \theta_E = 4\pi\sigma_v^2\frac{\zlo}{\zso}\,.
\eeq
We introduce the dimensionless transverse coordinates on the lens plane as
\be\label{eq:xy_def}
    \bm x = \frac{\bb}{\zsl\theta_E}, \qquad \bm y = \frac{\bo}{\zso\theta_E} \,.
\eeq
The vector $\bm x$ labels image positions on the lens plane, while $\bm y$ denotes the angular position of the source as seen by the observer, both measured in units of the Einstein angle. We discuss the axisymmetric case, and denote by
$ x=|\bm x|$,  $ y=|\bm y| $  the corresponding radial variables. The setup is controlled by two dimensionless parameters. The first is the standard diffraction parameter
\be\label{eq:nu_def}
    \nu = 16\pi^2\sigma_v^4 \lp \frac{\zlo\zsl}{\zso} \rp \omega \,,
\eeq
which measures the ratio between the gravitational-wave frequency and the characteristic lensing scale. $\nu$ also measures the size of the lensing scale relative to the coherent Fresnel region probed by the wave.   Large values $\nu\gg1$ correspond to geometric optics, while $\nu\lesssim1$ characterizes the wave-optics regime where diffraction and interference become important.

The second parameter controls the strength of the modified-gravity correction
\be\label{eq:epsilon_def}
    \varepsilon = 8\pi^2\sigma_v^4 \xi,
\eeq
with $\xi$ the curvature-coupling parameter appearing in Eq.~\eqref{eq_phengw}. In what follows we mainly focus on the regime $ \varepsilon>0  $,  for which the induced inverse-square interaction is effectively repulsive.
In these coordinates Eq.~\eqref{eq:diff_MG} becomes 
\be\label{eq:diff_MG_dimensionless}
    F^{\rm MG}_{\rm diff}(y) = \frac{\nu}{2\pi i} \int \dd^2 \bm x  \exp \llp \ii \nu \lp \frac{1}{2}|\bm x-\bm y|^2 -  x - \frac{\varepsilon}{x \nu^2} \rp  \rrp \,.
\eeq
Note that $F^{\rm MG}_{\rm diff}$ only depends on the distance of the source position on the lens plane from the lens $y$, and not its direction due to the rotational symmetry of the SIS lens model.

\subsection{GO: Images, magnifications and time delays}
For large $\nu$, the dominant contribution to the diffraction integral in Eq.~\eqref{eq:diff_MG_dimensionless} comes from stationary points of the effective dimensionless time-delay function
\be\label{eq:T_MG}
    T(\bm x, \bm y) \equiv \frac{1}{2}|\bm x-\bm y|^2 -  x - \frac{\varepsilon}{x \nu^2} \,.
\eeq
The first term corresponds to the geometrical propagation delay, the second is the standard SIS lensing potential, while the last term is induced by the curvature-dependent propagation effect.  Since we are formally taking the $\nu \gg 1$ limit, the last contribution is highly suppressed. 
We nevertheless keep this contribution, since it encodes the deviations from GR; otherwise, the GO regime would coincide with the GR result. In what follows we treat it as a small correction and expand perturbatively to leading order.
Using the axial symmetry of the lens, we can rotate the coordinate system of the 2D plane such that $\bm y$ is aligned with one of the axes. Using 2D polar coordinates in the lens plane, we can write $\bm x = x ( \cos \theta_x, \sin \theta_x)$ and   $\bm y = (y, 0)$, with $\theta_x$ the angle between the axis aligned with $\bm y$ and $\bm x$, and $\theta_x \in [ 0, 2 \pi)$. Then the lens equation gives 
\begin{align}
    \frac{\partial T}{\partial \theta_x} &= x y \sin \theta_x \,,\\
    \frac{\partial T}{\partial x} &= x - y \cos \theta_x - 1 + \frac{\varepsilon}{x^2 \nu^2}\,.
\end{align}
The first equation gives $\theta_x = 0 $ or $\theta_x = \pi$, thus the stationary point satisfies 
\begin{align}
    x - y - 1 + \frac{\varepsilon}{x^2 \nu^2} &= 0 \,, \qquad \theta_x = 0 \label{eq:stationary_positive}\\
    x + y - 1 + \frac{\varepsilon}{x^2 \nu^2} &= 0 \,, \qquad \theta_x = \pi \,.\label{eq:stationary_negative}
\end{align}
In standard SIS lensing there are at most two stationary points, $x = 1 \pm y$, which can be found by setting $\varepsilon=0$ in the previous equation.\footnote{These correspond to the usual $\bm x = (y \pm 1, 0)$ in our notation where $x$ is the modulus of the image vector in the 2D plane, and $\bm y = (y,0)$.} In particular, one has the second image only if $y<1$, since $x$ has to be positive, yielding the usual picture of having multiple images only if the source is position inside an Einstein ring radius from the lens. In the modified gravity case, the additional inverse-square contribution changes the structure of the effective Fermat potential and can generate extra extrema near the lens center. 

For nonzero $\varepsilon$, the stationary-point equations become cubic, and the image positions are shifted with respect to the GR SIS result. Treating the modified-gravity correction perturbatively, the two standard images are displaced according to
\begin{align}
    x^+_{{\rm out}} &= 1 + y - \frac{\varepsilon}{\nu^2 (1 + y)^2} + {\cal O} \lp \frac{\varepsilon^2}{\nu^4}\rp \qquad \theta_x = 0 \\
    x^-_{{\rm out}} &= 1 - y - \frac{\varepsilon}{\nu^2 (1 - y)^2} + {\cal O} \lp \frac{\varepsilon^2}{\nu^4}\rp \qquad \theta_x = \pi \,.
\end{align}
The correction is always negative for $\varepsilon>0$, so the images are shifted slightly toward the lens center. 
On top of these images, two more are formed close to the center at position
\begin{align}
    x^+_{ {\rm in}} &= \frac{1}{\nu}\sqrt{\frac{\varepsilon}{(1 + y)}} + {\cal O} \lp \frac{\varepsilon}{\nu^2}\rp  \qquad \theta_x = 0 \\
    x^-_{ {\rm in}} &= \frac{1}{\nu}\sqrt{\frac{\varepsilon}{ (1 - y)}} + {\cal O} \lp \frac{\varepsilon}{\nu^2}\rp \qquad \theta_x = \pi \,.
\end{align}
Thus $y \approx 1$ marks the beginning of the double image formation region, while another critical curve is formed at 
\be
    x^3_c = \frac{2 \varepsilon}{\nu^2} 
\eeq 
marking the onset of the 4-image formation region.
Substituting back into the lens equation for $\theta_c = \pi$ gives the associated caustic position in the source plane,
\be
y_c = 1 - x_c - \frac{\varepsilon}{\nu^2x_c^2} = 1 - \frac{3}{2} \lp \frac{2 \varepsilon}{\nu^2} \rp^{1/3} \,.
\eeq
Therefore, the modified propagation term changes the image multiplicity structure relative to the standard SIS lens.  The additional stationary points correspond to additional (as we will learn, highly demagnified) images localized close to the origin.

Evaluating Eq.~\eqref{eq:T_MG} on the image solutions, we can find the time delays, and time delay differences. 
The time delays of the two outer images  and two inner images are 
\begin{align}
    T^{+}_{\rm out} &= - y -\frac12 - \frac{\varepsilon}{\nu^2 (1 + y)} + {\cal O}\lp \frac{\varepsilon}{\nu^2}\rp^2 \,,\\
    T^{-}_{\rm out} &=  y -\frac12 - \frac{\varepsilon}{\nu^2 (1 - y)} + {\cal O} \lp \frac{\varepsilon}{\nu^2}\rp^2 \,,\\
    T^{+}_{\rm in} &= \frac{y^2}{2} - \frac{2}{\nu} \sqrt{\varepsilon(1 + y)} + \frac{\varepsilon}{2 (1 + y) \nu^2 } + {\cal O} \lp \frac{\varepsilon}{\nu^2}\rp^\frac32 \,,\\
    T^{-}_{\rm in} &= \frac{y^2}{2} - \frac{2}{\nu} \sqrt{\varepsilon(1 - y)} + \frac{\varepsilon}{2 (1 - y) \nu^2 } + {\cal O} \lp \frac{\varepsilon}{\nu^2}\rp^\frac32 \,.
\end{align}
The time delay difference between the two standard SIS-like images becomes
\be
T^{-}_{\rm out} - T^{+}_{\rm out} =  2y  - \frac{2 \varepsilon y}{\nu^2 (1 - y^2)} +{\cal O}\lp \frac{\varepsilon}{\nu^2}\rp^2\,.
\eeq 
Hence, the modified propagation equation changes the relative phase accumulated between interfering trajectories.
Finally, we can compute the magnifications of the four images. This is given by
\be 
    \mu^{-2} = \left| {\rm det} \lp \partial^2 T\rp \right| = \frac{1}{x^2} \left| \frac{\partial^2 T}{\partial x^2 }\frac{\partial^2 T}{\partial \theta_x^2 } - \lp \frac{\partial^2 T}{\partial x \partial \theta_x} \rp^2 \right|\,,
\eeq 
yielding 
\begin{align}
    \mu^{+}_{\rm out} &= \frac{y+1}{y} + \frac{\varepsilon}{\nu^2y(y+1)^2} + {\cal O} \lp \frac{\varepsilon}{\nu^2}\rp^2 \,,\\
    \mu^{-}_{\rm out} &= \frac{1- y}{y} + \frac{\varepsilon}{\nu^2y(y-1)^2} + {\cal O} \lp \frac{\varepsilon}{\nu^2}\rp^2\,, \\
    \mu^{+}_{\rm in} &=  \frac{\varepsilon}{2\nu^2 y(y+1)^2} + {\cal O} \lp \frac{\varepsilon}{\nu^2}\rp^\frac32 \,,\\
    \mu^{-}_{\rm in} &= \frac{\varepsilon}{2\nu^2 y(y-1)^2} + {\cal O} \lp \frac{\varepsilon}{\nu^2}\rp^\frac32 \,,
\end{align}
The magnification of the extra images produced by the modified gravity contribution is therefore negligible.  The main phenomenological role of modified gravity is expected to lie {\it not} in the production of bright additional images, but rather in the modification of the interference pattern in the wave-optics regime. Even relatively small changes in the time delays and stationary phases can produce sizeable distortions in the frequency dependence of the amplification factor, as these modifications accumulate with distance.

Thus, curvature-dependent propagation effects may remain perturbatively small in geometric optics, while -- as we now prove -- they become much more important in the wave-optics regime, where diffraction and interference are sensitive to the full infrared structure of the propagation kernel.

\subsection{WO: Numerical evaluation}
\begin{figure*}[t]
\centering
\includegraphics[width=1\linewidth]
{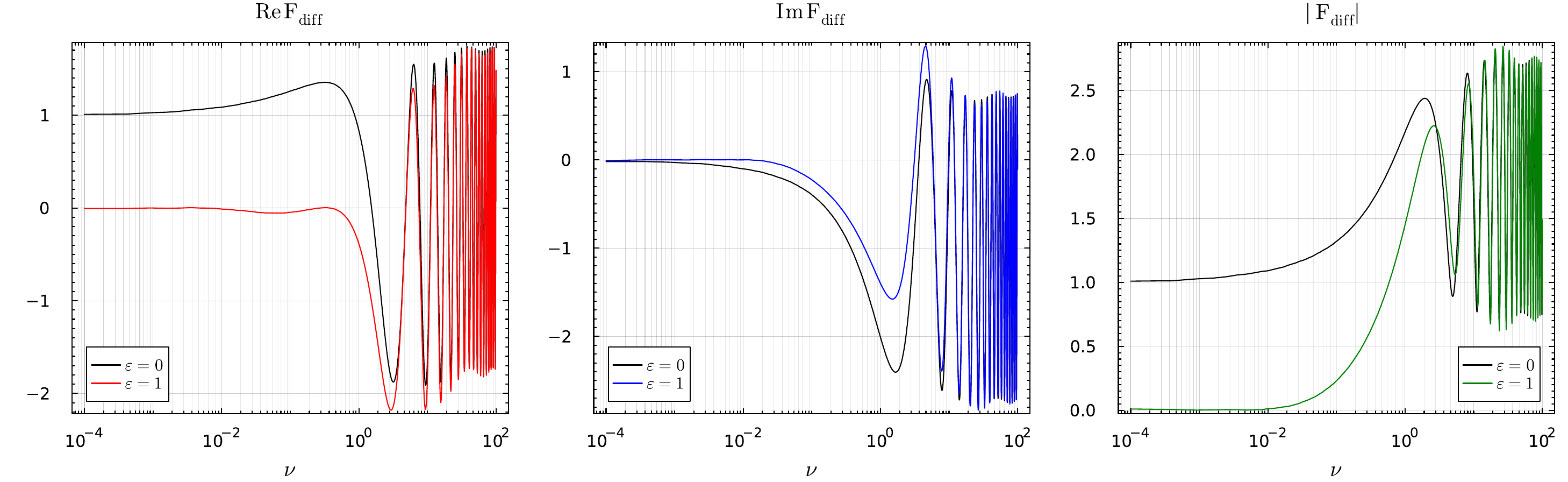}
\caption{\small Fresnel amplification factor for the SIS lens in GR (black lines, with $\varepsilon=0$) and in the curvature-coupled model (colored lines) with $\varepsilon=1$, for $y = 0.5$. From left to right: real part, imaginary part, absolute value. While the GR amplification factor's absolute value approaches a finite constant at low frequency (right plot, black line), the modified Fresnel result becomes strongly suppressed (right plot, green line). As we explain in the main text, this suppression is not the physical infrared limit, but rather signals the breakdown of the Fresnel approximation.
}
\label{fig:Fresnel_plot}
\end{figure*}

We start studying numerically the modified diffraction integral in Eq.~\eqref{eq:diff_MG_dimensionless}.  Given the lens symmetry, we can write the 2D integral of Eq. \eqref{eq:diff_GR} as 
\begin{align}
    F_{\rm diff}^{\rm MG}(y) = - \ii \nu \, e^{\frac{\ii \nu y^2}{2}} &\int^{\infty}_0 \dd x \, x  \, J_0 (\nu x y ) \times \nn \\
    &\times {\rm exp} \llp \ii \nu \lp \frac{x^2}{2} - x - \frac{\varepsilon}{\nu^2 x}\rp \rrp\,. \label{eq:F_diff_MG}
\end{align}
Here $J_0$ denotes the Bessel function of the first kind. The first two terms in the phase reproduce the standard SIS Fresnel kernel, while the last contribution proportional to $\varepsilon$ is entirely due to the curvature-dependent propagation effect. 
The diffraction integral is evaluated numerically by deforming the integration contour into the complex plane and adaptively constructing steepest-descent paths -- thereby converting the highly oscillatory WO integral into exponentially decaying contributions, which  can be integrated efficiently with standard quadrature methods~\cite{CarrilloGonzalez:2025gqm,Feldbrugge:2019fjs}. Representative examples are shown in Fig.~\ref{fig:Fresnel_plot}, where we notice sizeable distortions of the interference pattern at sufficiently low frequency.
The figure  shows that for $\nu \gg 1$, the curvature-dependent coupling becomes progressively less important, and the GR and MG predictions eventually coincide. However, significant deviations appear for $\nu \lesssim 1$. Importantly, the rightmost plot shows that $|F_{\rm diff}| \to 1$ as $\nu \to 0$ in GR, while $|F^{\rm MG}_{\rm diff}| \to 0$. Taken at face value, this would imply that the modified-gravity case predicts the disappearance of the wave in the low-frequency limit, a  clearly unphysical
conclusion. Rectifying this behavior will be the subject of the next Section, where we show that it originates from a breakdown of the Fresnel approximation in the modified-gravity theory.
Overall, this plot shows that there exists a broad class of theories with curvature-dependent interactions which become essentially indistinguishable from GR in the geometric-optics limit, while exhibiting large deviations in the wave-optics regime. This further motivates our analysis of the interplay between wave optics and modified gravity.

\section{Apparent low-frequency suppression of the amplification factor:
breakdown of the Fresnel approximation}
\label{sec:Fresnel_breakdown}

The numerical evaluation of the modified diffraction integral~\eqref{eq:F_diff_MG} illustrated in Fig.~\ref{fig:Fresnel_plot}, reveals a striking departure from the standard GR behavior at low frequency. In particular, when the Fresnel expression is extrapolated toward the infrared regime $\nu\to0$, the amplification factor appears to vanish:
 $   F_{\rm diff}^{\rm MG} \to0$.
At first sight, this result might suggest that curvature-dependent propagation effects suppress gravitational-wave propagation in the deep infrared. However, this interpretation would be incorrect. The vanishing of Eq.~\eqref{eq:F_diff_MG} is not the physical low-frequency limit of the amplification factor. Instead, it is a signal that the Fresnel/paraxial approximation itself has broken down, and does not provide physical results. The strong low-frequency suppression is the first indication that the infrared behavior of the system is qualitatively different from the GR case, and needs to be handled with theoretical care.

\subsection{Why the Fresnel solution vanishes as $\nu \to 0$.}

The origin of the suppression can be understood analytically by studying the static limit of the reduced Fresnel equation. Recall that the Fresnel/paraxial approximation is obtained by neglecting the second radial derivative term in Eq.~\eqref{eq:F_full_MG}, namely through the condition~\eqref{eq:fresnel_condition}. Taking the limit $\omega\to0$ \emph{after} this approximation --- which, as we will argue, is not justified in the present setup --- Eq.~\eqref{eq:F_full_MG} reduces to
\be\label{eq:StaticFresnelMG}
    \frac{1}{\tilde r^2} \nabla_{\tilde{\bm\theta}}^2 F_{\rm diff} = 2\xi\Delta U_{\rm L}\, F_{\rm diff} \,.
\eeq
For positive matter density and $\xi>0$, one has $\xi \Delta U_{\rm L} > 0$. To analyze the existence of regular solutions, multiply Eq.~\eqref{eq:StaticFresnelMG} by $F_{\rm diff}^\ast$ and integrate over the angular sphere,
\be
    \int \dd\Omega \, F_{\rm diff}^\ast \nabla_{\tilde{\bm\theta}}^2F_{\rm diff} =  2 \xi \int  \dd\Omega\, \Delta U_{\rm L} |F_{\rm diff}|^2 \,.
\eeq
Integrating by parts gives
\be\label{eq:positivity_argument}
    - \int \dd\Omega\, |\nabla_{\tilde{\bm\theta}}F_{\rm diff}|^2 =  2 \xi \int \dd\Omega\, \Delta U_{\rm L} |F_{\rm diff}|^2.
\eeq
The left-hand side is manifestly nonpositive, while the right-hand side is non-negative. Therefore Eq.~\eqref{eq:positivity_argument} can only be satisfied by the trivial solution
\be
 F_{\rm diff}=0 \,.
\eeq
The same conclusion follows directly from Eq.~\eqref{eq:StaticFresnelMG}: once the right-hand side is nonzero, a constant nonvanishing solution is no longer allowed.

This explains why the Fresnel integral appears to vanish in the low-frequency regime in the numerical analysis leading to Fig.~\ref{fig:Fresnel_plot}. Importantly, however, this is not a genuine physical prediction. Rather, it reveals that the reduced Fresnel equation becomes singular in the static limit. The term $\partial_{\tilde r}^2F$, discarded in the Fresnel approximation~\eqref{eq:fresnel_condition}, is no longer subleading once the curvature-induced interaction survives in the infrared. In other words, the low-frequency suppression of the Fresnel solution is not physical, but instead signals the breakdown of the eikonal/paraxial approximation.

\subsection{Breakdown of Fresnel approximation due to the curvature coupling}

The same conclusion can be reached from a scaling analysis of the paraxial approximation itself. Suppose that the amplification factor varies over a characteristic radial scale $L_r$. Then parametrically
\be
    \partial_rF \sim \frac{F}{L_r}\, , \qquad \partial_r^2F \sim \frac{F}{L_r^2}\,,
\eeq
and the eikonal approximation condition~\eqref{eq:fresnel_condition} becomes
\be 
 \omega L_r \gg 1 \,,
\eeq 
telling us that the envelope $F$ must vary slowly compared with the oscillation scale of the carrier wave. 

In a regime where the dynamics is dominated by the effective potential, Eq.~\eqref{eq:F_full_MG} schematically implies\footnote{\refrep{In this order of magnitude assessment, we neglect the angular Laplacian term compared to the effective potential term. This is a good approximation for large angular momenta. Indeed, since $ \nabla^2_{\tilde{\bm\theta}}F / \tilde r^2 \sim F / b^2$, where $b$ is the impact parameter, comparing with the effective potential term in Eq.~\eqref{eq:F_full_MG}, gives $ 1 / (\omega^2 b^2) \ll |U_{\rm eff}|$
}.}
\be
    2\ii\omega\partial_r F \sim 4\omega^2 \lp U_{\rm L} + \frac{\xi}{2\omega^2}\Delta U_{\rm L}\rp F.
\eeq
Hence,
\be\label{eq:fresnel_validity_2}
    L_r \sim \frac{1}{2\omega|U_{\rm eff}|}\,,
\eeq
where $|U_{\rm eff}|$ denotes the characteristic magnitude of $U_{\rm L} + \xi \Delta U_{\rm L}/2\omega^2$.
Substituting Eq.~\eqref{eq:fresnel_validity_2} into the eikonal condition gives
\be\label{eq:Ueff_small}
    |U_{\rm eff}| \ll 1 \,.
\eeq
For the modified SIS diffraction problem, in the low-frequency regime where the curvature-coupling dominates the effective potential, we find
\be 
    |U_{\rm eff}| \sim  \frac{\sigma_v^2 \xi}{r^2 \omega^2} \sim 32 \pi^2 \sigma_v^6  \lp \frac{\zlo \zsl}{r \zso} \rp^2 \, \frac{\varepsilon}{\nu^2} \,,
\eeq 
where we used Eqs.~\eqref{eq:nu_def} and~\eqref{eq:epsilon_def} and the form of the potential in Eq.~\eqref{eq:SIS_lap}. For cosmological source--lens--observer separations, the distance ratio is generically ${\cal O}(1)$, and therefore the condition~\eqref{eq:Ueff_small} parametrically requires 
\be 
\nu \gg \sigma_v^3 \sqrt{\varepsilon} \,.
\eeq 
The Fresnel approximation is thus reliable only in an intermediate/high-frequency regime where the curvature-induced contribution remains perturbative throughout the dominant integration region. Once the frequency becomes sufficiently small, the inverse-square interaction dominates the dynamics and the paraxial expansion ceases to be uniformly valid.

\smallskip

This observation has an important conceptual implication. In GR, the low-frequency limit is smooth because the lensing interaction switches off as $\omega\to0$. In the present case, the curvature-induced interaction survives in the infrared, qualitatively changing the structure of the propagation problem. The correct low-frequency description must therefore be formulated directly in terms of the full second-order wave equation rather than its Fresnel reduction. In the next Section we develop precisely such a description by studying the static Green's  function associated with the full infrared problem.

\section{Correct small-frequency limit from the full wave equation}
\label{sec:static}
In the previous Section we learned that the Fresnel/paraxial description becomes singular in the deep infrared regime. The origin of the problem is now clear: after the Fresnel reduction, the curvature-induced interaction survives in the limit $\omega\to0$, while the second radial derivative term has been discarded. Consequently, the reduced first-order equation no longer provides a self-consistent description of the dynamics. To determine the correct infrared behavior one must therefore return to the full second-order wave equation and take the static limit \emph{before} making any Fresnel reduction.

In this Section we therefore return to the full second-order wave equation. We first analyze the exact static problem and then extend the solution perturbatively to the regime $\omega \ll 1$.

\subsection{Static Green's  function}
\refrep{We now consider the static problem, namely Eq.~\eqref{eq:helmholtz_MG} at $\omega = 0$, 
\be
    \llp  \nabla^2_{\bmr} - 2\xi \Delta U_{\rm L} \rrp \phi_\omega(\bmr)=0 \,.
\eeq
For our purposes, it is more convenient to work directly with the Green's function rather than the amplification factor. We also exploit the spherical symmetry of the SIS profile by centering the coordinate system on the lens, rather than on the source as in the derivation of the diffraction integral. Accordingly, we use untilded coordinates here, in contrast to the tilded coordinates introduced in Eq.~\eqref{eq:F_full_MG}. For the SIS profile, where the Laplacian of the potential is given in Eq.~\eqref{eq:SIS_lap}, the static Green's  function satisfies
}
\be\label{eq:static_green_eq}
    \llp  \nabla^2_{\bmr} - \frac{4\sigma_v^2\xi}{r^2} \rrp G^{\rm MG}_0(\bm r,\bmr_s) = -4\pi \, \delta^{(3)}(\bm r-\bmr_s) \,,
\eeq
where the subscript $0$ stands for $\omega = 0$, and  where $\bmr_s$ denotes the source position relative to the lens center (see Fig.~\ref{fig:LensGeom}). \refrep{Thanks to the spherical symmetry of the potential, } we can solve Eq.~\eqref{eq:static_green_eq} in a partial wave expansion~\cite{1988sfbh.book.....F}.
We first expand the Green's function in spherical harmonics
\be\label{eq:pw_expansion}
    G^{\rm MG}_0(\bm r,\bmr_s) = \sum_{\ell=0}^{\infty} \sum_{m=-\ell}^{+\ell} g^{\rm MG}_{0, \ell} (r,r_s)\, Y_{\ell m}(\bmt) Y_{\ell m}^\ast(\bmt_S)\,,
\eeq
with the observer and source position decomposed as $\bmr = (r, \bmt)$ and $\bmr_s = (r_s, \bmt_S)$ respectively, in a coordinate system centered on the lens.
Substituting into Eq.~\eqref{eq:static_green_eq} gives the radial equation
\refrep{
\begin{align}\label{eq:radial_static}
    \llp  \frac{\dd^2}{\dd r^2} + \frac{2}{r}\frac{\dd}{\dd r} - \frac{\ell(\ell+1)+4\sigma_v^2\xi}{r^2} \rrp & g^{\rm MG}_{0, \ell} (r,r_s) = \nonumber\\
    & = -\frac{4\pi}{r^2}\delta(r-r_s) \,.
\end{align}
From this equation it is clear a particular feature of the SIS lens model:  the curvature-induced interaction appears as an inverse-square potential exactly like the angular momentum contribution [respectively the $4 \sigma^2_v \xi / r^2$ and $\ell(\ell + 1)/ r^2$ terms in Eq.~\eqref{eq:radial_static}]. Thus, we can define an effective angular momentum parameter $\lambda_\ell$ through 
\be \label{eq:lambda_relation}
    \lambda_\ell(\lambda_\ell+1) \equiv \ell(\ell+1)+4\sigma_v^2\xi \,.
\eeq 
This means that the effect of the curvature coupling is to shift the effective centrifugal barrier from $\ell(\ell+1)$ to $\lambda_\ell(\lambda_\ell+1)$. }
Solving for $\lambda_\ell$ gives
\be \label{eq:lambda_def}
    \lambda_\ell  = - \frac12 + \frac12 \sqrt{(2 \ell + 1 )^2 + 16 \sigma^2_v \xi} \,.
\eeq 
Physically, the inverse-square interaction modifies the propagation basis itself, rather than simply generating a perturbative correction to free propagation.~\footnote{
We mainly focus on positive values of $\xi$, for which $\lambda_\ell$ remains real and the inverse-square interaction is effectively repulsive. For sufficiently negative $\xi$, however, the quantity $ (2\ell+1)^2+16\sigma_v^2\xi $ can become negative, leading to complex $\lambda_\ell$. This corresponds to an attractive inverse-square problem, where the Hamiltonian may fail to be  self-adjoint and additional ultraviolet boundary conditions can become necessary. }

\smallskip

Away from the source position, Eq.~\eqref{eq:radial_static} is homogeneous and admits two independent solutions, scaling respectively as $r^{\lambda_\ell}$ and $r^{-\lambda_\ell-1}$. The choice of boundary conditions is dictated by the physical behavior required of the Green's  function. For $r<r_s$, regularity at the center excludes the $r^{-\lambda_\ell-1}$ branch, which is singular as $r\to0$, and therefore selects the $r^{\lambda_\ell}$ solution. Conversely, for $r>r_s$, we require the Green's  function to remain bounded and decay at spatial infinity, which selects the $r^{-\lambda_\ell-1}$ branch. The Green's  function is therefore constructed by matching these two branches at the source position $r=r_s$. Continuity of the Green's  function at $r=r_s$, together with the derivative discontinuity implied by integrating Eq.~\eqref{eq:radial_static} across the delta-function source, then fixes their relative normalization. Thus, imposing regularity at the lens center, decay at large radius, continuity at $r=r_s$, and the appropriate derivative jump, we obtain\footnote{We use the spherical-harmonic addition theorem,
\be 
\sum_{m=-\ell}^{+\ell}
Y_{\ell m}(\bmt) Y_{\ell m}^\ast(\bmt_S)  = \frac{2\ell+1}{4\pi} P_\ell(\cos\gamma),
\eeq 
with $ \cos\gamma = \bmt \cdot \bmt_s $.}
\be\label{eq:static_green_summed}
    G^{\rm MG}_0(r,\gamma;r_s)  = \sum_{\ell=0}^{\infty} \frac{2\ell+1}{2\lambda_\ell+1} \frac{ r_<^{\lambda_\ell} }{ r_>^{\lambda_\ell+1} } P_\ell(\cos\gamma),
\eeq
where \refrep{$\gamma \equiv \cos^{-1} (\bmt \cdot \bmt_s)$ is the angle between the observer and source positions (see Fig.~\ref{fig:LensGeom}), and}
\be\label{eq:rminusrplus}
    r_< \equiv \min(r,r_s ) \,, \qquad r_> \equiv \max(r,r_s).
\eeq
This expression provides the exact static Green's  function for the infrared, small-frequency limit associated with the curvature-coupled SIS lens.

\smallskip
The GR limit is recovered smoothly when $\xi\to0$. Then $ \lambda_\ell\to\ell $, and Eq.~\eqref{eq:static_green_summed} reduces to
\be
    G_0^{\rm GR}(\bm r,\bmr_s) = \sum_{\ell=0}^{\infty}  \frac{  r_<^\ell }{ r_>^{\ell+1} } P_\ell(\cos\gamma) = \frac{1}{|\bm r-\bmr_s|} \,,
\eeq
where the last resummation is the standard multipole expansion of the Coulomb Green's function. The important point is that, once $\xi\neq0$, the low-frequency Green's function is no longer free. The propagation basis itself becomes distorted by the inverse-square interaction.


Starting from these results, we might na\"ively -- but,
as we  discuss later, inconveniently -- define an amplification factor as the ratio of the static MG propagator over the GR one, 
\be \label{eq:F_static_def}
    F^{\rm MG}_{\omega = 0} (\bmr) = \frac{G^{\rm MG}_{0} (\bmr, \bmr_s)}{G^{\rm GR}_{0} (\bmr, \bmr_s)} = |\bm r-\bmr_s| \, G^{\rm MG}_{0} (\bmr, \bmr_s)\,.
\eeq 
Focusing on a distant observer, $ r\gg r_s$, the leading contribution comes from the monopole $\ell=0$, yielding
\be\label{eq:F_static_far}
    F^{\rm MG}_{\omega = 0} (\bmr)\simeq \frac{1}{2\lambda_0+1} \lp \frac{r_s}{r} \rp^{\lambda_0}\,.
\eeq
Since $\lambda_0=0$ in GR [but not in our modified
gravity setup, recall Eq.~\eqref{eq:lambda_def}] this formula seems to describe well the different regimes we are interested on. However, this approach has
limitations:
 %
%
it is not possible to extend this definition in the nonstatic limit and have a solution that interpolates smoothly for various values of $\nu$, as we have argued so far. In Sec.~\ref{sec:distorted_reference} we  discuss a different definition of the amplification factor which is more convenient as it allows us doing so.

\subsection{Exact Green's function}
\label{sec:partial_wave}
The previous Section established that the static problem is governed by a nontrivial  Green's function. However,  GW  lensing necessarily involves finite-frequency radiation with outgoing-wave boundary conditions. The next step is therefore to construct the full finite-frequency Green's function, and understand how it approaches the interacting infrared regime discussed above. 
For the SIS profile, the modified Helmoltz Eq.~\eqref{eq:helmholtz_MG}, with potentials given in Eqs.~\eqref{eq:SIS_lap} and~\eqref{eq:USIS}, takes the form 
\begin{align}
    \Big[ \nabla^2 + \omega^2 \lp 1 - 8 \sigma^2_v \ln \frac{r}{r_0} \rp   -  &\frac{4 \sigma^2_v \xi}{r^2} \Big] G^{\rm MG}_\omega (\bmr, \bmr_s ) = \nn \\
    &=- 4 \pi \delta^{(3)} (\bmr - \bmr_s) \,. \label{eq:full_green_finite}
\end{align}
The inverse-square term controls the infrared propagation basis, while the logarithmic potential governs the residual finite-frequency scattering. Note also that the SIS lens potential grows proportional to $ \log r$ starting from    the source position. This is clearly not physical, and one should think that this potential is matched to an outside region where the potential starts to decrease.\footnote{\refrep{We do not need to specify the detailed form or truncation scale of this outer profile, since the matching performed below to compute the Green's function only requires the existence of an overlap region in which the SIS description remains valid and the logarithmic contribution is perturbative. Under this interpretation, we simply require that the matching region lies sufficiently far from the lens core while remaining inside the regime where the SIS approximation is valid.}}
As in the static case of Sec.~\ref{sec:static}, the symmetry of the potential suggests performing a partial wave expansion
\be\label{eq:pw_expansion_MG}
    G^{\rm MG}_\omega(r, \gamma ; r_s) =  \sum_{\ell=0}^{\infty} \frac{2 \ell + 1 }{4 \pi}   g^{\rm MG}_{\omega \ell} (r,r_s)\, P_\ell (\cos \gamma)\,,
\eeq
with the radial Green's function satisfying
\begin{align}\label{eq:PartialWaveGMGw}
        \Big[ \frac{\dd^2}{\dd r^2} + \frac{2}{r}\frac{\dd}{\dd r} + \omega^2 \lp 1 - 8 \sigma^2_v \ln \frac{r}{r_0} \rp &-  \frac{\lambda_\ell(\lambda_\ell+1)}{r^2} \Big]  g^{\rm MG}_{\omega\ell}   = \nn \\
        &=-\frac{4\pi}{r^2}\delta(r-r_s) 
\end{align}
where $\lambda_\ell$ is the effective angular momentum introduced in Eq.~\eqref{eq:lambda_def}.
To construct the corresponding retarded Green's function, we introduce two independent homogeneous solutions: $\phi_{\omega\ell}^{\rm in}(r)$, regular at the lens center, and $\phi_{\omega\ell}^{\rm out}(r)$, satisfying outgoing-wave boundary conditions in the asymptotic radiative region.
Their asymptotic behaviors  are~\cite{Casals:2016soq,Leaver:1986gd}
\begin{align}
    \phi^{\rm in}_{\omega \ell} (r) &\sim 
    \begin{cases}
        A_{\rm tra} r^{\lambda_\ell}  &\qquad r \to 0 \\ 
        A_{\rm ref} \frac{e^{i \omega r}}{r} + A_{\rm inc}  \frac{e^{-i \omega r}}{r} &\qquad \omega r \to \infty     
    \end{cases}  \label{eq:IngoingSol}\\
    \phi^{\rm out}_{\omega \ell} (r) &\sim 
    \begin{cases}
        B_{\rm tra} r^{\lambda_\ell}  + B_{\rm inc} r^{-\lambda_\ell - 1 } &\qquad r \to 0 \\ 
        B_{\rm ref} \frac{e^{i \omega r}}{r}  &\qquad \omega r \to \infty 
    \end{cases}  \label{eq:OutgoingSol}
\end{align}
Then the radial Green's function takes the form 
\be \label{eq:GreenRadialFiniteFreq}
    g^{\rm MG}_{\omega \ell} (r,r_s) = 4\pi \, \frac{\phi_{\omega\ell}^{\rm in}(r_<) \phi_{\omega\ell}^{\rm out}(r_>)}{W_\ell} \,,
\eeq 
with $r_{<,>}$ defined as in Eq.~\eqref{eq:rminusrplus} and $W_\ell$ the associated Wronskian 
\be \label{eq:Wronskian}
    W_\ell = r^2  \llp \phi_{\omega\ell}^{\rm out} \frac{\dd\phi_{\omega\ell}^{\rm in}}{\dd r} - \phi_{\omega\ell}^{\rm in} \frac{\dd\phi_{\omega\ell}^{\rm out}}{\dd r} \rrp \,.
\eeq 

Using the homogeneous radial equation, one verifies immediately that ${\dd W_\ell}/{\dd r}=0$, so $W_\ell$ is constant in any source-free region.

\subsection{Small-frequency inner expansion: Matching with the reference solution}

For small but nonzero $\omega$, Eq.~\eqref{eq:PartialWaveGMGw} can be solved perturbatively in the inner region close to the lens. We expand the solutions of the associated homogeneous problem as
\be
    \phi_{\omega \ell}(r) = \phi_{0\ell}(r) + \omega^2 \phi_{\omega \ell}^{(2) } (r) +\mathcal O(\omega^4) \,,
\eeq
where $ \phi_{0\ell}(r) $ denotes the solution of the homogeneous static problem, namely $ \phi_{0\ell}(r) = A r^{\lambda_\ell} + B r^{-\lambda_\ell - 1} $. Defining
\be 
{\cal L}_{0\ell} \equiv \frac{\dd^2}{\dd r^2} + \frac{2}{r}\frac{\dd}{\dd r} -  \frac{\lambda_\ell(\lambda_\ell+1)}{r^2}
\eeq
the perturbative corrections of the regular/irregular solutions are defined by
\begin{align}
    {\cal L}_{0\ell} \: \phi_{\omega \ell}^{\rm reg, (2)}  &\approx - \llp 1-8\sigma_v^2\ln \lp \frac{r}{r_0} \rp \rrp \, r^{\lambda_\ell} \,,\\
    {\cal L}_{0\ell} \: \phi_{\omega \ell}^{\rm irreg, (2)}   &\approx - \llp 1-8\sigma_v^2\ln \lp \frac{r}{r_0} \rp \rrp \, r^{-\lambda_\ell - 1} \,.
\end{align}
Solving the equations perturbatively gives
\begin{align}
    \phi^{\rm reg}_{\omega \ell}  &\simeq r^{\lambda_\ell} \llp  1 + \omega^2 r^2 \lp  a^{\rm reg}_\ell+b^{\rm reg}_\ell\ln\frac{r}{r_0} \rp \rrp    + \mathcal O(\omega^4) \,, \\
    \phi^{\rm irreg}_{\omega \ell} &\simeq r^{- \lambda_\ell - 1} \llp  1 + \omega^2 r^2 \lp  a^{\rm irreg}_\ell+b^{\rm irreg}_\ell\ln\frac{r}{r_0} \rp \rrp + \mathcal O(\omega^4) \,,
\end{align}
for the regular and irregular solutions up to second order,  where the coefficients are
\begin{align}
     a^{\rm reg}_\ell &= -\frac{1}{2(3 + 2\lambda_\ell)} \llp 1 +  \frac{4\sigma_v^2(5 + 2\lambda_\ell)}{(3 + 2\lambda_\ell)} \rrp \,, \label{eq:areg}\\
     b^{\rm reg}_\ell &= \frac{4\sigma_v^2}{3 + 2\lambda_\ell} \,, \\ 
     a^{\rm irreg}_\ell &= -\frac{1}{2(1-2\lambda_\ell)} \llp 1 + \frac{4\sigma_v^2(2\lambda_\ell+5)}{(1-2\lambda_\ell)} \rrp \,, \\
     b^{\rm irreg}_\ell &= \frac{4\sigma_v^2}{1 - 2\lambda_\ell} \,. \label{eq:birreg}
\end{align}

These solutions define the distorted infrared basis associated with the inverse-square interaction. However, the retarded Green's function requires the ingoing and outgoing solutions appearing in Eqs.~\eqref{eq:IngoingSol} and~\eqref{eq:OutgoingSol}. While the globally regular solution is continuously connected to $\phi^{\rm reg}_{\omega\ell}$, the globally outgoing solution is not identified with the irregular branch alone. Instead, the dynamics mixes the two Frobenius branches, so that the outgoing-wave solution becomes a linear combination of the regular and irregular infrared modes. Determining this mixing requires matching the inner distorted basis to a second solution valid in an overlap region -- following an approach first developed in \cite{Unruh:1976fm} in the context of black hole absorption cross-sections.

The perturbative inner expansion is valid in the near-zone, where the frequency-dependent corrections remain small compared with the static distorted basis. Parametrically, this requires
\be 
    \omega r \ll 1 \,, \qquad \omega^2 r^2 \sigma^2_v  \left| \ln \lp \frac{r}{r_0} \rp \right| \ll 1 \,.
\eeq
On the other hand, we can define a reference problem by neglecting the logarithmic SIS contribution relative to the constant frequency term. This gives
\be\label{eq:PartialWaveRefSolution}
        \Big[ \frac{\dd^2}{\dd r^2} + \frac{2}{r}\frac{\dd}{\dd r} + \omega^2 -  \frac{\lambda_\ell(\lambda_\ell+1)}{r^2} \Big]  \phi^{\rm ref}_{\omega\ell}  = 0 \,.
\eeq
The approximation is valid whenever
\be
    8\sigma_v^2 \left| \ln \lp \frac{r}{r_0} \rp \right| \ll 1 \,.
\eeq
For realistic SIS lenses one typically has $\sigma_v^2 \ll 1$, so this condition is rather weak and allows a broad range of radii around $r_0$. Strictly speaking, the logarithmic growth of the SIS potential at asymptotic infinity is not physical and reflects the fact that the SIS profile is only an intermediate-scale approximation to realistic galactic halos. At sufficiently large distances one expects the profile to transition to an outer decaying distribution. Under this interpretation, the condition above simply requires that the matching region lies sufficiently far from the lens core while remaining inside the regime where the SIS approximation is valid.

Consequently, there exists a buffer region where both the perturbative expansion around the static solution and the reference-wave approximation are simultaneously valid. Physically, this region corresponds to scales still small compared with the wavelength, so that the infrared distorted basis remains appropriate, while the logarithmic SIS contribution only perturbs the local propagation.

We solve Eq.~\eqref{eq:PartialWaveRefSolution}, and the ingoing and outgoing basis of solutions of Eqs.~\eqref{eq:IngoingSol} and~\eqref{eq:OutgoingSol} is given by
\begin{align}\label{eq:IngoingOutgoingReferenceProblem}
    \phi^{{\rm ref}, {\rm in}}_{\omega \ell}   =  j_{\lambda_\ell} (\omega r) \,, \qquad  \phi^{{\rm ref}, {\rm out}}_{\omega \ell} =  h^{(1)}_{\lambda_\ell} (\omega r) \,,
\end{align}
where $ j_{\lambda_\ell} $ and $ h_{\lambda_\ell} $ are the spherical BesselJ and spherical Hankel functions of the first kind. Their asymptotic behavior makes their interpretation transparent. For $\omega r \ll 1$, the ingoing solution is regular as it behaves as
\begin{align}
    j_{\lambda_\ell} (\omega r) &\simeq \frac{\sqrt{\pi} }{2^{\lambda_\ell +1}  \Gamma \llp \lambda_\ell + 3/2 \rrp } (\omega r)^{\lambda_\ell}  ~~~\qquad \omega r \ll 1 \,,
\end{align}
which matches continuously onto the regular inner branch. At large radius one instead finds
\begin{align}
    j_{\lambda_\ell} (\omega r) &\simeq \frac{1}{ \omega r}  {\rm cos} \llp  \omega r -  \frac{\pi}{2} ( 1 + \lambda_\ell)\rrp  ~~ \omega r \gg 1 \,,
\end{align}
which corresponds to an equal superposition of outgoing and ingoing waves $e^{\pm \ii \omega r}$.
The outgoing solution instead behaves asymptotically as
\begin{align}
    h^{(1)}_{\lambda_\ell} (\omega r) &\simeq \frac{1}{ \omega r}  \, e^{\ii  \omega r -  \frac{\ii \pi}{2} ( 1 + \lambda_\ell)}  ~~ \omega r \gg 1 \,,
\end{align}
while in the inner region it contains both Frobenius branches,
\begin{align}
    h^{(1)}_{\lambda_\ell} (\omega r) &\simeq  B_{\rm tra} \, r^{\lambda_\ell}  + B_{\rm inc} \, r^{-\lambda_\ell-1 }~~~\qquad \omega r \ll 1 \,,
\end{align}
with coefficients given by
\begin{align}
    B_{\rm tra} &= \frac{\sqrt{\pi}\omega^{\lambda_\ell}}{2^{\lambda_\ell} \lp 1 + e^{2 \ii \pi \lambda_\ell}\rp \Gamma \llp \lambda_\ell + 3/2\rrp} \,, \\
    B_{\rm inc} &=  \frac{2^{\lambda_\ell} \Gamma \llp \lambda_\ell + 1/2\rrp}{\ii \sqrt{\pi}\omega^{\lambda_\ell + 1}} \,.
\end{align}

This explicitly shows that the physical outgoing-wave solution is not equal to the irregular branch alone. Instead, once propagated into the inner region, it becomes a linear combination of the regular and irregular infrared solutions. This is a standard feature of long-range scattering problems and reflects the fact that the basis naturally adapted to the origin differs from the basis naturally adapted to asymptotic radiation.  The hierarchy between the two branches is controlled by frequency. In particular, the coefficient of the irregular branch scales as $\omega^{-\lambda_\ell-1}$, while the regular branch scales as $\omega^{\lambda_\ell}$. Consequently, at sufficiently small frequency the outgoing solution is dominated by the irregular contribution, although the regular component remains nonvanishing.

Overall, in the near zone we can match the outgoing-wave solution onto the distorted infrared basis and obtain
\be
     \phi^{\rm out }_{\omega \ell} = B_{\rm tra} \phi^{\rm reg }_{\omega \ell} + B_{\rm inc} \phi^{\rm irreg }_{\omega \ell} \,,
\eeq
where $\phi^{{\rm reg} / {\rm irreg} }_{\omega \ell}$ are given before.  
The corresponding inner-region Green's function is then
\begin{align}
    G^{\rm MG}_\omega &(r, \gamma; r_s) = \sum_{\ell} \frac{2\ell+1}{2\lambda_\ell+1} P_\ell (\cos \gamma) \: \times \nn \\
    &\times r_<^{\lambda_\ell} r_>^{\lambda_\ell} \Bigg[   \, \frac{B_{\rm tra}}{B_{\rm inc}} \mathcal I^{\rm reg}_\ell(r_<,r_>)  + \frac{\mathcal I^{\rm irreg}_\ell(r_<,r_>) }{r_>^{2\lambda_\ell + 1} } \Bigg] \,,
\end{align}
where
\begin{align}
    \mathcal I^{\rm irreg}_\ell(r_<,r_>) =& 1 + \omega^2 r_<^2 \lp a^{\rm reg}_\ell + b^{\rm reg}_\ell \ln\frac{r_<}{r_0} \rp + \nn \\
    &+ \omega^2 r_>^2 \lp a^{\rm irreg}_\ell + b^{\rm irreg}_\ell \ln\frac{r_>}{r_0} \rp \,,
\end{align}
and
\begin{align}
    \mathcal I^{\rm reg}_\ell(r_<,r_>) =& 1 + \omega^2 r_<^2 \lp a^{\rm reg}_\ell + b^{\rm reg}_\ell \ln\frac{r_<}{r_0} \rp + \nn \\
    &+\omega^2 r_>^2 \lp a^{\rm reg}_\ell + b^{\rm reg}_\ell \ln\frac{r_>}{r_0} \rp \,.    
\end{align}
This expression should be interpreted as an inner-region expansion of the finite-frequency retarded Green's function rather than as an asymptotic radiative waveform.
Note that for $\omega \to 0$, the coefficient $ B_{\rm tra} / B_{\rm inc} \to 0$, as it is $ \propto \omega^{2 \lambda_\ell + 1} $, reproducing the correct static Green's function.

\refrep{Note that the ingoing and outgoing solutions should, in principle, receive a perturbative logarithmic SIS contribution, which can be incorporated into the matching procedure itself. In this case, the coefficients $B_{\rm tra}$ and $B_{\rm inc}$ would receive additional corrections of order $\sigma_v^2$, which, once inserted into the combination $B_{\rm tra}/B_{\rm inc}\propto\omega^{2\lambda_\ell+1}$  translate into a dimensionless correction of order $\sigma_v^2(\omega r)^{2\lambda_\ell+1}$ relative to the leading static term. For consistency, we compare this to the corrections already retained in $\phi^{\rm reg/irreg}_{\omega\ell}$: the coefficients $a^{\rm reg}_\ell$, $b^{\rm reg}_\ell$, $a^{\rm irreg}_\ell$, $b^{\rm irreg}_\ell$ in Eqs.~\eqref{eq:areg}--\eqref{eq:birreg} contain terms of explicit order $\sigma_v^2$, which multiply the overall $\omega^2 r^2$ prefactor already present in $\phi^{\rm reg/irreg}_{\omega\ell}$, and therefore represent a dimensionless  correction of order $\sigma_v^2(\omega r)^2$. Comparing these two dimensionless ratios, we see that neglecting ${\cal O}(\sigma_v^2(\omega r)^{2\lambda_\ell+1})$ corrections, while retaining those ${\cal O}(\sigma_v^2(\omega r)^{2})$ is consistent for $\ell\geq1$  in the near-zone $(\omega r \ll 1)$.  For the monopole, however, $2\lambda_0+1\approx1<2$:
the neglected correction is then of order
$O(\sigma_v^2\,\omega r)$, thus it should be included for consistency of the expansion. For simplicity, here we take the approach of \emph{not} including this corrections here, but to consider the expressions  of $a^{\rm reg}_\ell,a^{\rm irreg}_\ell$, $b^{\rm reg}_\ell,b^{\rm irreg}_\ell$ valid only up to ${\cal O}(\sigma^0_v)$. This way the matching is performed consistently for every multipole and all $\sigma_v^2$-dependence enters solely through the shifted angular momentum $\lambda_\ell$.}

\bigskip
The structure of the matching problem is closely analogous to the computation of tidal Love numbers (TLN)~\cite{Love1909}, which are the linear response coefficients quantifying how an object reacts to an external tidal field. In a standard tidal problem, one solves the static perturbation equations outside the object and decomposes the solution into two independent radial behaviors. One branch is identified with the externally applied tidal field, while the other is identified with the induced multipolar response of the object. The TLN is the coefficient of the induced response relative to the applied tidal field, after a conventional normalization has been chosen.
This notion has played an important role in black-hole perturbation theory. In four-dimensional asymptotically flat general relativity, the static TLNs of several black-hole families, including Schwarzschild, Kerr, and Reissner--Nordstrom black holes, vanish exactly~\cite{Binnington:2009bb,Damour:2009vw,Pani:2015hfa,Porto:2016zng,LeTiec:2020spy,Chia:2020yla,Charalambous:2021mea,Charalambous:2021kcz,Berens:2022ebl,Rai:2024lho,Charalambous:2022rre,Hui:2020xxx,Riva:2023rcm,DeLuca:2024ufn}. This result is physically important because tidal effects enter the gravitational waveform at fifth post-Newtonian order~\cite{Flanagan:2007ix,Hinderer:2007mb}. A nonzero black-hole TLN would therefore signal a departure from at least one of the assumptions underlying the standard four-dimensional GR result~\cite{Cardoso:2018ptl,Cardoso:2019vof,DeLuca:2021ite,DeLuca:2023mio}.
We can apply the same diagnostic to the present problem. The analog of the  static TLN computation is obtained from Eq.~\eqref{eq:PartialWaveGMGw} by setting $\omega=0$. As discussed above, the two independent static solutions are $r^{\lambda_\ell}$ and $r^{-\lambda_\ell-1}$. Therefore, after normalizing the regular branch, the most general static solution can be written in the form
\be
\label{eq:SIS_static_Love_def}
    \phi_{0\ell}(r)  \simeq  C_\ell \, r^{\lambda_\ell}  \llp 1   + \kappa_\ell^{\rm SIS, MG} \left(\frac{r_0}{r}\right)^{2\lambda_\ell+1} \rrp \,,
\eeq
where $\kappa_\ell^{\rm SIS, MG}$ is the coefficient of the singular branch relative to the regular branch, with the reference scale $r_0$ inserted to make the coefficient dimensionless. In this sense, $\kappa_\ell^{\rm SIS, MG}$ plays the role of a static TLN  coefficient for the inverse-square infrared problem of a SIS lens in the scalar theory considered. The value of $\kappa_\ell^{\rm SIS, MG}$ is fixed by the boundary condition at the lens. If the static solution is required to be regular at the lens center, the singular branch is excluded. Hence
\be
    \kappa_\ell^{\rm SIS, MG}=0\,.
\eeq
Thus, the static TLN  response vanishes in the present problem.  Note that the fact that a nonminimal coupling of the form $\propto R \phi^2$ does not produce a nonvanishing TLN  has also been found in the case of black holes for perturbations of any spin~\cite{Berens:2025okm}.

\refrep{We emphasize that the vanishing Love number follows directly from imposing regularity at the center of the SIS lens, which removes the singular branch. For a more realistic lens, one should consider a core and the boundary condition would instead be imposed at the core radius. In this situation a nonzero Love number is expected to arise, as for neutron stars.}

\section{Distorted-wave formulation of wave-optics lensing}
\label{sec:distorted_reference}

We learned  that the usual Fresnel approximation is invalid  in the present setup, since  curvature-dependent interactions qualitatively modify the infrared propagation problem. Extrapolating the standard Fresnel amplification factor beyond its regime of validity leads to the unphysical result $F^{\rm MG}_{\rm diff}\to0$ as $\nu\to0$, as shown in Fig.~\ref{fig:Fresnel_plot} and discussed in Sec.~\ref{sec:Fresnel_breakdown}. The infrared regime is intrinsically interacting, so the correct starting point is {\it not} free propagation plus perturbative corrections -- e.g. as in the (na\"ive) discussion around Eq.~\eqref{eq:F_static_def} -- but propagation in a curvature-distorted basis.

In this Section we reorganize the lensing problem accordingly. Rather than defining the amplification factor relative to a free spherical wave, as in Eq.~\eqref{eq:F_def}, we define it with respect to a reference solution which already incorporates the nonperturbative curvature-induced interaction. In particular, we will consider the reference solution of Eq.~\eqref{eq:PartialWaveRefSolution}.
This formulation provides a smooth interpolation between the $\nu\gg1$ and $\nu\ll1$ regimes, isolating the genuinely finite-frequency lensing effects.

\subsection{Distorted diffraction-integral formulation}
To construct a diffraction formalism which interpolates both at high and low frequency, we change the reference wave with respect to the usual GR treatment. Instead of defining the amplification factor relative to free propagation, we define it relative to the reference wave which already incorporates the curvature-induced interaction of Eq.~\eqref{eq:PartialWaveRefSolution}. 

We therefore decompose the full solution as
\be\label{eq:distorted_factorization}
	\phi_\omega(\bm r) \equiv F(\bmr , \omega ) \, \times \, \phi_\omega^{\rm ref}(\bm r) \,,
\eeq
instead of Eq.~\eqref{eq:F_def}, where the reference wave satisfies
\be
	\left[ \nabla^2+\omega^2-2\xi\Delta U_{\rm L} \right]\phi_\omega^{\rm ref}(\bm r)=0 \,.
\label{eq:RefFunction}
\eeq
For the SIS profile, the curvature term behaves as $\Delta U_{\rm L}\propto r^{-2}$ away from the origin, so the reference problem contains exactly the inverse-square interaction induced by the nonminimal coupling, while leaving the standard Newtonian lensing contribution proportional to $4\omega^2U_{\rm L}$ in the residual envelope equation. The reference solution therefore defines a curvature-dressed propagation basis.

\refrep{Note that the decomposition in Eq.~\eqref{eq:distorted_factorization} is not unique: a different choice of reference wave $\phi_\omega^{\rm ref}$ is compensated by a corresponding change in the amplification factor $F$, leaving the physical wave $\phi_\omega$ unchanged. We choose the specific reference wave in Eq.~\eqref{eq:RefFunction} because it reproduces the correct interacting static solution, and the solution is known analytically. As a consequence, the amplification factor defined relative to this distorted basis again approaches $F\to1$ in the static limit, allowing a consistent interpolation between the infrared and high-frequency regimes. In this way the infrared-sensitive part of the dynamics is treated nonperturbatively from the outset, while the remaining Newtonian lensing potential generates only residual finite-frequency phase shifts. Because of the arbitrariness of the decomposition in Eq.~\eqref{eq:distorted_factorization}, the amplification factor is now not an uniquely defined quantity. Thus, the key observable to take into account for future data analysis is the waveform itself, $\phi_\omega$.
}
Conceptually, this construction is analogous to distorted-wave scattering theory in quantum mechanics, where long-range interactions are absorbed into the reference problem, while the remaining interaction generates residual scattering phase shifts. In Coulomb scattering~\cite{Kthylwe1985}, this reorganization is required because the interaction decays too slowly for asymptotic states to be described by free plane waves. Here the situation is different: the curvature-induced interaction does decay spatially, but it survives in the infrared limit $\omega\to0$. The distorted basis therefore captures the nonperturbative infrared structure of the propagation problem from the beginning.

In a partial-wave expansion, the reference function satisfies Eq.~\eqref{eq:PartialWaveRefSolution}, whose ingoing and outgoing solutions are given in Eq.~\eqref{eq:IngoingOutgoingReferenceProblem}. The corresponding retarded Green's function is
\begin{align}\label{eq:ref_wave_partial}
	G_\omega^{\rm ref}(\bm r,\bmr_s)= &\ii\omega\sum_{\ell=0}^{\infty}(2\ell+1)   j_{\lambda_\ell}(\omega r_<)h_{\lambda_\ell}^{(1)}(\omega r_>)P_\ell(\cos\gamma) \,,
\end{align}
with $r_{<,>}$ given in Eq.~\eqref{eq:rminusrplus}. The shift $\ell\to\lambda_\ell$ prevents an exact resummation of the partial wave into a spherical wave, but in the eikonal regime ($\ell \gg 1$) one may approximate the reference Green's function as
\be\label{eq:GFRefEikonal}
	G_\omega^{\rm ref}(\bm r,\bmr_s) \simeq \frac{\ee^{\ii\omega |\bm r-\bmr_s|}}{|\bm r-\bmr_s|} \exp\left[\ii\Delta\Phi_{\rm ref}(b)\right],
\eeq
where $b$ is the impact parameter of the ray connecting $\bmr_s$ and $\bm r$. The reference phase is obtained from the large-$\ell$ shift of the Hankel phase,
\be\label{eq:PhaseShiftRef}
	\Delta\Phi_{\rm ref}(b) \simeq -\frac{\pi}{2} (\lambda_\ell-\ell)\bigg|_{\ell + 1/2\simeq\omega b}
	\simeq -\frac{\pi\sigma_v^2\xi}{\omega b}.
\eeq
Thus, in the geometric regime, the distorted Green's function is approximately the usual spherical propagator dressed by the phase induced by the curvature-dependent inverse-square interaction.
The advantage of this formulation is that the infrared-sensitive part of the propagation problem has been treated nonperturbatively from the outset. The remaining envelope $F$ therefore measures only the residual finite-frequency lensing effect relative to a curvature-distorted propagation basis.

\smallskip

Substituting Eq.~\eqref{eq:distorted_factorization} into the full wave equation gives the exact envelope equation
\be\label{eq:F_ref_exact}
	\partial_r^2F  + \frac{2}{r}\partial_r F + \frac{1}{r^2}\nabla_\Omega^2F  + 2\,\partial^i\ln\phi_\omega^{\rm ref}\,\partial_iF   = 4\omega^2U_{\rm L} \, F  \,. 
\eeq

We now show that, under suitable smoothness conditions on the distorted reference wave, the residual envelope obeys the standard Fresnel diffraction equation. The analog of the eikonal approximation of Eq.~\eqref{eq:fresnel_condition} corresponds
to requesting that the second radial derivative of the envelope is small compared with the first-order transport terms:
\be\label{eq:modified_eikonal_condition}
	\left|\partial_r^2 F\right| \ll \left| \frac{2}{r}\partial_r F  + 2\,\partial^i\ln\phi_\omega^{\rm ref}\, \partial_iF \right| \,. 
\eeq

For $r>r_s$, the radial dependence of the reference wave is carried by outgoing spherical Hankel functions. In the radiative regime, away from turning points and for $\omega r\gg \lambda_\ell^2$, we have 
\be
	\partial_r\ln h_{\lambda_\ell}^{(1)}(\omega r) \simeq \ii\omega-\frac{1}{r},
\eeq
so the radial transport operator reduces to the usual form,
\be
	\frac{2}{r}\partial_r F   + 2\,\partial_r\ln\phi_\omega^{\rm ref}\, \partial_rF \simeq 2\ii\omega\,\partial_rF \,.
\eeq

There remains, however, a possible transverse transport term induced by the distorted reference wave. In physical transverse coordinates $\bm b=r\bm\theta$, with
$\nabla_\Omega^2\simeq r^2 \,\nabla_b^2$, 
the envelope reduces to the standard Fresnel equation only if
\be\label{eq:transverse_ref_condition}
	\left| \partial_\perp^i\ln\phi_\omega^{\rm ref}\, \partial_i^\perp F \right| \ll \left| \omega\,\partial_rF \right|.
\eeq

When this condition is satisfied, Eq.~\eqref{eq:F_ref_exact} becomes
\be\label{eq:distorted_paraxial}
	2\ii\omega\,\partial_r F  + \frac{1}{r^2}  \nabla_\Omega^2F  - 4\omega^2U_{\rm L}F \simeq 0 \,.
\eeq
This equation has the same paraxial form as Eq.~\eqref{eq:schrodinger_like_GR}, but now for an amplification factor defined relative to the curvature-dressed reference wave rather than relative to a free wave.
Eq.~\eqref{eq:transverse_ref_condition} tells us that the reference wave must be approximately constant across the transverse region that contributes coherently to the diffraction integral. If $F$ varies over a longitudinal scale $L_r$ and a transverse Fresnel scale $b_F$, then
\be
	\partial_rF\sim\frac{F}{L_r} \,, \qquad \partial_\perp F\sim\frac{F}{b_F}.
\eeq
Condition~\eqref{eq:transverse_ref_condition} becomes
\be
	\frac{ |\partial_\perp\ln\phi_\omega^{\rm ref}|\,L_r }{b_F} \ll \omega.
\eeq
Using the paraxial scaling relation $ L_r\sim\omega b_F^2$
in a region where the potential is negligible (see
 Eq.~\eqref{eq:distorted_paraxial}), we find
\be
	|\partial_\perp\ln\phi_\omega^{\rm ref}| \ll \frac{1}{b_F}.
\label{eq:ref_smooth_fresnel}
\eeq
Equivalently, defining
\be
	b_{\rm MG}^{-1} \sim |\partial_\perp\ln\phi_\omega^{\rm ref}|,
\eeq
the required hierarchy becomes
\be
	b_F\ll b_{\rm MG}.
\eeq
Thus, the curvature-dressed reference wave must vary only on transverse scales larger than a coherent Fresnel patch. When this condition holds, the residual diffraction problem is described by the standard thin-lens integral, but with respect to the distorted propagation basis. An estimate of the transverse variation of the reference solution, in the eikonal regime using Eqs.~\eqref{eq:GFRefEikonal} and~\eqref{eq:PhaseShiftRef}, is  
\be
	|\partial_\perp\ln\phi_\omega^{\rm ref}| \sim |\partial_b\Phi_{\rm MG}| \sim  \frac{\sigma_v^2\xi}{\omega b^2}.
\eeq
The smooth-reference condition~\eqref{eq:ref_smooth_fresnel} 
corresponds to
\be
\label{eq_sigv2}
	\sigma_v^2\xi \ll \omega\,\frac{b^2}{b_F}.
\eeq
Within the coherent diffraction region we have  $b\sim b_F$, so Eq.~\eqref{eq_sigv2} further reduces to
\be
	\sigma_v^2\xi \ll \omega b_F.
\eeq
Equivalently, the transverse modulation scale of the distorted reference wave must remain much larger than the Fresnel scale. Under this condition, the distorted-wave formulation consistently reduces to the standard Fresnel diffraction problem for the residual envelope.

\subsection{Scattering amplitude formulation}

The  separation between the infrared interaction and the residual finite-frequency lensing effect, as
discussed in the previous Subsection, admits a natural formulation in scattering theory. In ordinary scattering theory we consider
an incoming plane wave, which interacts with a target and produces an outgoing scattered wave at large distance. The incoming plane wave admits the standard partial-wave decomposition
\be 
    e^{\ii k z} = \sum_{\ell=0}^{\infty}(2\ell+1)\ii^\ell j_{\ell}(\omega r)P_\ell(\cos\theta)\,,
\eeq 
where $\theta$ denotes the scattering angle between the optical axis and the observer.
To follow the same logic as in the modified diffraction-integral construction, we replace free propagation with the reference problem defined by Eq.~\eqref{eq:RefFunction}. 
We therefore consider the distorted analog of an incoming spherical wave
\be
	\phi_\omega^{\rm ref}(\bm r)= G_\omega^{\rm ref}(\bm r,\bmr_s)
\label{eq:dist_ref_plane_wave}
\eeq
with $G_\omega^{\rm ref}$ given in Eq.~\eqref{eq:ref_wave_partial}. This way the effect of the curvature-induced inverse-square interaction is already encoded in the propagation basis.

Using the reference Green's function, the full solution of Eq.~\eqref{eq:helmholtz_MG} can be expressed in Lippmann--Schwinger form,
\be\label{eq:distorted_LS}
	\phi_\omega(\bm r)=\phi_\omega^{\rm ref}(\bm r)-\frac{4 \omega^2}{4\pi}\int \dd^3\bm r'\, G^{\rm ref}_\omega (\bm r,\bm r') U_{\rm L}(\bm r') \phi_\omega(\bm r')\,,
\eeq
with $G^{\rm ref}_\omega$ corresponding to the reference Green's function of Eq.~\eqref{eq:ref_wave_partial}. Equation~\eqref{eq:distorted_LS} is implicit, since the full wave appears on both sides, but it naturally defines scattering relative to the distorted propagation basis.

To proceed, one must now invoke the infrared completion of the lens profile. For an SIS lens, $U_{\rm L}\sim \log(r/r_0)$ grows with distance and therefore does not define a standard asymptotic scattering problem by itself. We therefore assume that beyond some radius   the SIS profile is matched onto a profile whose potential decays sufficiently rapidly at infinity. In this way the residual potential acquires an effective finite support, allowing us to consider the asymptotic region $r\gg r'$ in Eq.~\eqref{eq:distorted_LS}.
In this regime the outgoing Hankel function appearing in the reference Green's function behaves asymptotically as
\be\label{eq:HankelAsymptotics}
	h_{\lambda_\ell}^{(1)}(\omega r) \sim \frac{1}{\omega r} \exp \llp \ii\omega r - \frac{\ii\pi\lambda_\ell}{2} \rrp \,.
\eeq
Hence, the scattered wave acquires the usual outgoing behavior proportional to $\ee^{\ii\omega r}/r$. We may therefore define the scattering amplitude from Eq.~\eqref{eq:distorted_LS} as
\be\label{eq:ScatteringAmplitudeDef}
	\phi_\omega(\bm r)=\phi_\omega^{\rm ref}(\bm r)+f^{\rm ref}(\bmt)\frac{e^{\ii \omega r}}{r}.
\eeq
This is the analog of the standard scattering decomposition, this time formulated relative to the distorted reference wave rather than relative to free propagation. The corresponding partial-wave expansion follows from the asymptotic behavior of the full and reference solutions. Outside the support of the residual potential, the full radial equation reduces again to the reference equation, so the asymptotic solutions are linear combinations of incoming and outgoing Hankel functions. Normalizing the incoming wave such that it coincides with the reference solution, the residual effect of the $4 \omega^2 U_{\rm L}$ potential is encoded in a phase shift $\delta_\ell^{\rm ref}$ of the outgoing component. The derivation then proceeds exactly as in  standard textbook treatments~\cite{1988sfbh.book.....F,Newton:1982qc}, with the replacement $\ell\rightarrow\lambda_\ell$, leading to
\be\label{eq:partial_wave_amp_full}
    f^{\rm ref}(\bmt)=\frac{1}{2 \ii \omega}\sum_{\ell=0}^{\infty}(2 \ell + 1 )\left(e^{2 \ii \delta_\ell^{\rm ref}} - 1 \right)P_\ell(\cos\theta)\,.
\eeq
The phase shifts $\delta_\ell^{\rm ref}$ are now defined relative to the distorted reference wave $j_{\lambda_\ell}(\omega r)$, instead of the incoming plane wave $j_{\ell}(\omega r)$. They may  be computed explicitly in the eikonal regime, where $\ell\gg1$ and $\theta\ll1$. In this limit we introduce the impact parameter
\be
	b=\frac{\lambda_\ell+1/2}{\omega},
\eeq
and use
\be
	P_\ell(\cos\theta)\simeq J_0(qb),\qquad q\simeq\omega\theta.
\eeq
At leading eikonal order, $\lambda_\ell=\ell+\mathcal O(\ell^{-1})$, so the difference between defining $b$ with $\lambda_\ell$ or with $\ell$ only affects subleading terms. We therefore use the standard leading-order replacement of the sum by an integral,
\be
	\sum_{\ell=0}^{\infty}(2\ell+1)\longrightarrow2\omega^2\int_0^\infty \dd b\, b,
\eeq
so that the distorted scattering amplitude becomes
\be
	f^{\rm ref}_{\rm eik}(\theta,\omega)\simeq-\ii\omega\int_0^\infty \dd b\, b\,J_0(qb)\left[\ee^{2\ii \delta_{\rm eik}^{\rm ref}(b,\omega)}-1\right].
\label{eq:eikonal_amp_ref}
\eeq

The residual phase shift may be computed directly from the radial equation in the WKB approximation. Since the curvature-induced inverse-square interaction has already been absorbed into the reference problem, the remaining phase shift is obtained by comparing the WKB phase of the full solution with that of the reference solution:
\be
	\delta^{\rm ref}_{\rm eik}(b,\omega)= \omega \int^\infty \dd z \left[\sqrt{1-4 U_{\rm L}-\frac{ b^2}{r^2}}-\sqrt{1-\frac{ b^2}{r^2}}\right].
\eeq
Expanding to first order in the SIS potential, we obtain
\be
	\delta_{\rm eik}^{\rm ref}(b,\omega)  \simeq -2\omega\int_{-\infty}^{+\infty}\dd z\,U_{\rm L}(b,z).
\eeq
This is the standard eikonal relation between the scattering phase and the projected lensing potential -- now formulated relative to the curvature-dressed propagation basis. Once the nonperturbative infrared interaction is absorbed into the reference wave, the residual finite-frequency scattering phase reduces to the usual projected Newtonian phase. For an SIS lens, $\psi(b)$ is logarithmic, so the distorted eikonal phase shift reproduces the standard SIS logarithmic phase, while the modified-gravity information is entirely carried by the distorted propagation basis through $\lambda_\ell$.

In~\cite{CarrilloGonzalez:2025gqm} it was shown that there is a direct connection between the eikonal scattering amplitude and the diffraction integral. Here we recover the same relation in the distorted-wave formulation, showing that it is independent of the particular reference function used to define the amplification factor or scattering amplitude.

The condition $b_{\rm MG}\gg b_F$ derived in the diffraction formulation does not appear explicitly in the scattering derivation. However, it is the corresponding consistency condition for the eikonal approximation in the diffraction regime. It ensures that the distorted reference wave does not develop appreciable transverse structure across a coherent Fresnel patch, so that the eikonal phase may still be computed along the usual approximately straight trajectory.

This construction therefore separates a nonperturbative modification of the propagation basis -- encoded in $\lambda_\ell$ and surviving in the infrared limit $\omega\to0$ --  from a genuine finite-frequency scattering effect -- encapsulated  in the residual phase shifts $\delta_\ell^{\rm ref}(\omega)$. This is the scattering-theory counterpart of the statement that the low-frequency limit is governed by an interacting Green's function rather than by free propagation.

\section{Numerical comparisons and estimates for LISA}
\label{sec_numest}

We now compare numerically the size of the correction induced by the curvature-dressed reference Green's function. For this purpose we need to evaluate the reference Green's function in Eq.~\eqref{eq:ref_wave_partial}. We assume $r>r_s$ for concreteness. The partial-wave sum is effectively supported up to $\ell_{\rm max}\sim \omega r_s$, since $j_\ell(x)$ suppresses modes with $\ell\gg x$. For LISA sources with frequency in the millihertz range,
\be 
 f= 10^{-3} {\rm Hz} \,, \qquad \omega = \frac{2 \pi f}{c} \simeq 10^6 {\rm pc}^{-1}
\eeq 
Thus, even for a source relatively close to the lens, $r_s \sim 1 {\rm kpc}$, one has 
\be 
\ell_{\rm max} \sim \omega r_s \simeq 6 \times 10^8\,.
\eeq 
The partial-wave sum in Eq.~\eqref{eq:ref_wave_partial} is therefore dominated by large-$\ell$ modes, which motivates using the eikonal form in Eq.~\eqref{eq:GFRefEikonal}. In this regime the effect of the modified propagation is encoded in the additional reference phase $\Delta\Phi_{\rm ref}$.

We first isolate this effect by comparing the SIS solution in GR with the corresponding solution in the modified theory, keeping the same SIS amplification factor and changing only the reference Green's function. In particular, we define
\be
     \Delta^{\rm MG - GR} \equiv \frac{\phi_\omega - \phi^{\rm GR}_{\omega}}{\phi^{\rm GR}_{\omega}} =
    e^{i  \Delta \Phi_{\rm ref}(b)} - 1 
\eeq 
where $\phi_\omega = F^{\rm SIS} G^{\rm ref}_\omega$ from Eq.~\eqref{eq:distorted_factorization}, while $\phi^{\rm GR}_{\omega}$ is the GR solution discussed in Sec.~\ref{sec_GR}, obtained by setting $\xi=0$. The phase difference is given in Eq.~\eqref{eq:PhaseShiftRef}. In the dimensionless variables used for the SIS diffraction integral, it becomes
\be\label{eq:PhaseShiftRefdimensionless}
	\Delta\Phi_{\rm ref}(b) = -\frac{ \varepsilon}{ 2 \nu x}
\eeq
where $b = \zsl \theta_E x$. This expression shows that the correction is enhanced at low dimensionless frequency and at small impact parameter. Therefore, for fixed $x$, decreasing $\nu$ increases the accumulated reference phase, while decreasing $\varepsilon$ shifts the same effect to lower frequencies.

For LISA-band sources, the SIS wave-optics regime is naturally associated with sub-galactic effective lens masses~\cite{Caliskan:2022hbu}, roughly $M_{\rm SIS}\sim 10^6-10^8M_\odot$, where
\be
    M_{\rm SIS}= 4\pi^2\sigma_v^4\frac{\zlo\zsl}{\zso}\,.
\eeq
For $f\sim10^{-3}{\rm Hz}$, this gives
\be
    \nu = 4M_{\rm SIS}\omega  
    \simeq 1.2\times10^{-7}\left(\frac{M_{\rm SIS}}{M_\odot}\right) \,,
\eeq
and therefore $\nu \in (10^{-1}, 10)$ over the mass range above.

\begin{figure}
    \centering
    \includegraphics[width=\linewidth]{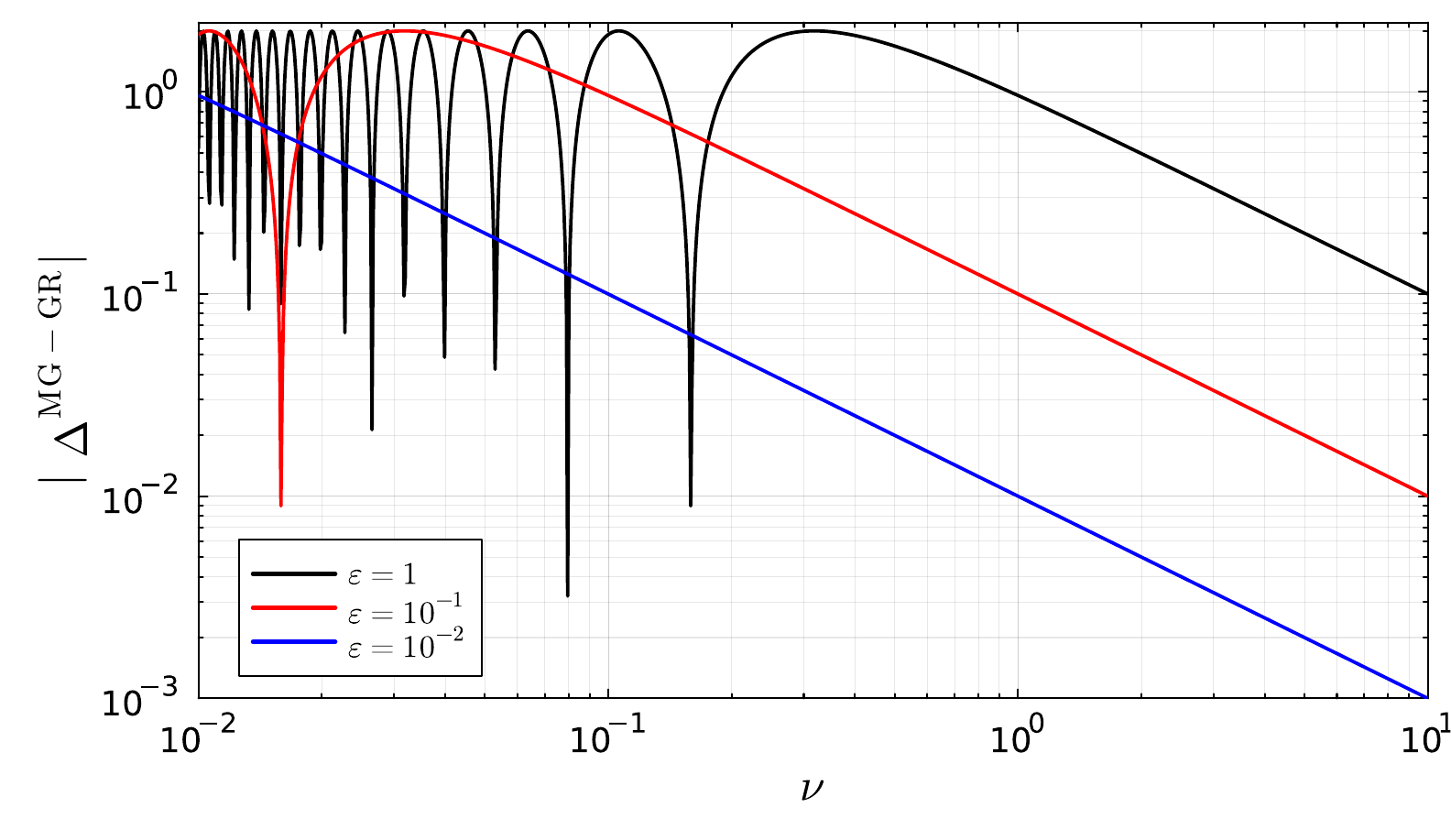}
    \caption{Relative correction induced by the curvature-dressed reference phase for fixed impact parameter $x=0.5$. We plot $|\Delta^{\rm MG-GR}|=|e^{i\Delta\Phi_{\rm ref}}-1|$ as a function of the dimensionless frequency $\nu$, for different values of the curvature-coupling parameter $\varepsilon$. The correction grows toward small $\nu$, as expected from $\Delta\Phi_{\rm ref}=-\varepsilon/(2\nu x)$, while decreasing $\varepsilon$ shifts the onset of the effect to lower frequencies.}
    \label{fig:RelativeDifference_y05}
\end{figure}

Figure~\ref{fig:RelativeDifference_y05} shows the reference-wave correction at fixed observer position $x=0.5$. The figure plots $|\Delta^{\rm MG-GR}|$, or equivalently $|e^{i\Delta\Phi_{\rm ref}}-1|$, as a function of $\nu$. For $\varepsilon=1$, the correction is already large throughout most of the plotted range and displays rapid oscillations at low $\nu$, reflecting the fact that $\Delta\Phi_{\rm ref}\propto 1/\nu$. For smaller $\varepsilon$, the same pattern is shifted to lower frequencies: the $\varepsilon=10^{-1}$ and $\varepsilon=10^{-2}$ curves show that the onset of an order-one phase correction occurs when $\nu x\sim \varepsilon$. This plot therefore provides a direct diagnostic of the modified reference phase, independently of the detailed structure of the SIS diffraction integral.

\begin{figure}
    \centering
    \includegraphics[width=\linewidth]{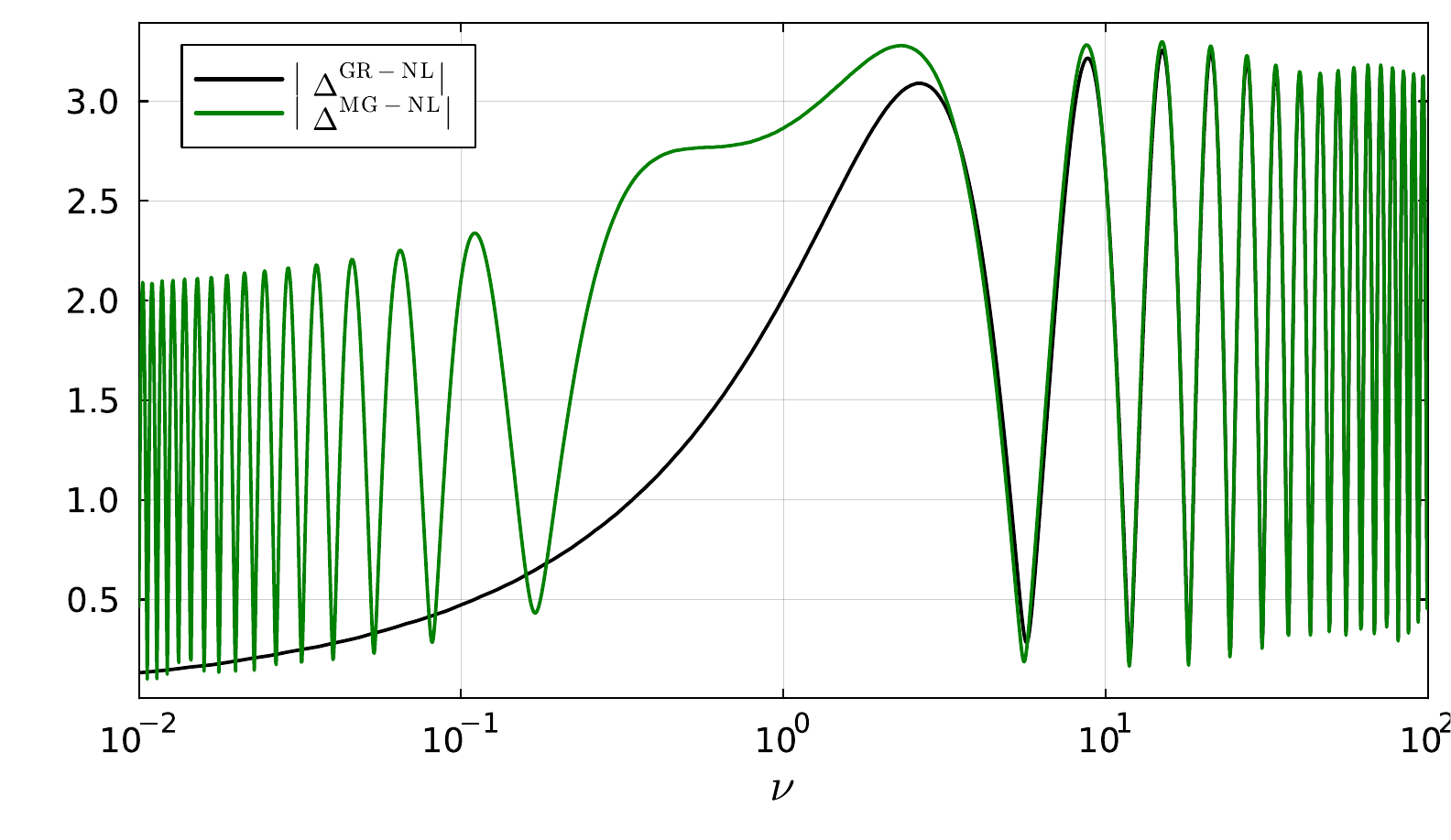}
    \caption{Comparison between the ordinary SIS lensing distortion and the modified-gravity distortion relative to the unlensed waveform, for fixed $x=0.5$ and $\varepsilon=1$. The black curve shows $|\Delta^{\rm GR-NL}|=|F^{\rm SIS}-1|$, while the Green's curve shows $|\Delta^{\rm MG-NL}|=|F^{\rm SIS}e^{i\Delta\Phi_{\rm ref}}-1|$. At large $\nu$, the reference phase becomes small and the MG result approaches the GR one. At lower $\nu$, the extra phase shifts the SIS interference pattern and produces rapid oscillations.}
    \label{fig:RelativeDifferenceMGGRNL_y05}
\end{figure}

We then compare the lensed waveforms directly with the corresponding unlensed waveform. This is a more physical diagnostic, because it keeps the usual SIS diffraction pattern and asks how much the lensed signal differs from the no-lens result in GR and in the modified theory. We define
\begin{align}
    \Delta^{\rm GR - NL} &\equiv \frac{\phi^{\rm GR}_\omega - \phi^{\rm NL}_{\omega}}{\phi^{\rm NL}_{\omega}} =  F^{\rm SIS}  - 1\,,\\
    \Delta^{\rm MG - NL} &\equiv \frac{\phi_\omega - \phi^{\rm NL}_{\omega}}{\phi^{\rm NL}_{\omega}} =  F^{\rm SIS}   e^{i  \Delta \Phi_{\rm ref}(b)} - 1\,,
\end{align}
where $\phi^{\rm NL}_{\omega}$ is the unlensed waveform.
Here the GR curve measures the ordinary SIS lensing distortion relative to the unlensed wave, while the MG curve includes both the SIS diffraction factor and the additional curvature-dressed reference phase. The absolute values of these quantities are shown in Fig.~\ref{fig:RelativeDifferenceMGGRNL_y05} for $\varepsilon=1$.

The comparison shows that the modified reference phase can change the apparent lensing distortion in a way that is not simply an overall rescaling of the GR result. At high $\nu$, the phase $\Delta\Phi_{\rm ref}$ becomes small, and the MG curve approaches the GR curve. In this regime the modified propagation has little impact on the standard SIS diffraction pattern. At intermediate frequencies, however, the additional phase shifts the interference pattern, changing both the height and the position of the peaks and troughs. At low $\nu$, the phase varies rapidly as a function of frequency, producing the oscillatory behavior visible in the MG curve. This reflects the fact that, for $\varepsilon=1$ and $x=0.5$, one has $\Delta\Phi_{\rm ref}=-1/\nu$, so small changes in $\nu$ correspond to large changes in the reference phase.

For this benchmark value, the consistency of the eikonal resummation is controlled by the smallness of the angular-momentum shift relative to the angular momentum itself. In the dimensionless variables this amounts to $\varepsilon/(\nu^2x^2)\ll1$. Thus the high-frequency part of Fig.~\ref{fig:RelativeDifferenceMGGRNL_y05} lies in the controlled regime of the eikonal reference Green's function, while the low-frequency part should be interpreted as illustrating the continuation of the same phase model into the strongly oscillatory regime. This is useful for visualizing the trend, but the quantitatively controlled comparison is the approach of the MG curve to the GR curve at larger $\nu$.

\begin{figure}
    \centering
    \includegraphics[width=\linewidth]{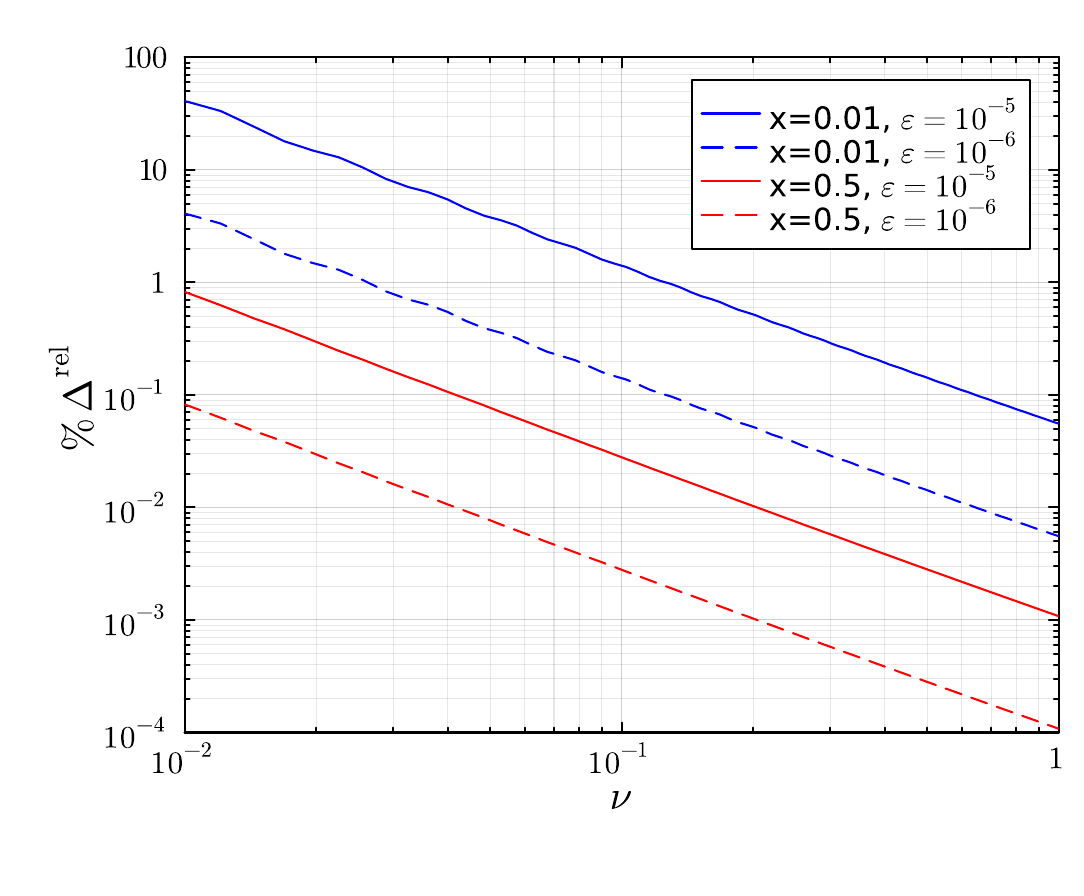}
    \caption{Percentage difference between the MG and GR lensing distortions, normalized to the GR lensed distortion [see Eq.~\eqref{eq:DeltaRef}], as a function of the dimensionless frequency $\nu$. The curves show two impact-parameter values, $x=0.01$ and $x=0.5$, and two values of the curvature-coupling parameter, $\varepsilon=10^{-5}$ and $\varepsilon=10^{-6}$. The correction is largest at low frequency and small impact parameter, while it is rapidly suppressed for larger $x$ or smaller $\varepsilon$.}
    \label{fig:MultiEpsilonDifference}
\end{figure}

One may wonder what values of $\varepsilon$ correspond to reasonable values of the coupling $\xi$. Using the effective SIS mass scale, the dimensionless curvature-coupling parameter can be written as
\be
 \varepsilon = 8 \pi^2 \sigma_v^4 \xi
 = 2 \xi \left( \frac{M^{\rm SIS} \zso}{\zsl \zlo} \right) .
\eeq
Inverting this relation gives
\begin{align}\label{eq:XiEpsilon}
    \xi \simeq
    \varepsilon
    \left( \frac{M_\odot}{M_{\rm SIS}} \right)
    \left( \frac{\zsl}{\rm pc} \right)
    \left( \frac{\zlo}{\zso} \right)
    \times 10^{13} .
\end{align}
For a representative cosmological lensing configuration, with
$\zso,\zsl,\zlo \sim 100\,{\rm Mpc}$, and for the SIS masses considered above, this estimate gives
\begin{align}
    \xi \simeq \varepsilon \,[10^{13}-10^{15}] .
\end{align}
Thus, values $\varepsilon \sim 1$ would require an extremely large coupling $\xi$.
Conversely, keeping $\xi$ in a more moderate range forces $\varepsilon$ to be very small, suppressing the corresponding phenomenological effect.
This is illustrated in Fig.~\ref{fig:MultiEpsilonDifference}, where we show the percentage difference
\begin{align}\label{eq:DeltaRef}
    \%\,\Delta_{\rm rel} \equiv 100 \times  \left| \frac{\Delta^{\rm GR-NL}-\Delta^{\rm MG-NL}} {\Delta^{\rm GR-NL}}  \right| \,.
\end{align}
The correction is largest at low frequency and for small impact parameters. In particular, for $x=0.01$ and $\varepsilon=10^{-5}$ the relative difference can reach the level of tens of percent at $\nu \sim 10^{-2}$, whereas it is strongly reduced for $x=0.5$ or for $\varepsilon=10^{-6}$.

A possible way to reduce the required value of $\xi$ is to consider configurations in which the source lies much closer to the lens. Such a hierarchy can arise in compact hierarchical triples, where the binary source may be separated from the lens by only $\zsl \sim 10^{-3}-1\,{\rm pc}$, while the system as a whole remains at cosmological distance, $\zlo,\zso \sim 100\,{\rm Mpc}$. Although these estimates are usually formulated for black-hole lenses~\cite{Antonini:2012ad,Berczik_2024,Cardoso:2026ugm,Pijnenburg:2024btj}, a formal extrapolation to the SIS case gives
\begin{align}\label{eq:XiEpsilonLISA}
    \xi \simeq \varepsilon \,[10^{2}-10^{7}] \,.
\end{align}
In this regime, values of $\varepsilon$ large enough to generate visible deviations in Fig.~\ref{fig:MultiEpsilonDifference} could correspond to much less extreme values of $\xi$.
Close source-lens configurations therefore provide a more favorable regime in which the modified-gravity correction may become relevant without requiring an unnaturally large coupling.

\refrep{One could also consider ground based detectors, where the relevant frequency is $f \sim 100 {\rm Hz}$. In this case, one has $\nu \in (10^{-1}, 10)$ for much smaller SIS masses, of the order  $M_{\rm SIS}\sim 10-10^3M_\odot$, reflecting the five orders of magnitude difference in the frequency range. Considering the same hierarchical-triple configuration as
before, Eq.~\eqref{eq:XiEpsilon} then gives $\xi \simeq \varepsilon \,[10^{7}-10^{12}]$, i.e. five orders of magnitude larger than the LISA estimate in Eq.~\eqref{eq:XiEpsilonLISA}.for the same $\varepsilon$.
Ground-based detectors can therefore only probe this effect for correspondingly larger values of the curvature coupling $\xi$, making space-based detectors the more sensitive probe of curvature-induced propagation effects.}

\section{Discussion}

Modifications of General Relativity remain of considerable theoretical interest in connection with the physics of the dark sector, cosmology, and the effective-field-theory description of gravity. Several modified-gravity theories predict departures from standard GR in the propagation of gravitational waves, making propagation effects particularly valuable observational probes of gravity beyond Einstein theory.

In this work we initiated the study of wave-optics gravitational lensing in modified gravity. We considered a simple phenomenological model involving a curvature-dependent coupling in the propagation equation, which we used as a proxy for more general curvature-induced propagation effects. The motivation for focusing on wave optics is straightforward: curvature-dependent corrections are generically negligible in geometric optics, since the wavelength of a high-frequency wave is much shorter than the characteristic curvature scale of the background. Wave optics therefore provides a natural regime in which these effects can accumulate coherently and become observable.

We first showed that the low-frequency behavior of wave-optics lensing is not universal, but depends on the dynamical structure of the propagation equation. In standard GR, the lensing interaction is proportional to $\omega^2$, so the interaction switches off as $\omega\to0$ and the static problem reduces to free propagation. In the curvature-coupled model considered here, this argument no longer applies. The curvature-induced interaction survives in the infrared limit, implying that the $\omega\to0$ problem remains interacting.

This has an immediate consequence for the standard Fresnel treatment. If the usual diffraction integral is extrapolated to sufficiently low frequency in the modified theory, it predicts that the amplification factor tends to zero (see Fig.~\ref{fig:Fresnel_plot}). We showed that this behavior is not physical. Rather, it signals the breakdown of the Fresnel/paraxial approximation itself: the second radial derivative discarded in the Fresnel reduction is no longer subleading once the curvature-induced interaction remains active in the infrared.

To determine the correct infrared behavior, we returned to the full second-order wave equation. For a singular isothermal sphere lens, the curvature coupling generates an inverse-square interaction. This makes the problem analytically tractable, since the radial equation can be written in terms of an effective angular momentum parameter. The static limit is then governed by an interacting Green's function rather than by the free Green's function of GR. This provides the correct infrared reference problem and makes explicit that the propagation basis itself is modified.

We then constructed the corresponding finite-frequency description. In the small-frequency regime, the finite-frequency solutions can be matched onto the distorted static basis. In this language one can define response coefficients analogous to tidal Love numbers for the SIS infrared problem. For regular boundary conditions at the lens center these coefficients vanish, but the propagation basis remains nontrivially distorted by the curvature-induced inverse-square interaction.

Motivated by this structure, we proposed a distorted-wave formulation of the lensing problem. Instead of defining the amplification factor relative to a free spherical wave, we define it relative to a curvature-dressed reference wave which already contains the infrared interaction. The remaining Newtonian lensing potential then produces residual finite-frequency phase shifts. This reorganization restores a smooth interpretation of the infrared limit: the amplification factor measures lensing relative to the correct interacting propagation basis, rather than relative to a free wave which is no longer the natural reference solution.

We also showed that the same logic admits a scattering-amplitude interpretation. The distorted reference wave plays the role of the incoming propagation basis, while the residual scattering amplitude is generated by phase shifts defined relative to this basis. In the eikonal limit, this construction reproduces the same projected Newtonian phase appearing in the diffraction formulation. Thus the relation between the eikonal scattering amplitude and the diffraction integral persists, but it must be formulated relative to the appropriate curvature-dressed reference problem.

Overall, the picture that emerges is simple. Curvature-dependent propagation effects may be invisible in geometric optics, but they can reorganize the infrared structure of wave propagation. In the SIS example this reorganization is captured by the shifted angular momentum parameter of the inverse-square reference problem. The residual amplification factor retains the usual SIS form only after the correct distorted reference wave has been factored out.

The numerical estimates of Sec.~\ref{sec_numest} illustrate the size and qualitative form of the effect. Figure~\ref{fig:RelativeDifference_y05} isolates the reference-wave correction and shows that it grows toward small dimensionless frequency, consistently with $\Delta\Phi_{\rm ref}\propto1/\nu$. Decreasing the effective coupling $\varepsilon$ shifts the onset of the correction to lower frequencies, without changing the qualitative behavior. Figure~\ref{fig:RelativeDifferenceMGGRNL_y05} then compares the lensed waveform with the unlensed one, both in GR and in MG. The modified reference phase does not simply rescale the GR SIS signal: it shifts the interference pattern and changes both the height and the position of the oscillatory features around $\nu \sim 1$. At high $\nu$, the modified curve approaches the GR one, as expected from the suppression of the curvature-induced phase in the geometric-optics regime.
Figure~\ref{fig:MultiEpsilonDifference} further shows that even much smaller values, such as $\varepsilon=10^{-5}$, can lead to percent-level or tens-of-percent differences at low $\nu$ for small impact parameter, while the effect is rapidly suppressed for larger $x$ and smaller $\varepsilon$.

These estimates also clarify the observational regime where the effect is most relevant. For LISA-band sources, with $f\sim10^{-3}{\rm Hz}$, sub-galactic effective SIS masses $M_{\rm SIS}\sim10^6-10^8M_\odot$ correspond to $\nu\sim10^{-1}-10$, precisely the range where wave-optics interference is important. For ground-based detectors, taking instead $f\sim10^2{\rm Hz}$, the same lenses would give much larger values, $\nu\sim10^4-10^6$. They are therefore deep in the geometric-optics regime, where the curvature-induced correction to the diffraction factor is suppressed, and the solution approaches the GR result.

Our results illustrate the power of wave-optics lensing as a probe of modified gravity. Effects that are suppressed, or entirely degenerate with GR, in the geometric-optics regime can become visible once the finite wavelength of the gravitational wave resolves the infrared structure of the propagation problem. The present work should therefore be regarded as a first step toward a broader theory of wave-optics lensing beyond GR. It would be important to extend the analysis to different lens profiles, beyond the SIS model considered here, and to lens populations relevant for LISA. 
A dedicated parameter-estimation analysis for LISA would be needed to determine how well wave-optics lensing can constrain the modified-gravity parameters, and whether these constraints can be competitive with other cosmological tests of gravity. 
It would also be interesting to study other modified-gravity theories, in which the propagation equation may contain different curvature operators, additional polarizations, or nontrivial mixing between tensor and scalar modes. These extensions would clarify how general the mechanism identified here is, and whether wave-optics lensing can provide a systematic observational window on modified gravitational-wave propagation.

\acknowledgments
It is a pleasure to thank Valerio De Luca for discussions. A.G. is supported by funds provided by the Center for Particle Cosmology at the University of Pennsylvania.
G.T. is partially funded by the STFC Grant No. ST/X000648/1.

\bibliography{bibliography}

@article{Hui:2016ltb,
    author = "Hui, Lam and Ostriker, Jeremiah P. and Tremaine, Scott and Witten, Edward",
    title = "{Ultralight scalars as cosmological dark matter}",
    eprint = "1610.08297",
    archivePrefix = "arXiv",
    primaryClass = "astro-ph.CO",
    doi = "10.1103/PhysRevD.95.043541",
    journal = "Phys. Rev. D",
    volume = "95",
    number = "4",
    pages = "043541",
    year = "2017"
}

@article{Zumalacarregui:2026uqs,
    author = "Zumalac{\'a}rregui, Miguel and Shan, Xikai",
    title = "{Effective description of lensed gravitational waves diffracted by stellar fields}",
    eprint = "2606.17765",
    archivePrefix = "arXiv",
    primaryClass = "astro-ph.HE",
    month = "6",
    year = "2026"
}

@article{Liu:2024xxn,
    author = "Liu, Anna and Chandramouli, Rohit S. and Hannuksela, Otto A. and Yunes, Nicol{\'a}s and Li, Tjonnie G. F.",
    title = "{Millilensing induced systematic biases in parametrized tests of general relativity}",
    eprint = "2410.21738",
    archivePrefix = "arXiv",
    primaryClass = "gr-qc",
    doi = "10.1103/bsml-2rdh",
    journal = "Phys. Rev. D",
    volume = "112",
    number = "6",
    pages = "064043",
    year = "2025"
}

@article{Chan:2024qmb,
    author = "Chan, Juno C. L. and Seo, Eungwang and Li, Alvin K. Y. and Fong, Heather and Ezquiaga, Jose M.",
    title = "{Detectability of lensed gravitational waves in matched-filtering searches}",
    eprint = "2411.13058",
    archivePrefix = "arXiv",
    primaryClass = "gr-qc",
    doi = "10.1103/PhysRevD.111.084019",
    journal = "Phys. Rev. D",
    volume = "111",
    number = "8",
    pages = "084019",
    year = "2025"
}

@article{Mishra:2023ddt,
    author = "Mishra, Anuj and Meena, Ashish Kumar and More, Anupreeta and Bose, Sukanta",
    title = "{Exploring the impact of microlensing on gravitational wave signals: Biases, population characteristics, and prospects for detection}",
    eprint = "2306.11479",
    archivePrefix = "arXiv",
    primaryClass = "astro-ph.CO",
    doi = "10.1093/mnras/stae836",
    journal = "Mon. Not. Roy. Astron. Soc.",
    volume = "531",
    number = "1",
    pages = "764--787",
    year = "2024"
}

@article{Antonini:2012ad,
    author = "Antonini, Fabio and Perets, Hagai B.",
    title = "{Secular evolution of compact binaries near massive black holes: Gravitational wave sources and other exotica}",
    eprint = "1203.2938",
    archivePrefix = "arXiv",
    primaryClass = "astro-ph.GA",
    doi = "10.1088/0004-637X/757/1/27",
    journal = "Astrophys. J.",
    volume = "757",
    pages = "27",
    year = "2012"
}

@article{Berczik_2024,
   title={Dynamical evolution timescales for the triple supermassive black hole system in NGC 6240},
   volume={687},
   ISSN={1432-0746},
   url={http://dx.doi.org/10.1051/0004-6361/202450141},
   DOI={10.1051/0004-6361/202450141},
   journal={Astronomy and Astrophysics},
   publisher={EDP Sciences},
   author={Berczik, P. and Sobolenko, M. and Ishchenko, M.},
   year={2024},
   month=jul, pages={L18} }

@article{Cardoso:2026ugm,
    author = "Cardoso, Vitor and Ficarra, Giuseppe and Redondo-Yuste, Jaime and Santos, Jo{\~a}o Sieiro dos",
    title = "{The third wheel: ringdown and lensing of triple systems}",
    eprint = "2605.20320",
    archivePrefix = "arXiv",
    primaryClass = "gr-qc",
    month = "5",
    year = "2026"
}

@article{LIGOScientific:2025rsn,
    author = "Abac, A. G. and others",
    collaboration = "LIGO Scientific, VIRGO, KAGRA",
    title = "{GW231123: A Binary Black Hole Merger with Total Mass 190 -265 M$_{\odot}$}",
    eprint = "2507.08219",
    archivePrefix = "arXiv",
    primaryClass = "astro-ph.HE",
    reportNumber = "DCC: P2500026-v6, DCC: P2500026-v8",
    doi = "10.3847/2041-8213/ae0c9c",
    journal = "Astrophys. J. Lett.",
    volume = "993",
    number = "1",
    pages = "L25",
    year = "2025"
}

@article{Unruh:1976fm,
    author = "Unruh, W. G.",
    title = "{Absorption Cross-Section of Small Black Holes}",
    doi = "10.1103/PhysRevD.14.3251",
    journal = "Phys. Rev. D",
    volume = "14",
    pages = "3251--3259",
    year = "1976"
}

@article{ET:2025xjr,
    author = "Abac, Adrian and others",
    collaboration = "ET",
    title = "{The Science of the Einstein Telescope}",
    eprint = "2503.12263",
    archivePrefix = "arXiv",
    primaryClass = "gr-qc",
    reportNumber = "ET-0036C-25",
    doi = "10.1088/1475-7516/2026/03/081",
    journal = "JCAP",
    volume = "03",
    pages = "081",
    year = "2026"
}

@article{Barack:2018yly,
    author = "Barack, Leor and others",
    title = "{Black holes, gravitational waves and fundamental physics: a roadmap}",
    eprint = "1806.05195",
    archivePrefix = "arXiv",
    primaryClass = "gr-qc",
    doi = "10.1088/1361-6382/ab0587",
    journal = "Class. Quant. Grav.",
    volume = "36",
    number = "14",
    pages = "143001",
    year = "2019"
}

@article{LISA:2024hlh,
    author = "Colpi, Monica and others",
    collaboration = "LISA",
    title = "{LISA Definition Study Report}",
    eprint = "2402.07571",
    archivePrefix = "arXiv",
    primaryClass = "astro-ph.CO",
    month = "2",
    year = "2024"
}

@article{Nair:2019iur,
    author = "Nair, Remya and Perkins, Scott and Silva, Hector O. and Yunes, Nicol{\'a}s",
    title = "{Fundamental Physics Implications for Higher-Curvature Theories from Binary Black Hole Signals in the LIGO-Virgo Catalog GWTC-1}",
    eprint = "1905.00870",
    archivePrefix = "arXiv",
    primaryClass = "gr-qc",
    doi = "10.1103/PhysRevLett.123.191101",
    journal = "Phys. Rev. Lett.",
    volume = "123",
    number = "19",
    pages = "191101",
    year = "2019"
}

@article{Joyce:2014kja,
    author = "Joyce, Austin and Jain, Bhuvnesh and Khoury, Justin and Trodden, Mark",
    title = "{Beyond the Cosmological Standard Model}",
    eprint = "1407.0059",
    archivePrefix = "arXiv",
    primaryClass = "astro-ph.CO",
    doi = "10.1016/j.physrep.2014.12.002",
    journal = "Phys. Rept.",
    volume = "568",
    pages = "1--98",
    year = "2015"
}

@article{Sennett:2019bpc,
    author = "Sennett, Noah and Brito, Richard and Buonanno, Alessandra and Gorbenko, Victor and Senatore, Leonardo",
    title = "{Gravitational-Wave Constraints on an Effective Field-Theory Extension of General Relativity}",
    eprint = "1912.09917",
    archivePrefix = "arXiv",
    primaryClass = "gr-qc",
    doi = "10.1103/PhysRevD.102.044056",
    journal = "Phys. Rev. D",
    volume = "102",
    number = "4",
    pages = "044056",
    year = "2020"
}

@article{Will:2014kxa,
    author = "Will, Clifford M.",
    title = "{The Confrontation between General Relativity and Experiment}",
    eprint = "1403.7377",
    archivePrefix = "arXiv",
    primaryClass = "gr-qc",
    doi = "10.12942/lrr-2014-4",
    journal = "Living Rev. Rel.",
    volume = "17",
    pages = "4",
    year = "2014"
}

@article{LISACosmologyWorkingGroup:2026zah,
    author = "Akama, Shingo and others",
    collaboration = "LISA Cosmology Working Group",
    title = "{Testing gravitational wave polarizations with LISA}",
    eprint = "2603.03165",
    archivePrefix = "arXiv",
    primaryClass = "astro-ph.CO",
    month = "3",
    year = "2026"
}

@article{Endlich:2017tqa,
    author = "Endlich, Solomon and Gorbenko, Victor and Huang, Junwu and Senatore, Leonardo",
    title = "{An effective formalism for testing extensions to General Relativity with gravitational waves}",
    eprint = "1704.01590",
    archivePrefix = "arXiv",
    primaryClass = "gr-qc",
    doi = "10.1007/JHEP09(2017)122",
    journal = "JHEP",
    volume = "09",
    pages = "122",
    year = "2017"
}

@article{Camanho:2014apa,
    author = "Camanho, Xian O. and Edelstein, Jose D. and Maldacena, Juan and Zhiboedov, Alexander",
    title = "{Causality Constraints on Corrections to the Graviton Three-Point Coupling}",
    eprint = "1407.5597",
    archivePrefix = "arXiv",
    primaryClass = "hep-th",
    doi = "10.1007/JHEP02(2016)020",
    journal = "JHEP",
    volume = "02",
    pages = "020",
    year = "2016"
}

@article{Burgess:2003jk,
    author = "Burgess, C. P.",
    title = "{Quantum gravity in everyday life: General relativity as an effective field theory}",
    eprint = "gr-qc/0311082",
    archivePrefix = "arXiv",
    doi = "10.12942/lrr-2004-5",
    journal = "Living Rev. Rel.",
    volume = "7",
    pages = "5--56",
    year = "2004"
}

@article{Donoghue:1994dn,
    author = "Donoghue, John F.",
    title = "{General relativity as an effective field theory: The leading quantum corrections}",
    eprint = "gr-qc/9405057",
    archivePrefix = "arXiv",
    reportNumber = "UMHEP-408",
    doi = "10.1103/PhysRevD.50.3874",
    journal = "Phys. Rev. D",
    volume = "50",
    pages = "3874--3888",
    year = "1994"
}

@article{Goyal:2025eqo,
    author = "Goyal, Srashti and Villarrubia-Rojo, Hector and Zumalacarregui, Miguel",
    title = "{Across the Universe: GW231123 as a magnified and diffracted black hole merger}",
    eprint = "2512.17631",
    archivePrefix = "arXiv",
    primaryClass = "astro-ph.GA",
    month = "12",
    year = "2025"
}

@article{Chan:2025kyu,
    author = "Chan, Juno C. L. and Ezquiaga, Jose Mar{\'\i}a and Lo, Rico K. L. and Bowman, Joey and Maga{\~n}a Zertuche, Lorena and Vujeva, Luka",
    title = "{Discovering gravitational waveform distortions from lensing: a deep dive into GW231123}",
    eprint = "2512.16916",
    archivePrefix = "arXiv",
    primaryClass = "gr-qc",
    month = "12",
    year = "2025"
}

@article{Hu:2025lhv,
    author = "Hu, Qian and Narola, Harsh and Heynen, Jef and Wright, Mick and Veitch, John and Janquart, Justin and Van Den Broeck, Chris",
    title = "{GW231123: Overlapping Gravitational Wave Signals?}",
    eprint = "2512.17550",
    archivePrefix = "arXiv",
    primaryClass = "gr-qc",
    reportNumber = "LIGO-P2500697",
    month = "12",
    year = "2025"
}

@article{Chakraborty:2025pxt,
    author = "Chakraborty, Aniruddha and Mukherjee, Suvodip",
    title = "{The First Model-Independent Upper Bound on Micro-lensing Signature of the Highest Mass Binary Black Hole Event GW231123}",
    eprint = "2512.19077",
    archivePrefix = "arXiv",
    primaryClass = "gr-qc",
    month = "12",
    year = "2025"
}

@article{Li:2018prc,
    author = "Li, Shun-Sheng and Mao, Shude and Zhao, Yuetong and Lu, Youjun",
    title = "{Gravitational lensing of gravitational waves: A statistical perspective}",
    eprint = "1802.05089",
    archivePrefix = "arXiv",
    primaryClass = "astro-ph.CO",
    doi = "10.1093/mnras/sty411",
    journal = "Mon. Not. Roy. Astron. Soc.",
    volume = "476",
    number = "2",
    pages = "2220--2229",
    year = "2018"
}

@article{Wierda:2021upe,
    author = "Wierda, A. Renske A. C. and Wempe, Ewoud and Hannuksela, Otto A. and Koopmans, L. {\'e}on V. E. and Van Den Broeck, Chris",
    title = "{Beyond the Detector Horizon: Forecasting Gravitational-Wave Strong Lensing}",
    eprint = "2106.06303",
    archivePrefix = "arXiv",
    primaryClass = "astro-ph.HE",
    doi = "10.3847/1538-4357/ac1bb4",
    journal = "Astrophys. J.",
    volume = "921",
    number = "2",
    pages = "154",
    year = "2021"
}

@article{Ng:2017yiu,
    author = "Ng, Ken K. Y. and Wong, Kaze W. K. and Broadhurst, Tom and Li, Tjonnie G. F.",
    title = "{Precise LIGO Lensing Rate Predictions for Binary Black Holes}",
    eprint = "1703.06319",
    archivePrefix = "arXiv",
    primaryClass = "astro-ph.CO",
    doi = "10.1103/PhysRevD.97.023012",
    journal = "Phys. Rev. D",
    volume = "97",
    number = "2",
    pages = "023012",
    year = "2018"
}

@article{Sereno:2010dr,
    author = "Sereno, M. and Sesana, A. and Bleuler, A. and Jetzer, Ph. and Volonteri, M. and Begelman, M. C.",
    title = "{Strong lensing of gravitational waves as seen by LISA}",
    eprint = "1011.5238",
    archivePrefix = "arXiv",
    primaryClass = "astro-ph.CO",
    doi = "10.1103/PhysRevLett.105.251101",
    journal = "Phys. Rev. Lett.",
    volume = "105",
    pages = "251101",
    year = "2010"
}

@article{Gutierrez:2025ymd,
    author = "Guti{\'e}rrez, Juan and Lagos, Macarena",
    title = "{Strong-lensing rates of massive black hole binaries in LISA}",
    eprint = "2510.02061",
    archivePrefix = "arXiv",
    primaryClass = "astro-ph.CO",
    doi = "10.1103/yd5h-ql5f",
    journal = "Phys. Rev. D",
    volume = "112",
    number = "12",
    pages = "123512",
    year = "2025"
}

@article{Hannuksela:2019kle,
    author = "Hannuksela, O. A. and Haris, K. and Ng, K. K. Y. and Kumar, S. and Mehta, A. K. and Keitel, D. and Li, T. G. F. and Ajith, P.",
    title = "{Search for gravitational lensing signatures in LIGO-Virgo binary black hole events}",
    eprint = "1901.02674",
    archivePrefix = "arXiv",
    primaryClass = "gr-qc",
    reportNumber = "LIGO Document P1800297, LIGO-P1800297",
    doi = "10.3847/2041-8213/ab0c0f",
    journal = "Astrophys. J. Lett.",
    volume = "874",
    number = "1",
    pages = "L2",
    year = "2019"
}

@article{LIGOScientific:2021izm,
    author = "Abbott, R. and others",
    collaboration = "LIGO Scientific, VIRGO",
    title = "{Search for Lensing Signatures in the Gravitational-Wave Observations from the First Half of LIGO--Virgo's Third Observing Run}",
    eprint = "2105.06384",
    archivePrefix = "arXiv",
    primaryClass = "gr-qc",
    reportNumber = "LIGO-P2000400",
    doi = "10.3847/1538-4357/ac23db",
    journal = "Astrophys. J.",
    volume = "923",
    number = "1",
    pages = "14",
    year = "2021"
}

@article{LIGOScientific:2023bwz,
    author = "Abbott, R. and others",
    collaboration = "LIGO Scientific, KAGRA, VIRGO",
    title = "{Search for Gravitational-lensing Signatures in the Full Third Observing Run of the LIGO--Virgo Network}",
    eprint = "2304.08393",
    archivePrefix = "arXiv",
    primaryClass = "gr-qc",
    reportNumber = "LIGO-P2200031",
    doi = "10.3847/1538-4357/ad3e83",
    journal = "Astrophys. J.",
    volume = "970",
    number = "2",
    pages = "191",
    year = "2024"
}

@article{Takahashi:2005sxa,
    author = "Takahashi, Ryuichi and Suyama, Teruaki and Michikoshi, Shugo",
    title = "{Scattering of gravitational waves by the weak gravitational fields of lens objects}",
    eprint = "astro-ph/0503343",
    archivePrefix = "arXiv",
    doi = "10.1051/0004-6361:200500140",
    journal = "Astron. Astrophys.",
    volume = "438",
    pages = "L5",
    year = "2005"
}

@article{Takahashi:2005ug,
    author = "Takahashi, Ryuichi",
    title = "{Amplitude and phase fluctuations for gravitational waves propagating through inhomogeneous mass distribution in the universe}",
    eprint = "astro-ph/0511517",
    archivePrefix = "arXiv",
    doi = "10.1086/503323",
    journal = "Astrophys. J.",
    volume = "644",
    pages = "80--85",
    year = "2006"
}

@article{Oguri:2020ldf,
    author = "Oguri, Masamune and Takahashi, Ryuichi",
    title = "{Probing Dark Low-mass Halos and Primordial Black Holes with Frequency-dependent Gravitational Lensing Dispersions of Gravitational Waves}",
    eprint = "2007.01936",
    archivePrefix = "arXiv",
    primaryClass = "astro-ph.CO",
    doi = "10.3847/1538-4357/abafab",
    journal = "Astrophys. J.",
    volume = "901",
    number = "1",
    pages = "58",
    year = "2020"
}

@article{Yarimoto:2024uew,
    author = "Yarimoto, Hirotaka and Oguri, Masamune",
    title = "{Born approximation in wave optics of gravitational lensing revisited}",
    eprint = "2412.07272",
    archivePrefix = "arXiv",
    primaryClass = "astro-ph.CO",
    doi = "10.1103/PhysRevD.111.083541",
    journal = "Phys. Rev. D",
    volume = "111",
    number = "8",
    pages = "083541",
    year = "2025"
}

@article{Garoffolo:2022usx,
    author = "Garoffolo, Alice",
    title = "{Wave-optics limit of the stochastic gravitational wave background}",
    eprint = "2210.05718",
    archivePrefix = "arXiv",
    primaryClass = "astro-ph.CO",
    doi = "10.1016/j.dark.2024.101475",
    journal = "Phys. Dark Univ.",
    volume = "44",
    pages = "101475",
    year = "2024"
}

@article{Jung:2017flg,
    author = "Jung, Sunghoon and Shin, Chang Sub",
    title = "{Gravitational-Wave Fringes at LIGO: Detecting Compact Dark Matter by Gravitational Lensing}",
    eprint = "1712.01396",
    archivePrefix = "arXiv",
    primaryClass = "astro-ph.CO",
    doi = "10.1103/PhysRevLett.122.041103",
    journal = "Phys. Rev. Lett.",
    volume = "122",
    number = "4",
    pages = "041103",
    year = "2019"
}

@article{yeung2024wolensing,
  title={wolensing: A Python package for computing the amplification factor for gravitational waves with wave-optics effects},
  author={Yeung, Simon and Cheung, Mark HY and Zumalacarregui, Miguel and Hannuksela, Otto A},
  journal={arXiv preprint arXiv:2410.19804},
  year={2024}
}

@article{Jow:2022pux,
    author = "Jow, Dylan L. and Pen, Ue-Li and Feldbrugge, Job",
    title = "{Regimes in astrophysical lensing: refractive optics, diffractive optics, and the Fresnel scale}",
    eprint = "2204.12004",
    archivePrefix = "arXiv",
    primaryClass = "astro-ph.HE",
    doi = "10.1093/mnras/stad2332",
    journal = "Mon. Not. Roy. Astron. Soc.",
    volume = "525",
    number = "2",
    pages = "2107--2124",
    year = "2023"
}

@article{Diego:2019lcd,
    author = "Diego, J. M. and Hannuksela, O. A. and Kelly, P. L. and Broadhurst, T. and Kim, K. and Li, T. G. F. and Smoot, G. F. and Pagano, G.",
    title = "{Observational signatures of microlensing in gravitational waves at LIGO/Virgo frequencies}",
    eprint = "1903.04513",
    archivePrefix = "arXiv",
    primaryClass = "astro-ph.CO",
    doi = "10.1051/0004-6361/201935490",
    journal = "Astron. Astrophys.",
    volume = "627",
    pages = "A130",
    year = "2019"
}

@article{Takahashi:2004mc,
    author = "Takahashi, Ryuichi",
    title = "{Quasigeometrical optics approximation in gravitational lensing}",
    eprint = "astro-ph/0402165",
    archivePrefix = "arXiv",
    doi = "10.1051/0004-6361:20040212",
    journal = "Astron. Astrophys.",
    volume = "423",
    pages = "787--792",
    year = "2004"
}

@article{Grespan:2023cpa,
    author = "Grespan, Margherita and Biesiada, Marek",
    title = "{Strong Gravitational Lensing of Gravitational Waves: A Review}",
    doi = "10.3390/universe9050200",
    journal = "Universe",
    volume = "9",
    number = "5",
    pages = "200",
    year = "2023"
}

@article{Lai:2018rto,
    author = "Lai, Kwun-Hang and Hannuksela, Otto A. and Herrera-Mart{\'\i}n, Antonio and Diego, Jose M. and Broadhurst, Tom and Li, Tjonnie G. F.",
    title = "{Discovering intermediate-mass black hole lenses through gravitational wave lensing}",
    eprint = "1801.07840",
    archivePrefix = "arXiv",
    primaryClass = "gr-qc",
    doi = "10.1103/PhysRevD.98.083005",
    journal = "Phys. Rev. D",
    volume = "98",
    number = "8",
    pages = "083005",
    year = "2018"
}

@article{Suyama:2025gbh,
    author = "Suyama, Teruaki and Kapadia, Shasvath J.",
    title = "{Phase-magnitude relation in gravitational lensing: Reformulation and applications of the Kramers-Kronig relation}",
    eprint = "2506.02430",
    archivePrefix = "arXiv",
    primaryClass = "gr-qc",
    doi = "10.1103/mwy4-wspz",
    journal = "Phys. Rev. D",
    volume = "112",
    number = "6",
    pages = "063529",
    year = "2025"
}

@article{Tanaka:2023mvy,
    author = "Tanaka, So and Suyama, Teruaki",
    title = "{Kramers-Kronig relation in gravitational lensing}",
    eprint = "2303.05650",
    archivePrefix = "arXiv",
    primaryClass = "gr-qc",
    doi = "10.1103/PhysRevD.108.044015",
    journal = "Phys. Rev. D",
    volume = "108",
    number = "4",
    pages = "044015",
    year = "2023"
}

@book{Schneider:1992bmb,
    author = {Schneider, Peter and Ehlers, J{\"u}rgen and Falco, Emilio E.},
    title = "{Gravitational Lenses}",
    doi = "10.1007/978-3-662-03758-4",
    isbn = "978-3-540-66506-9, 978-3-662-03758-4",
    publisher = "Springer",
    series = "Astronomy and Astrophysics Library",
    year = "1992"
}

@article{Nakamura:1999uwi,
    author = "Nakamura, Takahiro T. and Deguchi, Shuji",
    title = "{Wave Optics in Gravitational Lensing}",
    doi = "10.1143/ptps.133.137",
    journal = "Prog. Theor. Phys. Suppl.",
    volume = "133",
    pages = "137--153",
    year = "1999"
}

@article{Nakamura:1997sw,
    author = "Nakamura, Takahiro T.",
    title = "{Gravitational lensing of gravitational waves from inspiraling binaries by a point mass lens}",
    reportNumber = "UTAP-272-97, YITP-97-61",
    doi = "10.1103/PhysRevLett.80.1138",
    journal = "Phys. Rev. Lett.",
    volume = "80",
    pages = "1138--1141",
    year = "1998"
}

@article{Takahashi:2003ix,
    author = "Takahashi, Ryuichi and Nakamura, Takashi",
    title = "{Wave effects in gravitational lensing of gravitational waves from chirping binaries}",
    eprint = "astro-ph/0305055",
    archivePrefix = "arXiv",
    doi = "10.1086/377430",
    journal = "Astrophys. J.",
    volume = "595",
    pages = "1039--1051",
    year = "2003"
}

@article{Urrutia:2024pos,
    author = "Urrutia, Juan and Vaskonen, Ville",
    title = "{Dark timbre of gravitational waves}",
    eprint = "2402.16849",
    archivePrefix = "arXiv",
    primaryClass = "gr-qc",
    doi = "10.1103/PhysRevD.111.123047",
    journal = "Phys. Rev. D",
    volume = "111",
    number = "12",
    pages = "123047",
    year = "2025"
}

@article{Urrutia:2023mtk,
    author = {Urrutia, Juan and Vaskonen, Ville and Veerm{\"a}e, Hardi},
    title = "{Gravitational wave microlensing by dressed primordial black holes}",
    eprint = "2303.17601",
    archivePrefix = "arXiv",
    primaryClass = "astro-ph.CO",
    doi = "10.1103/PhysRevD.108.023507",
    journal = "Phys. Rev. D",
    volume = "108",
    number = "2",
    pages = "023507",
    year = "2023"
}

@article{Urrutia:2021qak,
    author = "Urrutia, Juan and Vaskonen, Ville",
    title = "{Lensing of gravitational waves as a probe of compact dark matter}",
    eprint = "2109.03213",
    archivePrefix = "arXiv",
    primaryClass = "astro-ph.CO",
    doi = "10.1093/mnras/stab3118",
    journal = "Mon. Not. Roy. Astron. Soc.",
    volume = "509",
    number = "1",
    pages = "1358--1365",
    year = "2021"
}

@article{Leung:2023lmq,
    author = "Leung, Calvin and Jow, Dylan and Saha, Prasenjit and Dai, Liang and Oguri, Masamune and Koopmans, L{\'e}on V. E.",
    title = "{Wave Optics, Interference, and Decoherence in Strong Gravitational Lensing}",
    eprint = "2304.01202",
    archivePrefix = "arXiv",
    primaryClass = "astro-ph.HE",
    doi = "10.1007/s11214-025-01157-7",
    journal = "Space Sci. Rev.",
    volume = "221",
    number = "2",
    pages = "29",
    year = "2025"
}

@article{Suvorov:2021uvd,
    author = "Suvorov, Arthur G.",
    title = "{Wave-optical Effects in the Microlensing of Continuous Gravitational Waves by Star Clusters}",
    eprint = "2112.01670",
    archivePrefix = "arXiv",
    primaryClass = "astro-ph.HE",
    doi = "10.3847/1538-4357/ac5f45",
    journal = "Astrophys. J.",
    volume = "930",
    number = "1",
    pages = "13",
    year = "2022"
}

@article{Braga:2024pik,
    author = "Braga, Ginevra and Garoffolo, Alice and Ricciardone, Angelo and Bartolo, Nicola and Matarrese, Sabino",
    title = "{Proper time path integrals for gravitational waves: an~improved wave optics framework}",
    eprint = "2405.20208",
    archivePrefix = "arXiv",
    primaryClass = "astro-ph.CO",
    doi = "10.1088/1475-7516/2024/11/031",
    journal = "JCAP",
    volume = "11",
    pages = "031",
    year = "2024"
}

@article{Savastano:2023spl,
    author = "Savastano, Stefano and Tambalo, Giovanni and Villarrubia-Rojo, Hector and Zumalacarregui, Miguel",
    title = "{Weakly lensed gravitational waves: Probing cosmic structures with wave-optics features}",
    eprint = "2306.05282",
    archivePrefix = "arXiv",
    primaryClass = "gr-qc",
    doi = "10.1103/PhysRevD.108.103532",
    journal = "Phys. Rev. D",
    volume = "108",
    number = "10",
    pages = "103532",
    year = "2023"
}

@article{Caliskan:2023zqm,
    author = "{\c{C}}al{\i}{\c{s}}kan, Mesut and Anil Kumar, Neha and Ji, Lingyuan and Ezquiaga, Jose M. and Cotesta, Roberto and Berti, Emanuele and Kamionkowski, Marc",
    title = "{Probing wave-optics effects and low-mass dark matter halos with lensing of gravitational waves from massive black holes}",
    eprint = "2307.06990",
    archivePrefix = "arXiv",
    primaryClass = "astro-ph.CO",
    doi = "10.1103/PhysRevD.108.123543",
    journal = "Phys. Rev. D",
    volume = "108",
    number = "12",
    pages = "123543",
    year = "2023"
}

@article{Caliskan:2022hbu,
    author = "{\c{C}}al{\i}{\c{s}}kan, Mesut and Ji, Lingyuan and Cotesta, Roberto and Berti, Emanuele and Kamionkowski, Marc and Marsat, Sylvain",
    title = "{Observability of lensing of gravitational waves from massive black hole binaries with LISA}",
    eprint = "2206.02803",
    archivePrefix = "arXiv",
    primaryClass = "astro-ph.CO",
    doi = "10.1103/PhysRevD.107.043029",
    journal = "Phys. Rev. D",
    volume = "107",
    number = "4",
    pages = "043029",
    year = "2023"
}

@article{Tambalo:2022wlm,
    author = "Tambalo, Giovanni and Zumalac{\'a}rregui, Miguel and Dai, Liang and Cheung, Mark Ho-Yeuk",
    title = "{Gravitational wave lensing as a probe of halo properties and dark matter}",
    eprint = "2212.11960",
    archivePrefix = "arXiv",
    primaryClass = "astro-ph.CO",
    doi = "10.1103/PhysRevD.108.103529",
    journal = "Phys. Rev. D",
    volume = "108",
    number = "10",
    pages = "103529",
    year = "2023"
}

@article{Yeung:2021chy,
    author = "Yeung, Simon M. C. and Cheung, Mark H. Y. and Seo, Eungwang and Gais, Joseph A. J. and Hannuksela, Otto A. and Li, Tjonnie G. F.",
    title = "{Detectability of microlensed gravitational waves}",
    eprint = "2112.07635",
    archivePrefix = "arXiv",
    primaryClass = "gr-qc",
    doi = "10.1093/mnras/stad2772",
    journal = "Mon. Not. Roy. Astron. Soc.",
    volume = "526",
    number = "2",
    pages = "2230--2240",
    year = "2023"
}

@article{Feldbrugge:2019fjs,
    author = "Feldbrugge, Job and Pen, Ue-Li and Turok, Neil",
    title = "{Oscillatory path integrals for radio astronomy}",
    eprint = "1909.04632",
    archivePrefix = "arXiv",
    primaryClass = "astro-ph.HE",
    doi = "10.1016/j.aop.2023.169255",
    journal = "Annals Phys.",
    volume = "451",
    pages = "169255",
    year = "2023"
}

@article{Bonga:2024orc,
    author = "Bonga, B{\'e}atrice and Feldbrugge, Job and Ribes Metidieri, Ariadna",
    title = "{Wave optics for rotating stars}",
    eprint = "2410.03828",
    archivePrefix = "arXiv",
    primaryClass = "gr-qc",
    doi = "10.1103/PhysRevD.111.063061",
    journal = "Phys. Rev. D",
    volume = "111",
    number = "6",
    pages = "063061",
    year = "2025"
}

@article{Sugiyama:2019dgt,
    author = "Sugiyama, Sunao and Kurita, Toshiki and Takada, Masahiro",
    title = "{On the wave optics effect on primordial black hole constraints from optical microlensing search}",
    eprint = "1905.06066",
    archivePrefix = "arXiv",
    primaryClass = "astro-ph.CO",
    doi = "10.1093/mnras/staa407",
    journal = "Mon. Not. Roy. Astron. Soc.",
    volume = "493",
    number = "3",
    pages = "3632--3641",
    year = "2020"
}

@article{Villarrubia-Rojo:2024xcj,
    author = "Villarrubia-Rojo, Hector and Savastano, Stefano and Zumalac{\'a}rregui, Miguel and Choi, Lyla and Goyal, Srashti and Dai, Liang and Tambalo, Giovanni",
    title = "{Gravitational lensing of waves: Novel methods for wave-optics phenomena}",
    eprint = "2409.04606",
    archivePrefix = "arXiv",
    primaryClass = "gr-qc",
    doi = "10.1103/PhysRevD.111.103539",
    journal = "Phys. Rev. D",
    volume = "111",
    number = "10",
    pages = "103539",
    year = "2025"
}

@article{Tambalo:2022plm,
    author = "Tambalo, Giovanni and Zumalac{\'a}rregui, Miguel and Dai, Liang and Cheung, Mark Ho-Yeuk",
    title = "{Lensing of gravitational waves: Efficient wave-optics methods and validation with symmetric lenses}",
    eprint = "2210.05658",
    archivePrefix = "arXiv",
    primaryClass = "gr-qc",
    doi = "10.1103/PhysRevD.108.043527",
    journal = "Phys. Rev. D",
    volume = "108",
    number = "4",
    pages = "043527",
    year = "2023"
}

@article{Singh:2025uvp,
    author = "Singh, Shashwat and Brando de Oliveira, Guilherme and Savastano, Stefano and Zumalac{\'a}rregui, Miguel",
    title = "{Gravitational wave lensing: probing Fuzzy Dark Matter with LISA}",
    eprint = "2502.10758",
    archivePrefix = "arXiv",
    primaryClass = "astro-ph.CO",
    doi = "10.1088/1475-7516/2025/07/025",
    journal = "JCAP",
    volume = "07",
    pages = "025",
    year = "2025"
}

@article{Macquart:2004sh,
    author = "Macquart, Jean-Pierre",
    title = "{Scattering of gravitational radiation: Second order moments of the wave amplitude}",
    eprint = "astro-ph/0402661",
    archivePrefix = "arXiv",
    doi = "10.1051/0004-6361:20034512",
    journal = "Astron. Astrophys.",
    volume = "422",
    pages = "761--775",
    year = "2004"
}

@article{Li:2022izh,
    author = "Li, Zhao and Qiao, Jin and Zhao, Wen and Er, Xinzhong",
    title = "{Gravitational Faraday Rotation of gravitational waves by a Kerr black hole}",
    eprint = "2204.10512",
    archivePrefix = "arXiv",
    primaryClass = "gr-qc",
    doi = "10.1088/1475-7516/2022/10/095",
    journal = "JCAP",
    volume = "10",
    pages = "095",
    year = "2022"
}

@article{Mehrabi:2012dy,
    author = "Mehrabi, Ahmad and Rahvar, Sohrab",
    title = "{Studying wave optics in exoplanet microlensing light curves}",
    eprint = "1207.4034",
    archivePrefix = "arXiv",
    primaryClass = "astro-ph.EP",
    doi = "10.1093/mnras/stt243",
    journal = "Mon. Not. Roy. Astron. Soc.",
    volume = "431",
    pages = "1264",
    year = "2013"
}

@article{Diego:2019rzc,
    author = "Diego, Jose M.",
    title = "{Constraining the abundance of primordial black holes with gravitational lensing of gravitational waves at LIGO frequencies}",
    eprint = "1911.05736",
    archivePrefix = "arXiv",
    primaryClass = "astro-ph.CO",
    doi = "10.1103/PhysRevD.101.123512",
    journal = "Phys. Rev. D",
    volume = "101",
    number = "12",
    pages = "123512",
    year = "2020"
}

@article{Ramesh:2021nnl,
    author = "Ramesh, Rahul and Meena, Ashish Kumar and Bagla, Jasjeet Singh",
    title = "{Wave effects in double-plane lensing}",
    eprint = "2109.09998",
    archivePrefix = "arXiv",
    primaryClass = "astro-ph.CO",
    doi = "10.1007/s12036-022-09821-y",
    journal = "J. Astrophys. Astron.",
    volume = "43",
    number = "2",
    pages = "38",
    year = "2022"
}

@article{Feldbrugge:2020ycp,
    author = "Feldbrugge, Job and Turok, Neil",
    title = "{Gravitational lensing of binary systems in wave optics}",
    eprint = "2008.01154",
    archivePrefix = "arXiv",
    primaryClass = "gr-qc",
    month = "8",
    year = "2020"
}

@article{Feldbrugge:2020tti,
    author = "Feldbrugge, Job",
    title = "{Multiplane lensing in wave optics}",
    eprint = "2010.03089",
    archivePrefix = "arXiv",
    primaryClass = "astro-ph.CO",
    doi = "10.1093/mnras/stad349",
    journal = "Mon. Not. Roy. Astron. Soc.",
    volume = "520",
    number = "2",
    pages = "2995--3006",
    year = "2023"
}

@article{Yamamoto:2003cd,
    author = "Yamamoto, Kazuhiro",
    title = "{Path integral formulation for wave effect in multi-lens system}",
    eprint = "astro-ph/0309696",
    archivePrefix = "arXiv",
    month = "9",
    year = "2003"
}

@article{Amoruso:2026txw,
    author = "Amoruso, Ripalta and Braga, Ginevra and Garoffolo, Alice and Lopez, Francescopaolo and Bartolo, Nicola and Matarrese, Sabino",
    title = "{Gravitational-wave lensing beyond rays: a disordered-system approach}",
    eprint = "2604.15313",
    archivePrefix = "arXiv",
    primaryClass = "astro-ph.CO",
    month = "4",
    year = "2026"
}

@article{Saketh:2025cwf,
    author = "Saketh, M. V. S. and Ghosh, Rajes and Mishra, Anuj",
    title = "{Strong-field gravitational-wave lensing in the Kerr background}",
    eprint = "2511.23110",
    archivePrefix = "arXiv",
    primaryClass = "gr-qc",
    doi = "10.1103/9bxg-sqn5",
    journal = "Phys. Rev. D",
    volume = "113",
    number = "8",
    pages = "084056",
    year = "2026"
}

@article{Kthylwe1985,
doi = {10.1088/0305-4470/18/15/022},
url = {https://doi.org/10.1088/0305-4470/18/15/022},
year = {1985},
month = {oct},
publisher = {},
volume = {18},
number = {15},
pages = {2957},
author = {K -E Thylwe and J N L Connor},
title = {A complex angular momentum theory of modified Coulomb scattering},
journal = {Journal of Physics A: Mathematical and General}
}

@article{Turyshev:2018gjj,
    author = "Turyshev, Slava G. and Toth, Viktor T.",
    title = "{Diffraction of light by the gravitational field of the Sun and the solar corona}",
    eprint = "1810.06627",
    archivePrefix = "arXiv",
    primaryClass = "gr-qc",
    doi = "10.1103/PhysRevD.99.024044",
    journal = "Phys. Rev. D",
    volume = "99",
    number = "2",
    pages = "024044",
    year = "2019"
}

@article{Li:2025lvl,
    author = "Li, Zhao and Zhao, Wen",
    title = "{Rigorous calculation of scalar scattering in the Schwarzschild background: The convergence of the partial-wave series and the Poisson spot}",
    eprint = "2508.17253",
    archivePrefix = "arXiv",
    primaryClass = "gr-qc",
    doi = "10.1103/xsxj-9vdw",
    journal = "Phys. Rev. D",
    volume = "112",
    number = "8",
    pages = "083030",
    year = "2025"
}

@article{Motohashi:2021zyv,
    author = "Motohashi, Hayato and Noda, Sousuke",
    title = "{Exact solution for wave scattering from black holes: Formulation}",
    eprint = "2103.10802",
    archivePrefix = "arXiv",
    primaryClass = "gr-qc",
    doi = "10.1093/ptep/ptac020",
    journal = "PTEP",
    volume = "2021",
    number = "8",
    pages = "083E03",
    year = "2021"
}

@article{Kubota:2024zkv,
    author = "Kubota, Kei-ichiro and Arai, Shun and Motohashi, Hayato and Mukohyama, Shinji",
    title = "{Spin wave optics for gravitational waves lensed by a Kerr black hole}",
    eprint = "2408.03289",
    archivePrefix = "arXiv",
    primaryClass = "gr-qc",
    reportNumber = "YITP-24-89, IPMU24-0031",
    doi = "10.1103/PhysRevD.110.124011",
    journal = "Phys. Rev. D",
    volume = "110",
    number = "12",
    pages = "124011",
    year = "2024"
}

@article{Kubota:2023dlz,
    author = "Kubota, Kei-ichiro and Arai, Shun and Mukohyama, Shinji",
    title = "{Spin optics for gravitational waves lensed by a rotating object}",
    eprint = "2309.11024",
    archivePrefix = "arXiv",
    primaryClass = "gr-qc",
    reportNumber = "YITP-23-113, IPMU23-0032",
    doi = "10.1103/PhysRevD.109.044027",
    journal = "Phys. Rev. D",
    volume = "109",
    number = "4",
    pages = "044027",
    year = "2024"
}

@article{Nambu:2015aea,
    author = "Nambu, Yasusada and Noda, Sousuke",
    title = "{Wave Optics in Black Hole Spacetimes: Schwarzschild Case}",
    eprint = "1502.05468",
    archivePrefix = "arXiv",
    primaryClass = "gr-qc",
    doi = "10.1088/0264-9381/33/7/075011",
    journal = "Class. Quant. Grav.",
    volume = "33",
    pages = "075011",
    year = "2016"
}

@article{Nambu:2019sqn,
    author = "Nambu, Yasusada and Noda, Sousuke and Sakai, Yuichiro",
    title = "{Wave Optics in Spacetimes with Compact Gravitating Object}",
    eprint = "1905.01793",
    archivePrefix = "arXiv",
    primaryClass = "gr-qc",
    doi = "10.1103/PhysRevD.100.064037",
    journal = "Phys. Rev. D",
    volume = "100",
    number = "6",
    pages = "064037",
    year = "2019"
}

@article{Pijnenburg:2024btj,
    author = "Pijnenburg, Martin and Cusin, Giulia and Pitrou, Cyril and Uzan, Jean-Philippe",
    title = "{Wave optics lensing of gravitational waves: Theory and phenomenology of triple systems in the LISA band}",
    eprint = "2404.07186",
    archivePrefix = "arXiv",
    primaryClass = "gr-qc",
    doi = "10.1103/PhysRevD.110.044054",
    journal = "Phys. Rev. D",
    volume = "110",
    number = "4",
    pages = "044054",
    year = "2024"
}

@article{Chan:2025wgz,
    author = "Chan, Juno C. L. and Dyson, Conor and Garcia, Matilde and Redondo-Yuste, Jaime and Vujeva, Luka",
    title = "{Lensing and wave optics in the strong field of a black hole}",
    eprint = "2502.14073",
    archivePrefix = "arXiv",
    primaryClass = "gr-qc",
    doi = "10.1103/6h6r-46cd",
    journal = "Phys. Rev. D",
    volume = "112",
    number = "6",
    pages = "064009",
    year = "2025"
}

@article{CarrilloGonzalez:2025gqm,
    author = "Carrillo Gonzalez, Mariana and De Luca, Valerio and Garoffolo, Alice and Parra-Martinez, Julio and Trodden, Mark",
    title = "{Scattering perspective on gravitational lensing}",
    eprint = "2511.15797",
    archivePrefix = "arXiv",
    primaryClass = "hep-th",
    doi = "10.1103/w2z2-974w",
    journal = "Phys. Rev. D",
    volume = "113",
    number = "2",
    pages = "024024",
    year = "2026"
}

@article{Casals:2016soq,
    author = "Casals, Marc and Kavanagh, Chris and Ottewill, Adrian C.",
    title = "{High-order late-time tail in a Kerr spacetime}",
    eprint = "1608.05392",
    archivePrefix = "arXiv",
    primaryClass = "gr-qc",
    doi = "10.1103/PhysRevD.94.124053",
    journal = "Phys. Rev. D",
    volume = "94",
    number = "12",
    pages = "124053",
    year = "2016"
}

@article{Leaver:1986gd,
    author = "Leaver, Edward W.",
    title = "{Spectral decomposition of the perturbation response of the Schwarzschild geometry}",
    doi = "10.1103/PhysRevD.34.384",
    journal = "Phys. Rev. D",
    volume = "34",
    pages = "384--408",
    year = "1986"
}

@article{Vishveshwara:1970zz,
    author = "Vishveshwara, C. V.",
    title = "{Scattering of Gravitational Radiation by a Schwarzschild Black-hole}",
    doi = "10.1038/227936a0",
    journal = "Nature",
    volume = "227",
    pages = "936--938",
    year = "1970"
}

@article{Bai:2016ivl,
    author = "Bai, Dong and Huang, Yue",
    title = "{More on the Bending of Light in Quantum Gravity}",
    eprint = "1612.07629",
    archivePrefix = "arXiv",
    primaryClass = "hep-th",
    doi = "10.1103/PhysRevD.95.064045",
    journal = "Phys. Rev. D",
    volume = "95",
    number = "6",
    pages = "064045",
    year = "2017"
}

@article{Chi:2019owc,
    author = "Chi, Huan-Hang",
    title = "{Graviton Bending in Quantum Gravity from One-Loop Amplitudes}",
    eprint = "1903.07944",
    archivePrefix = "arXiv",
    primaryClass = "hep-th",
    reportNumber = "SLAC--PUB--17410, SLAC-PUB-17410",
    doi = "10.1103/PhysRevD.99.126008",
    journal = "Phys. Rev. D",
    volume = "99",
    number = "12",
    pages = "126008",
    year = "2019"
}

@article{Bjerrum-Bohr:2014zsa,
    author = "Bjerrum-Bohr, N. E. J. and Donoghue, John F. and Holstein, Barry R. and Plant{\'e}, Ludovic and Vanhove, Pierre",
    title = "{Bending of Light in Quantum Gravity}",
    eprint = "1410.7590",
    archivePrefix = "arXiv",
    primaryClass = "hep-th",
    reportNumber = "IPHT-T14-108, IHES-P-14-33, ACFI-T14-21",
    doi = "10.1103/PhysRevLett.114.061301",
    journal = "Phys. Rev. Lett.",
    volume = "114",
    number = "6",
    pages = "061301",
    year = "2015"
}

@article{Bastianelli:2021nbs,
    author = "Bastianelli, Fiorenzo and Comberiati, Francesco and de la Cruz, Leonardo",
    title = "{Light bending from eikonal in worldline quantum field theory}",
    eprint = "2112.05013",
    archivePrefix = "arXiv",
    primaryClass = "hep-th",
    doi = "10.1007/JHEP02(2022)209",
    journal = "JHEP",
    volume = "02",
    pages = "209",
    year = "2022"
}

@article{Comberiati:2024uuc,
    author = "Comberiati, Francesco and de la Cruz, Leonardo",
    title = "{Gravitational lensing in a plasma from worldlines}",
    eprint = "2412.14126",
    archivePrefix = "arXiv",
    primaryClass = "hep-th",
    doi = "10.1103/PhysRevD.111.054030",
    journal = "Phys. Rev. D",
    volume = "111",
    number = "5",
    pages = "054030",
    year = "2025"
}

@article{Cangemi:2023bpe,
    author = "Cangemi, Lucile and Chiodaroli, Marco and Johansson, Henrik and Ochirov, Alexander and Pichini, Paolo and Skvortsov, Evgeny",
    title = "{Compton Amplitude for Rotating Black Hole from QFT}",
    eprint = "2312.14913",
    archivePrefix = "arXiv",
    primaryClass = "hep-th",
    reportNumber = "UUITP--40/23, NORDITA 2023-117",
    doi = "10.1103/PhysRevLett.133.071601",
    journal = "Phys. Rev. Lett.",
    volume = "133",
    number = "7",
    pages = "071601",
    year = "2024"
}

@article{Ivanov:2024sds,
    author = "Ivanov, Mikhail M. and Li, Yue-Zhou and Parra-Martinez, Julio and Zhou, Zihan",
    title = "{Gravitational Raman Scattering in Effective Field Theory: A Scalar Tidal Matching at O(G3)}",
    eprint = "2401.08752",
    archivePrefix = "arXiv",
    primaryClass = "hep-th",
    reportNumber = "MIT-CTP/5664",
    doi = "10.1103/PhysRevLett.132.131401",
    journal = "Phys. Rev. Lett.",
    volume = "132",
    number = "13",
    pages = "131401",
    year = "2024",
    note = "[Erratum: Phys.Rev.Lett. 134, 159901 (2025)]"
}

@article{Correia:2024jgr,
    author = "Correia, Miguel and Isabella, Giulia",
    title = "{The Born regime of gravitational amplitudes}",
    eprint = "2406.13737",
    archivePrefix = "arXiv",
    primaryClass = "hep-th",
    doi = "10.1007/JHEP03(2025)144",
    journal = "JHEP",
    volume = "03",
    pages = "144",
    year = "2025"
}

@article{Caron-Huot:2025tlq,
    author = "Caron-Huot, Simon and Correia, Miguel and Isabella, Giulia and Solon, Mikhail",
    title = "{Gravitational Wave Scattering via the Born Series: Scalar Tidal Matching to O(G7) and Beyond}",
    eprint = "2503.13593",
    archivePrefix = "arXiv",
    primaryClass = "hep-th",
    doi = "10.1103/qd3c-nfz6",
    journal = "Phys. Rev. Lett.",
    volume = "135",
    number = "19",
    pages = "191601",
    year = "2025"
}

@article{Chen:2022clh,
    author = "Chen, Wei-Ming and Chung, Ming-Zhi and Huang, Yu-tin and Kim, Jung-Wook",
    title = "{Gravitational Faraday effect from on-shell amplitudes}",
    eprint = "2205.07305",
    archivePrefix = "arXiv",
    primaryClass = "hep-th",
    reportNumber = "KOBE-COSMO-22-04, QMUL-PH-22-16",
    doi = "10.1007/JHEP12(2022)058",
    journal = "JHEP",
    volume = "12",
    pages = "058",
    year = "2022"
}

@article{Bautista:2021wfy,
    author = "Bautista, Yilber Fabian and Guevara, Alfredo and Kavanagh, Chris and Vines, Justin",
    title = "{Scattering in black hole backgrounds and higher-spin amplitudes. Part I}",
    eprint = "2107.10179",
    archivePrefix = "arXiv",
    primaryClass = "hep-th",
    doi = "10.1007/JHEP03(2023)136",
    journal = "JHEP",
    volume = "03",
    pages = "136",
    year = "2023"
}

@article{Bautista:2022wjf,
    author = "Bautista, Yilber Fabian and Guevara, Alfredo and Kavanagh, Chris and Vines, Justin",
    title = "{Scattering in black hole backgrounds and higher-spin amplitudes. Part II}",
    eprint = "2212.07965",
    archivePrefix = "arXiv",
    primaryClass = "hep-th",
    doi = "10.1007/JHEP05(2023)211",
    journal = "JHEP",
    volume = "05",
    pages = "211",
    year = "2023"
}

@article{Bjerrum-Bohr:2025bqg,
    author = "Bjerrum-Bohr, N. Emil J. and Chen, Gang and Eriksen, Carl Jordan and Shah, Nabha",
    title = "{The gravitational Compton amplitude from flat and curved spacetimes at second post-Minkowskian order}",
    eprint = "2506.19705",
    archivePrefix = "arXiv",
    primaryClass = "hep-th",
    doi = "10.1007/JHEP10(2025)235",
    journal = "JHEP",
    volume = "10",
    pages = "235",
    year = "2025"
}

@article{Akpinar:2025byi,
    author = "Akpinar, Dogan",
    title = "{Scattering gravitons off general spinning compact objects to O(G2S4)}",
    eprint = "2511.10280",
    archivePrefix = "arXiv",
    primaryClass = "hep-th",
    doi = "10.1103/1zs7-kj4f",
    journal = "Phys. Rev. D",
    volume = "113",
    number = "4",
    pages = "045003",
    year = "2026"
}

@article{DiVecchia:2023frv,
    author = "Di Vecchia, Paolo and Heissenberg, Carlo and Russo, Rodolfo and Veneziano, Gabriele",
    title = "{The gravitational eikonal: From particle, string and brane collisions to black-hole encounters}",
    eprint = "2306.16488",
    archivePrefix = "arXiv",
    primaryClass = "hep-th",
    reportNumber = "CERN-TH-2023-108, NORDITA 2023-026, QMUL-PH-23-09, UUITP-14/23",
    doi = "10.1016/j.physrep.2024.06.002",
    journal = "Phys. Rept.",
    volume = "1083",
    pages = "1--169",
    year = "2024"
}

@article{KoemansCollado:2019ggb,
    author = "Koemans Collado, Arnau and Di Vecchia, Paolo and Russo, Rodolfo",
    title = "{Revisiting the second post-Minkowskian eikonal and the dynamics of binary black holes}",
    eprint = "1904.02667",
    archivePrefix = "arXiv",
    primaryClass = "hep-th",
    reportNumber = "QMUL-PH-19-08",
    doi = "10.1103/PhysRevD.100.066028",
    journal = "Phys. Rev. D",
    volume = "100",
    number = "6",
    pages = "066028",
    year = "2019"
}

@article{Aoude:2022thd,
    author = "Aoude, Rafael and Haddad, Kays and Helset, Andreas",
    title = "{Classical Gravitational Spinning-Spinless Scattering at O(G2S{\ensuremath{\infty}})}",
    eprint = "2205.02809",
    archivePrefix = "arXiv",
    primaryClass = "hep-th",
    reportNumber = "CP3-22-32, UUITP-24/22, CALT-TH-2022-018",
    doi = "10.1103/PhysRevLett.129.141102",
    journal = "Phys. Rev. Lett.",
    volume = "129",
    number = "14",
    pages = "141102",
    year = "2022"
}

@article{Chiodaroli:2021eug,
    author = "Chiodaroli, Marco and Johansson, Henrik and Pichini, Paolo",
    title = "{Compton black-hole scattering for s {\ensuremath{\leq}} 5/2}",
    eprint = "2107.14779",
    archivePrefix = "arXiv",
    primaryClass = "hep-th",
    reportNumber = "UUITP-34/21, NORDITA 2021-013",
    doi = "10.1007/JHEP02(2022)156",
    journal = "JHEP",
    volume = "02",
    pages = "156",
    year = "2022"
}

@article{Bellazzini:2022wzv,
    author = "Bellazzini, Brando and Isabella, Giulia and Riva, Massimiliano Maria",
    title = "{Classical vs quantum eikonal scattering and its causal structure}",
    eprint = "2211.00085",
    archivePrefix = "arXiv",
    primaryClass = "hep-th",
    doi = "10.1007/JHEP04(2023)023",
    journal = "JHEP",
    volume = "04",
    pages = "023",
    year = "2023"
}

@article{Bini:2025ltr,
    author = "Bini, Donato and Di Russo, Giorgio",
    title = "{Topological stars and scalar wave equation: Exact resummation of the renormalized angular momentum in the eikonal limit}",
    eprint = "2506.14442",
    archivePrefix = "arXiv",
    primaryClass = "gr-qc",
    doi = "10.1103/dw5y-4pv8",
    journal = "Phys. Rev. D",
    volume = "112",
    number = "6",
    pages = "064008",
    year = "2025"
}

@article{Bini:2025bll,
    author = "Bini, Donato and Di Russo, Giorgio and Geralico, Andrea",
    title = "{Kerr spacetime and scalar wave equation: Exact resummation of the renormalized angular momentum in the eikonal limit}",
    eprint = "2508.12046",
    archivePrefix = "arXiv",
    primaryClass = "gr-qc",
    doi = "10.1103/mzqw-wbvf",
    journal = "Phys. Rev. D",
    volume = "112",
    number = "6",
    pages = "064077",
    year = "2025"
}

@article{Aoude:2022trd,
    author = "Aoude, Rafael and Haddad, Kays and Helset, Andreas",
    title = "{Searching for Kerr in the 2PM amplitude}",
    eprint = "2203.06197",
    archivePrefix = "arXiv",
    primaryClass = "hep-th",
    reportNumber = "CP3-22-15, UUITP-12/22, CALT-TH-2022-009",
    doi = "10.1007/JHEP07(2022)072",
    journal = "JHEP",
    volume = "07",
    pages = "072",
    year = "2022"
}

@book{Newton:1982qc,
  author    = {Newton, Roger G.},
  title     = {Scattering Theory of Waves and Particles},
  year      = {1982},
  publisher = {Springer-Verlag},
  address   = {New York, U.S.A},
  series    = {Theoretical and Mathematical Physics},
}

@book{1988sfbh.book.....F,
    author = {{Futterman}, J.~A.~H. and {Handler}, F.~A. and {Matzner}, R.~A.},
    title = "{Scattering from black holes}",
    year = 1988,
    publisher = "Cambridge University Press"
}

@article{Ivanov:2025ozg,
    author = "Ivanov, Mikhail M. and Li, Yue-Zhou and Parra-Martinez, Julio and Zhou, Zihan",
    title = "{Resummation of Universal Tails in Gravitational Waveforms}",
    eprint = "2504.07862",
    archivePrefix = "arXiv",
    primaryClass = "hep-th",
    reportNumber = "MIT-CTP/5863",
    doi = "10.1103/jzd1-qzkt",
    journal = "Phys. Rev. Lett.",
    volume = "135",
    number = "14",
    pages = "141401",
    year = "2025"
}

@incollection{Glauber1959,
  author    = {Glauber, R. J.},
  title     = {High Energy Collision Theory},
  booktitle = {Lectures in Theoretical Physics},
  editor    = {Brittin, W. E. and Dunham, L. G.},
  volume    = {1},
  pages     = {315--414},
  publisher = {Wiley-Interscience},
  address   = {New York},
  year      = {1959}
}

@article{Wallace1973,
title = {Eikonal expansion},
journal = {Annals of Physics},
volume = {78},
number = {1},
pages = {190-257},
year = {1973},
issn = {0003-4916},
doi = {https://doi.org/10.1016/0003-4916(73)90008-0},
url = {https://www.sciencedirect.com/science/article/pii/0003491673900080},
author = {Stephen J Wallace}
}

@article{Swift1974,
  title = {Eikonal expansion as the high-energy limit of the Born series},
  author = {Swift, Arthur R.},
  journal = {Phys. Rev. D},
  volume = {9},
  issue = {6},
  pages = {1740--1749},
  numpages = {0},
  year = {1974},
  month = {Mar},
  publisher = {American Physical Society},
  doi = {10.1103/PhysRevD.9.1740},
  url = {https://link.aps.org/doi/10.1103/PhysRevD.9.1740}
}

@article{Levy1969,
  title = {Eikonal Approximation in Quantum Field Theory},
  author = {L\'evy, Maurice and Sucher, Joseph},
  journal = {Phys. Rev.},
  volume = {186},
  issue = {5},
  pages = {1656--1670},
  numpages = {0},
  year = {1969},
  month = {Oct},
  publisher = {American Physical Society},
  doi = {10.1103/PhysRev.186.1656},
  url = {https://link.aps.org/doi/10.1103/PhysRev.186.1656}
}

@article{Chen1984,
  title = {Combining the Born approximation with the eikonal method},
  author = {Chen, T. W.},
  journal = {Phys. Rev. D},
  volume = {29},
  issue = {8},
  pages = {1839--1841},
  numpages = {0},
  year = {1984},
  month = {Apr},
  publisher = {American Physical Society},
  doi = {10.1103/PhysRevD.29.1839},
  url = {https://link.aps.org/doi/10.1103/PhysRevD.29.1839}
}

@article{Byron1973,
  title = {Potential Scattering in the Eikonal Approximation},
  author = {Byron, F. W. and Joachain, C. J. and Mund, E. H.},
  journal = {Phys. Rev. D},
  volume = {8},
  issue = {8},
  pages = {2622--2639},
  numpages = {0},
  year = {1973},
  month = {Oct},
  publisher = {American Physical Society},
  doi = {10.1103/PhysRevD.8.2622},
  url = {https://link.aps.org/doi/10.1103/PhysRevD.8.2622}
}

@article{AccettulliHuber:2020oou,
    author = "Accettulli Huber, Manuel and Brandhuber, Andreas and De Angelis, Stefano and Travaglini, Gabriele",
    title = "{Eikonal phase matrix, deflection angle and time delay in effective field theories of gravity}",
    eprint = "2006.02375",
    archivePrefix = "arXiv",
    primaryClass = "hep-th",
    reportNumber = "QMUL-PH-20-05, SAGEX-20-04-E",
    doi = "10.1103/PhysRevD.102.046014",
    journal = "Phys. Rev. D",
    volume = "102",
    number = "4",
    pages = "046014",
    year = "2020"
}

@article{Nie:2024pby,
    author = "Nie, Wen-Kai and Tan, Lin-Tao and Zhang, Jun and Zhou, Shuang-Yong",
    title = "{Scalar-Gauss-Bonnet gravity: infrared causality and detectability of GW observations}",
    eprint = "2410.10973",
    archivePrefix = "arXiv",
    primaryClass = "hep-th",
    reportNumber = "USTC-ICTS/PCFT-24-35",
    doi = "10.1088/1475-7516/2025/08/086",
    journal = "JCAP",
    volume = "08",
    pages = "086",
    year = "2025"
}

@article{Brandhuber:2024bnz,
    author = "Brandhuber, Andreas and Brown, Graham R. and Pichini, Paolo and Travaglini, Gabriele and Vives Matasan, Pablo",
    title = "{Spinning binary dynamics in cubic effective field theories of gravity}",
    eprint = "2405.13826",
    archivePrefix = "arXiv",
    primaryClass = "hep-th",
    reportNumber = "QMUL-PH-24-09",
    doi = "10.1007/JHEP08(2024)188",
    journal = "JHEP",
    volume = "08",
    pages = "188",
    year = "2024"
}

@article{Brandhuber:2024qdn,
    author = "Brandhuber, Andreas and Brown, Graham R. and Chen, Gang and Travaglini, Gabriele and Vives Matasan, Pablo",
    title = "{Spinning waveforms in cubic effective field theories of gravity}",
    eprint = "2408.00587",
    archivePrefix = "arXiv",
    primaryClass = "hep-th",
    reportNumber = "QMUL-PH-24-11",
    doi = "10.1007/JHEP12(2024)039",
    journal = "JHEP",
    volume = "12",
    pages = "039",
    year = "2024"
}

@article{Carrillo-Gonzalez:2021mqj,
    author = "Carrillo-Gonz{\'a}lez, Mariana and de Rham, Claudia and Tolley, Andrew J.",
    title = "{Scattering amplitudes for binary systems beyond GR}",
    eprint = "2107.11384",
    archivePrefix = "arXiv",
    primaryClass = "hep-th",
    reportNumber = "Imperial/TP/2021/MC/02",
    doi = "10.1007/JHEP11(2021)087",
    journal = "JHEP",
    volume = "11",
    pages = "087",
    year = "2021"
}

@article{Dalang:2021qhu,
    author = "Dalang, Charles and Cusin, Giulia and Lagos, Macarena",
    title = "{Polarization distortions of lensed gravitational waves}",
    eprint = "2104.10119",
    archivePrefix = "arXiv",
    primaryClass = "gr-qc",
    doi = "10.1103/PhysRevD.105.024005",
    journal = "Phys. Rev. D",
    volume = "105",
    number = "2",
    pages = "024005",
    year = "2022"
}

@article{Isaacson:1968hbi,
    author = "Isaacson, Richard A.",
    title = "{Gravitational Radiation in the Limit of High Frequency. I. The Linear Approximation and Geometrical Optics}",
    doi = "10.1103/PhysRev.166.1263",
    journal = "Phys. Rev.",
    volume = "166",
    pages = "1263--1271",
    year = "1968"
}

@article{Isaacson:1968zza,
    author = "Isaacson, Richard A.",
    title = "{Gravitational Radiation in the Limit of High Frequency. II. Nonlinear Terms and the Ef fective Stress Tensor}",
    doi = "10.1103/PhysRev.166.1272",
    journal = "Phys. Rev.",
    volume = "166",
    pages = "1272--1279",
    year = "1968"
}

@article{Bertacca:2017vod,
    author = "Bertacca, Daniele and Raccanelli, Alvise and Bartolo, Nicola and Matarrese, Sabino",
    title = "{Cosmological perturbation effects on gravitational-wave luminosity distance estimates}",
    eprint = "1702.01750",
    archivePrefix = "arXiv",
    primaryClass = "gr-qc",
    doi = "10.1016/j.dark.2018.03.001",
    journal = "Phys. Dark Univ.",
    volume = "20",
    pages = "32--40",
    year = "2018"
}

@article{Laguna:2009re,
    author = "Laguna, Pablo and Larson, Shane L. and Spergel, David and Yunes, Nicolas",
    title = "{Integrated Sachs-Wolfe Effect for Gravitational Radiation}",
    eprint = "0905.1908",
    archivePrefix = "arXiv",
    primaryClass = "gr-qc",
    doi = "10.1088/2041-8205/715/1/L12",
    journal = "Astrophys. J. Lett.",
    volume = "715",
    pages = "L12",
    year = "2010"
}

@article{Ezquiaga:2020dao,
    author = "Ezquiaga, Jose Mar{\'\i}a and Zumalac{\'a}rregui, Miguel",
    title = "{Gravitational wave lensing beyond general relativity: birefringence, echoes and shadows}",
    eprint = "2009.12187",
    archivePrefix = "arXiv",
    primaryClass = "gr-qc",
    doi = "10.1103/PhysRevD.102.124048",
    journal = "Phys. Rev. D",
    volume = "102",
    number = "12",
    pages = "124048",
    year = "2020"
}

@article{Streibert:2024cuf,
    author = "Streibert, Julius and Silva, Hector O. and Zumalac{\'a}rregui, Miguel",
    title = "{Eikonal gravitational-wave lensing in Einstein-aether theory}",
    eprint = "2404.07782",
    archivePrefix = "arXiv",
    primaryClass = "gr-qc",
    doi = "10.1103/jrsp-1xll",
    journal = "Phys. Rev. D",
    volume = "112",
    number = "2",
    pages = "024073",
    year = "2025"
}

@article{Menadeo:2024uoq,
    author = "Menadeo, Nicola and Zumalac{\'a}rregui, Miguel",
    title = "{Gravitational wave propagation beyond general relativity: Geometric optic expansion and lens-induced dispersion}",
    eprint = "2411.07164",
    archivePrefix = "arXiv",
    primaryClass = "gr-qc",
    doi = "10.1103/PhysRevD.111.104022",
    journal = "Phys. Rev. D",
    volume = "111",
    number = "10",
    pages = "104022",
    year = "2025"
}

@article{Garoffolo:2019mna,
    author = "Garoffolo, Alice and Tasinato, Gianmassimo and Carbone, Carmelita and Bertacca, Daniele and Matarrese, Sabino",
    title = "{Gravitational waves and geometrical optics in scalar-tensor theories}",
    eprint = "1912.08093",
    archivePrefix = "arXiv",
    primaryClass = "gr-qc",
    doi = "10.1088/1475-7516/2020/11/040",
    journal = "JCAP",
    volume = "11",
    pages = "040",
    year = "2020"
}

@article{Flanagan:2008kz,
    author = "Flanagan, Eanna E. and Rosenthal, Eran and Wasserman, Ira M.",
    title = "{Modification to the Luminosity Distance Redshift Relation in Modified Gravity Theories}",
    eprint = "0810.0535",
    archivePrefix = "arXiv",
    primaryClass = "astro-ph",
    doi = "10.1103/PhysRevD.79.044032",
    journal = "Phys. Rev. D",
    volume = "79",
    pages = "044032",
    year = "2009"
}

@article{Balaudo:2022znx,
    author = "Balaudo, Anna and Garoffolo, Alice and Martinelli, Matteo and Mukherjee, Suvodip and Silvestri, Alessandra",
    title = "{Prospects of testing late-time cosmology with weak lensing of gravitational waves and galaxy surveys}",
    eprint = "2210.06398",
    archivePrefix = "arXiv",
    primaryClass = "astro-ph.CO",
    doi = "10.1088/1475-7516/2023/06/050",
    journal = "JCAP",
    volume = "06",
    pages = "050",
    year = "2023"
}

@article{Balaudo:2023klo,
    author = "Balaudo, Anna and Pantiri, Mattia and Silvestri, Alessandra",
    title = "{Number count of gravitational waves and supernovae in luminosity distance space for $\Lambda$CDM and scalar-tensor theories}",
    eprint = "2311.17904",
    archivePrefix = "arXiv",
    primaryClass = "astro-ph.CO",
    doi = "10.1088/1475-7516/2024/02/023",
    journal = "JCAP",
    volume = "02",
    pages = "023",
    year = "2024"
}

@article{Garoffolo:2020vtd,
    author = "Garoffolo, Alice and Raveri, Marco and Silvestri, Alessandra and Tasinato, Gianmassimo and Carbone, Carmelita and Bertacca, Daniele and Matarrese, Sabino",
    title = "{Detecting Dark Energy Fluctuations with Gravitational Waves}",
    eprint = "2007.13722",
    archivePrefix = "arXiv",
    primaryClass = "astro-ph.CO",
    doi = "10.1103/PhysRevD.103.083506",
    journal = "Phys. Rev. D",
    volume = "103",
    number = "8",
    pages = "083506",
    year = "2021"
}

@article{Dalang:2019rke,
    author = "Dalang, Charles and Fleury, Pierre and Lombriser, Lucas",
    title = "{Horndeski gravity and standard sirens}",
    eprint = "1912.06117",
    archivePrefix = "arXiv",
    primaryClass = "gr-qc",
    reportNumber = "IFT-UAM/CSIC-168",
    doi = "10.1103/PhysRevD.102.044036",
    journal = "Phys. Rev. D",
    volume = "102",
    number = "4",
    pages = "044036",
    year = "2020"
}

@article{Dalang:2020eaj,
    author = "Dalang, Charles and Fleury, Pierre and Lombriser, Lucas",
    title = "{Scalar and tensor gravitational waves}",
    eprint = "2009.11827",
    archivePrefix = "arXiv",
    primaryClass = "gr-qc",
    reportNumber = "IFT-UAM/CSIC-20-136",
    doi = "10.1103/PhysRevD.103.064075",
    journal = "Phys. Rev. D",
    volume = "103",
    number = "6",
    pages = "064075",
    year = "2021"
}

@article{Tasinato:2021wol,
    author = "Tasinato, Gianmassimo and Garoffolo, Alice and Bertacca, Daniele and Matarrese, Sabino",
    title = "{Gravitational-wave cosmological distances in scalar-tensor theories of gravity}",
    eprint = "2103.00155",
    archivePrefix = "arXiv",
    primaryClass = "gr-qc",
    doi = "10.1088/1475-7516/2021/06/050",
    journal = "JCAP",
    volume = "06",
    pages = "050",
    year = "2021"
}

@article{Mpetha:2022xqo,
    author = "Mpetha, Charlie T. and Congedo, Giuseppe and Taylor, Andy",
    title = "{Future prospects on testing extensions to {\ensuremath{\Lambda}}CDM through the weak lensing of gravitational waves}",
    eprint = "2208.05959",
    archivePrefix = "arXiv",
    primaryClass = "astro-ph.CO",
    doi = "10.1103/PhysRevD.107.103518",
    journal = "Phys. Rev. D",
    volume = "107",
    number = "10",
    pages = "103518",
    year = "2023"
}

@article{Robertson:2020mfh,
    author = "Robertson, Andrew and Smith, Graham P. and Massey, Richard and Eke, Vincent and Jauzac, Mathilde and Bianconi, Matteo and Ryczanowski, Dan",
    title = "{What does strong gravitational lensing? The mass and redshift distribution of high-magnification lenses}",
    eprint = "2002.01479",
    archivePrefix = "arXiv",
    primaryClass = "astro-ph.CO",
    doi = "10.1093/mnras/staa1429",
    journal = "Mon. Not. Roy. Astron. Soc.",
    volume = "495",
    number = "4",
    pages = "3727--3739",
    year = "2020"
}

@ARTICLE{Love1909,
       author = {{Love}, A.~E.~H.},
        title = "{Earth, the yielding of the, to disturbing forces}",
      journal = {Mont. Not. Roy. Astr. Soc.},
         year = 1909,
        month = apr,
       volume = {69},
        pages = {476},
          doi = {10.1093/mnras/69.6.476},
       adsurl = {https://ui.adsabs.harvard.edu/abs/1909MNRAS..69..476L}
}

@article{Charalambous:2022rre,
    author = "Charalambous, Panagiotis and Dubovsky, Sergei and Ivanov, Mikhail M.",
    title = "{Love symmetry}",
    eprint = "2209.02091",
    archivePrefix = "arXiv",
    primaryClass = "hep-th",
    doi = "10.1007/JHEP10(2022)175",
    journal = "JHEP",
    volume = "10",
    pages = "175",
    year = "2022"
}

@article{DeLuca:2024ufn,
    author = "De Luca, Valerio and Garoffolo, Alice and Khoury, Justin and Trodden, Mark",
    title = "{Tidal Love numbers and Green{\textquoteright}s functions in black hole spacetimes}",
    eprint = "2407.07156",
    archivePrefix = "arXiv",
    primaryClass = "gr-qc",
    doi = "10.1103/PhysRevD.110.064081",
    journal = "Phys. Rev. D",
    volume = "110",
    number = "6",
    pages = "064081",
    year = "2024"
}

@article{Hui:2020xxx,
    author = "Hui, Lam and Joyce, Austin and Penco, Riccardo and Santoni, Luca and Solomon, Adam R.",
    title = "{Static response and Love numbers of Schwarzschild black holes}",
    eprint = "2010.00593",
    archivePrefix = "arXiv",
    primaryClass = "hep-th",
    doi = "10.1088/1475-7516/2021/04/052",
    journal = "JCAP",
    volume = "04",
    pages = "052",
    year = "2021"
}

@article{Berens:2025okm,
    author = "Berens, Roman and Hui, Lam and McLoughlin, Daniel and Penco, Riccardo and Staunton, John",
    title = "{Geometric Symmetries for the Vanishing of the Black Hole Tidal Love Numbers}",
    eprint = "2510.18952",
    archivePrefix = "arXiv",
    primaryClass = "hep-th",
    month = "10",
    year = "2025"
}

@article{Flanagan:2007ix,
    author = "Flanagan, Eanna E. and Hinderer, Tanja",
    title = "{Constraining neutron star tidal Love numbers with gravitational wave detectors}",
    eprint = "0709.1915",
    archivePrefix = "arXiv",
    primaryClass = "astro-ph",
    doi = "10.1103/PhysRevD.77.021502",
    journal = "Phys. Rev. D",
    volume = "77",
    pages = "021502",
    year = "2008"
}

@article{Hinderer:2007mb,
    author = "Hinderer, Tanja",
    title = "{Tidal Love numbers of neutron stars}",
    eprint = "0711.2420",
    archivePrefix = "arXiv",
    primaryClass = "astro-ph",
    doi = "10.1086/533487",
    journal = "Astrophys. J.",
    volume = "677",
    pages = "1216--1220",
    year = "2008",
    note = "[Erratum: Astrophys.J. 697, 964 (2009)]"
}

@article{Binnington:2009bb,
    author = "Binnington, Taylor and Poisson, Eric",
    title = "{Relativistic theory of tidal Love numbers}",
    eprint = "0906.1366",
    archivePrefix = "arXiv",
    primaryClass = "gr-qc",
    doi = "10.1103/PhysRevD.80.084018",
    journal = "Phys. Rev. D",
    volume = "80",
    pages = "084018",
    year = "2009"
}

@article{Damour:2009vw,
    author = "Damour, Thibault and Nagar, Alessandro",
    title = "{Relativistic tidal properties of neutron stars}",
    eprint = "0906.0096",
    archivePrefix = "arXiv",
    primaryClass = "gr-qc",
    doi = "10.1103/PhysRevD.80.084035",
    journal = "Phys. Rev. D",
    volume = "80",
    pages = "084035",
    year = "2009"
}

@article{Pani:2015hfa,
    author = "Pani, Paolo and Gualtieri, Leonardo and Maselli, Andrea and Ferrari, Valeria",
    title = "{Tidal deformations of a spinning compact object}",
    eprint = "1503.07365",
    archivePrefix = "arXiv",
    primaryClass = "gr-qc",
    doi = "10.1103/PhysRevD.92.024010",
    journal = "Phys. Rev. D",
    volume = "92",
    number = "2",
    pages = "024010",
    year = "2015"
}

@article{Porto:2016zng,
    author = "Porto, Rafael A.",
    title = "{The Tune of Love and the Nature(ness) of Spacetime}",
    eprint = "1606.08895",
    archivePrefix = "arXiv",
    primaryClass = "gr-qc",
    doi = "10.1002/prop.201600064",
    journal = "Fortsch. Phys.",
    volume = "64",
    number = "10",
    pages = "723--729",
    year = "2016"
}

@article{LeTiec:2020spy,
    author = "Le Tiec, Alexandre and Casals, Marc",
    title = "{Spinning Black Holes Fall in Love}",
    eprint = "2007.00214",
    archivePrefix = "arXiv",
    primaryClass = "gr-qc",
    doi = "10.1103/PhysRevLett.126.131102",
    journal = "Phys. Rev. Lett.",
    volume = "126",
    number = "13",
    pages = "131102",
    year = "2021"
}

@article{Chia:2020yla,
    author = "Chia, Horng Sheng",
    title = "{Tidal deformation and dissipation of rotating black holes}",
    eprint = "2010.07300",
    archivePrefix = "arXiv",
    primaryClass = "gr-qc",
    doi = "10.1103/PhysRevD.104.024013",
    journal = "Phys. Rev. D",
    volume = "104",
    number = "2",
    pages = "024013",
    year = "2021"
}

@article{Charalambous:2021mea,
    author = "Charalambous, Panagiotis and Dubovsky, Sergei and Ivanov, Mikhail M.",
    title = "{On the Vanishing of Love Numbers for Kerr Black Holes}",
    eprint = "2102.08917",
    archivePrefix = "arXiv",
    primaryClass = "hep-th",
    reportNumber = "INR-TH-2021-001",
    doi = "10.1007/JHEP05(2021)038",
    journal = "JHEP",
    volume = "05",
    pages = "038",
    year = "2021"
}

@article{Charalambous:2021kcz,
    author = "Charalambous, Panagiotis and Dubovsky, Sergei and Ivanov, Mikhail M.",
    title = "{Hidden Symmetry of Vanishing Love Numbers}",
    eprint = "2103.01234",
    archivePrefix = "arXiv",
    primaryClass = "hep-th",
    reportNumber = "INR-TH-2021-003",
    doi = "10.1103/PhysRevLett.127.101101",
    journal = "Phys. Rev. Lett.",
    volume = "127",
    number = "10",
    pages = "101101",
    year = "2021"
}

@article{Berens:2022ebl,
    author = "Berens, Roman and Hui, Lam and Sun, Zimo",
    title = "{Ladder symmetries of black holes and de Sitter space: love numbers and quasinormal modes}",
    eprint = "2212.09367",
    archivePrefix = "arXiv",
    primaryClass = "hep-th",
    doi = "10.1088/1475-7516/2023/06/056",
    journal = "JCAP",
    volume = "06",
    pages = "056",
    year = "2023"
}

@article{Rai:2024lho,
    author = "Rai, Mudit and Santoni, Luca",
    title = {{Ladder symmetries and Love numbers of Reissner-Nordstr{\"o}m black holes}},
    eprint = "2404.06544",
    archivePrefix = "arXiv",
    primaryClass = "gr-qc",
    doi = "10.1007/JHEP07(2024)098",
    journal = "JHEP",
    volume = "07",
    pages = "098",
    year = "2024"
}

@article{Riva:2023rcm,
    author = "Riva, Massimiliano Maria and Santoni, Luca and Savi{\'c}, Nikola and Vernizzi, Filippo",
    title = "{Vanishing of nonlinear tidal Love numbers of Schwarzschild black holes}",
    eprint = "2312.05065",
    archivePrefix = "arXiv",
    primaryClass = "gr-qc",
    doi = "10.1016/j.physletb.2024.138710",
    journal = "Phys. Lett. B",
    volume = "854",
    pages = "138710",
    year = "2024"
}

@article{Cardoso:2018ptl,
    author = "Cardoso, Vitor and Kimura, Masashi and Maselli, Andrea and Senatore, Leonardo",
    title = "{Black Holes in an Effective Field Theory Extension of General Relativity}",
    eprint = "1808.08962",
    archivePrefix = "arXiv",
    primaryClass = "gr-qc",
    doi = "10.1103/PhysRevLett.121.251105",
    journal = "Phys. Rev. Lett.",
    volume = "121",
    number = "25",
    pages = "251105",
    year = "2018",
    note = "[Erratum: Phys.Rev.Lett. 131, 109903 (2023)]"
}

@article{Cardoso:2019vof,
    author = "Cardoso, Vitor and Gualtieri, Leonardo and Moore, Christopher J.",
    title = "{Gravitational waves and higher dimensions: Love numbers and Kaluza-Klein excitations}",
    eprint = "1910.09557",
    archivePrefix = "arXiv",
    primaryClass = "gr-qc",
    doi = "10.1103/PhysRevD.100.124037",
    journal = "Phys. Rev. D",
    volume = "100",
    number = "12",
    pages = "124037",
    year = "2019"
}

@article{DeLuca:2021ite,
    author = "De Luca, Valerio and Pani, Paolo",
    title = "{Tidal deformability of dressed black holes and tests of ultralight bosons in extended mass ranges}",
    eprint = "2106.14428",
    archivePrefix = "arXiv",
    primaryClass = "gr-qc",
    doi = "10.1088/1475-7516/2021/08/032",
    journal = "JCAP",
    volume = "08",
    pages = "032",
    year = "2021"
}

@article{DeLuca:2023mio,
    author = "De Luca, Valerio and Khoury, Justin and Wong, Sam S. C.",
    title = "{Nonlinearities in the tidal Love numbers of black holes}",
    eprint = "2305.14444",
    archivePrefix = "arXiv",
    primaryClass = "gr-qc",
    doi = "10.1103/PhysRevD.108.024048",
    journal = "Phys. Rev. D",
    volume = "108",
    number = "2",
    pages = "024048",
    year = "2023"
}

@book{poisson_will_2014, place={Cambridge}, title={Gravity: Newtonian, Post-Newtonian, Relativistic}, DOI={10.1017/CBO9781139507486}, publisher={Cambridge University Press}, author={Poisson, Eric and Will, Clifford M.}, year={2014}}

\end{document}